\documentclass{elsart}

\usepackage{times}
\usepackage[T1]{fontenc}
\usepackage[latin1]{inputenc}
\usepackage{amsmath}
\usepackage{amssymb}
\usepackage{amsfonts}
\usepackage{stmaryrd}
\usepackage{enumerate}
\usepackage{txfonts}
\usepackage{pxfonts}
\usepackage{epsfig}
\usepackage{color}
\usepackage{array}
\usepackage{bussproofs}
\usepackage{xypic}
\xyoption{all}
\usepackage{natbib}

\newcommand{\mc}{\mathcal}
\newcommand{\restr}{\upharpoonright}
\newcommand{\fl}{\to}

\newcommand{\fref}{\textit{ref}}
\newcommand{\und}{\underline}
\newcommand{\ov}{\overline}
\newcommand{\eq}{\simeq_\varepsilon}
\newcommand{\reec}{\Rightarrow}
\newcommand{\cqfd}{\hfill\ensuremath{\Box}}
\newcommand{\finex}{\hfill\ensuremath{\diamond}}
\newcommand{\bec}{\begin{center}}
\newcommand{\eec}{\end{center}}
\newcommand{\btab}{\begin{tabular}}
\newcommand{\etab}{\end{tabular}}
\newcommand{\GG}{\mc{G}}
\newcommand{\LL}{\llbracket}
\newcommand{\RR}{\rrbracket}
\newcommand{\trad}[1]{\LL#1\RR}
\newcommand{\XX}{\vec{X}}
\newcommand{\DD}{\mc{D}}
\newcommand{\RRR}{\mc{R}}
\newcommand{\id}{\textit{id}}
\newcommand{\ppar}{\invamp}

\newcommand{\ie}{i.e.}
\newcommand{\quant}{\textit{quant}}

\newcommand{\rank}{\textit{rank}}
\newcommand{\var}{\textit{var}}
\newcommand{\origin}{\textit{origin}}
\newcommand{\ld}{\textit{ld}}

\newcommand{\hh}{\mathbf{H}}
\newcommand{\hhh}{|\mathbf{H}|}

\newcommand{\barre}{\ | \ }
\newcommand{\barreD}{\ | \ \Delta}
\newcommand{\debinf}{\XX;\Gamma\vdash}
\newcommand{\flt}[1]{\xrightarrow{#1}}
\newcommand{\lmc}{\lambda\mu2}

\makeatletter
\renewcommand\paragraph{\@startsection{paragraph}{4}{\z@}%
                                      {1.25ex \@plus.5ex \@minus.2ex}%
                                      {-1em}%
                                      {\normalfont\normalsize\bfseries}}

\makeatother

\newenvironment{minilist}{\begin{list}{$\bullet$}{
  \setlength{\parsep}{0pt}
  \setlength{\topsep}{0pt}
  \setlength{\itemsep}{-\parsep}
  \setlength{\labelsep}{0.5em}}}{\end{list}}

\begin{document}


\newtheorem{Def}{Definition}
\newtheorem{Th}{Theorem}
\newtheorem{Ex}{Example}
\newtheorem{Prop}{Proposition}
\newtheorem{Lemme}{Lemma}
\newtheorem{Corollaire}{Corollary}

\begin{frontmatter}
  \title{Second-order Type Isomorphisms through Game Semantics}
  \author{Joachim de Lataillade}
  \address{Équipe Preuves, Programmes et Systèmes\\
    CNRS -- Université Paris 7 Denis Diderot}
  \begin{abstract}

The characterization of second-order type isomorphisms is a purely syntactical
problem that we propose to study under
the enlightenment of game semantics. We study this question in the case of second-order $\lambda\mu$-calculus, which can be seen as
an extension of system F to classical logic, and for which we define a
categorical framework: control hyperdoctrines.

Our game model of $\lambda\mu$-calculus is based on polymorphic arenas 
(closely related to Hughes' hyperforests) which evolve during the play 
(following the ideas of Murawski-Ong).
We show that type isomorphisms coincide with the "equality" on 
arenas associated with types. Finally we deduce the equational
characterization of type isomorphisms from this equality.
We also recover from the same model Roberto Di Cosmo's characterization of type isomorphisms for system F.

This approach leads to a geometrical comprehension on the question of second order type isomorphisms, which can be easily extended to
some other polymorphic calculi including additional programming features.

  \end{abstract}

  \begin{keyword}
    Types Isomorphisms, Second-order $\lambda\mu$-calculus, Game Semantics, Hyperdoctrines, Control Categories
  \end{keyword}
\end{frontmatter}

\maketitle
\newpage

\section{Introduction}

\textbf{Denotational semantics}\qquad Defining a semantic for a language is a fundamental tool for understanding the way this language works. 
Thus, semantics is a very active domain of research in theoretical computer science: in particular, there has been an important 
investigation on semantics which could modelize a language as precisely as possible; this has led to the emergence of game 
semantics in the early 90s, whose success is due to the deep adequation of its models with the syntax.
The present work illustrates the ability of game semantics to modelize a language precisely: consequently, 
it is possible to extract from the model some properties of the language. So, this work has to be understood as an example of 
accomplishment of the original
goal of denotational semantics: using abstract tools to prove concrete properties
on a programming language. In this article, the property we extract concerns a 
non-trivial problem, the characterization of type isomorphisms for second-order languages.

\textbf{Type isomorphisms.}\quad The problem of type isomorphisms is a
syntactical question: two types $A$ and $B$ are isomorphic
($A\simeq B$) if there exist two terms $f:A\fl B$ and $g:B\fl A$
such that $f\circ g=id_B$ and $g\circ f=id_A$. This equivalence
relation on data types allows to translate a program from one type
to the other without any change on the calculatory meaning of the
program. Thus, a search in a library up to type isomorphism will
help the programmer to find all the functions that can potentially
serve his purpose, and to reuse them in the new typing
context~\cite{rittri}. This is particularly appealing with
functional languages, because in this case the type can really be
seen as a partial specification of the program: such a library
search up to isomorphisms has been implemented in particular for Caml
Light by Jérôme Vouillon. It can also be used in proof assistants
to help finding proofs in libraries and reusing them~\cite{isoreusedep} (for more
details on the use of type isomorphisms in computer science,
see~\cite{isotypes}).

\begin{figure}
\epsfig{file=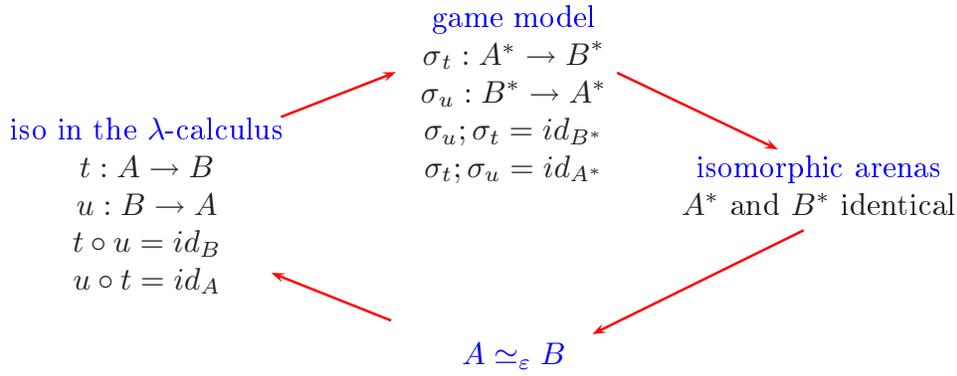,width=5cm}
\caption{Steps of the proof of Olivier Laurent in his work on type isomorphisms}
\label{prooflaurent}
\end{figure}

When dealing with type isomorphisms, the key problem, given a
programming language, is to find a characterization of isomorphic
types through an equational system. This can be done either
syntactically (by working directly on terms) or semantically (by using
an adequate model of the calculus, \ie\ such that there are no more
isomorphisms in the model than in the language).  For the
$\lambda$-calculus, the problem has been solved semantically as early
as in 1981~\cite{isosoloviev}, but Olivier Laurent has recently
proposed a new approach based on game semantics~\cite{classisos}:
taking the usual HON game model for $\lambda$-calculus (which we call
the \textit{propositional game model}), he proved that the equality
modulo isomorphism in the syntax corresponds to the notion of equality
between forests, and proved the equational characterization of
isomorphisms by this means. The main steps of his proof are summed up
on figure~\ref{prooflaurent}.  The advantage of this point of view is
that it immediately gave him a characterization of type isomorphisms
for the $\lambda\mu$-calculus, requiring no additional work.

\textbf{$\lambda\mu2$-calculus.}\quad The calculus we consider in this work is the call-by-name disjunctive second-order 
$\lambda\mu$-calculus (shortly, $\lambda\mu2$).
The $\lambda\mu$-calculus has been introduced by Parigot~\cite{lambdamu} as a way to add the notion of \textit{control} to 
$\lambda$-calculus, and hence to associate a calculus to classical logic. There are a call-by-value version and a call-by-name version 
of this calculus, that Peter Selinger proved to be isomorphic one to the other~\cite{selinger}. The $\lambda\mu2$-calculus is just an 
extension of this calculus to second order: here we will consider a Church-style presentation of second-order terms.

As far as we know, the characterization of type isomorphisms for $\lambda\mu2$ has not been done yet. However, using the results of 
Roberto Di Cosmo concerning system F~\cite{isotypes} and of Olivier Laurent concerning the $\lambda\mu$-calculus~\cite{classisos}, one
can suggest that the equational system that characterizes type isomorphisms for $\lambda\mu2$ is the system $\eq$ 
presented on figure~\ref{isolm2}
(note that we have now an equality corresponding the interaction between $\forall$ and $\ppar$).
We propose in this paper a semantic demonstration of this result, in the spirit of the work of Olivier Laurent.

\begin{figure}
\[\begin{array}{c@{\qquad}c@{\qquad}c}
A× B\eq B× A& A×\top\eq A& \forall X.\forall Y.A\eq\forall Y.\forall X.A\\
A× (B× C)\eq(A× B)× C&\forall X.\top\eq\top& \forall X.(A× B)\eq\forall X.A×\forall X.B\\
A\fl (B\fl C)\eq(A× B)\fl C& \top\fl A\eq A&A\ppar B\eq B\ppar A\\
(A\fl B)\ppar C\eq A\fl(B\ppar C)&A\fl\top\eq\top&A\ppar (B\ppar C)\eq(A\ppar B)\ppar C\\
(A× B)\ppar C\eq(A\ppar C)×(B\ppar C)&\top\ppar A\eq \top&\\
&\bot\ppar A\eq A& 
\end{array}\]
\begin{equation*}
A\ppar \forall X.B\eq\forall X.(A\ppar B)\qquad\text{ if $X$ does not appear free in $A$}
\end{equation*}
\caption{Equational system for type isomorphisms in $\lambda\mu2$}
\label{isolm2}
\end{figure}

\textbf{Categorical models for $\lambda\mu2$.}\quad The first part of this work is dedicated to the description of a 
categorical structure
which generates models of $\lmc$. This construction is essentially a mix between the structure of hyperdoctrines, introduced by 
Lawvere~\cite{hyperdoc1},
which have been proved to be a categorical model of system F, and the control categories, invented by Peter Selinger~\cite{selinger}
to give a categorical characterization of models of the $\lambda\mu$-calculus. The
only points that require more caution are at the interface between the two structures, \ie\ at the interaction between the functor $\Pi_I$
that models quantification in a hyperdoctrine and the binoidal functor $\ppar$ of control categories.

\textbf{Game semantics.}\quad Models of second order calculi do not come about easily due to
impredicativity. Among the different possibilities, we choose
models based on game semantics because of their high degree of
adequation with the syntax: indeed, game semantics has been widely
used to construct fully complete models for various calculi, such
as PCF~\cite{ajm,ho}, $\mu$PCF~\cite{controlgames}, Idealized
Algol~\cite{iagames}, etc. This means that this semantics gives a
very faithful description of the behavior of the syntax modulo
reduction rules in the system. And this is precisely what we need
to deal semantically with type isomorphisms: a model which is so
precise that it contains no more isomorphisms than the syntax.

The first game model of system F was a complete HON-style game
model by Hughes~\cite{phdhughes} from which we inherit the notion
of hyperforests (i.e. forests with more structure); unfortunately
the complex mechanism for interaction in this
model prevents us from calculating isomorphisms efficiently.
Murawski and Ong developed an alternative model (for affine
polymorphism) based on the notion of evolving games~\cite{muraw}:
we will reuse this idea in the context of a HON-style game.
Finally, Abramsky and Jagadeesan built a
model dedicated to generic polymorphism~\cite{gamesajf}, and thus
their model is not appropriate for our objectives.

\textbf{The model.}\quad
The second part of this paper presents polymorphic arenas and strategies on these arenas:
polymorphic arenas are forests with a precise structure for nodes that make them very closed to second-order formulas.
A structure of hyperforest can be extracted from these arenas (however, note that hyperforests are not \textit{the} basic structure used
to define arenas). The notion of move in a polymorphic arena is more sophisticated than in propositional game semantics, but 
these moves carry all the second-order structure, so that the definitions of plays, views, strategies, etc, will not change.

We prove that we have obtained a model for $\lambda\mu2$ by using the tools defined in the first part. In this model, the two
players $\mathbf{O}$ and $\mathbf{P}$ have a very symmetrical
behavior, so that interaction is easy to define. But this symmetry
is paid by the fact that this model, being very liberal, is far
from being complete (which is not a problem by itself in our
perspective), and in particular it has too many isomorphisms
compared to our language.

\begin{figure}
\epsfig{file=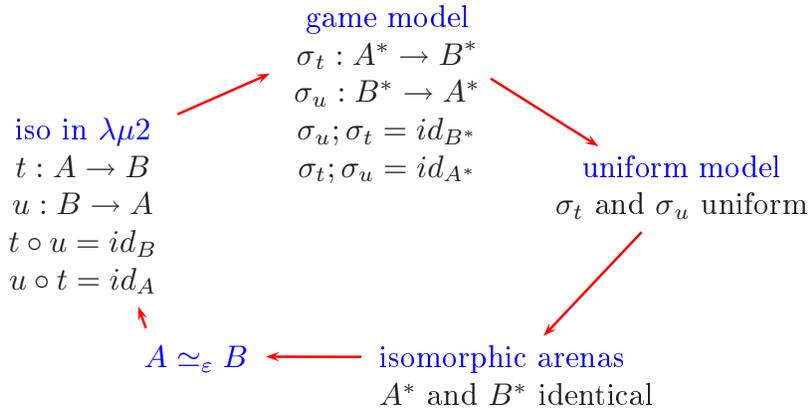,width=5cm}
\caption{Steps of our demonstration}
\label{proofjosh}
\end{figure}

\textbf{Uniformity.}\quad That is why we add a new
property for strategies, uniformity (also inspired partly
by~\cite{muraw}), which breaks this symmetry between players and
gives raise to a sub-model (which is also far from being complete,
but we do not care for that) where the isomorphisms will happen to
have exactly the same form as in $\lambda\mu2$: uniformity is just an
ad hoc property, precisely defined to retrieve exactly $\lambda\mu2$
isomorphisms.

The core theorem of our work on isomorphisms consists in proving that, in the
uniform model, the existence of a game isomorphism between two polymorphic
arenas ($A\simeq_g B$) induces that these two arenas are equal in
the most natural sense ($A\simeq_a B$). Then we can conclude on
the characterization of type isomorphisms: if we denote $A^*$ the
interpretation of a type $A$ in the uniform model, then we have:
$$A\simeq B\Leftrightarrow A^*\simeq_g B^*\Leftrightarrow A^*\simeq_a B^*\Leftrightarrow A\simeq_\epsilon B$$

The main steps of this reasoning are summed up on
figure~\ref{proofjosh}. As an easy corollary of this result, one is
able to retrieve the characterization of type isomorphisms for
Church-style system F, proved syntactically by Roberto Di
Cosmo~\cite{isotypes}. Moreover, the results can also be extended
easily to some little extensions of the calculus, like a calculus with
a fixpoint operator. Finally, the geometrical aspect of this work
leads us to an interesting remark: hyperforests, which naturally carry
the equivalence corresponding to type isomorphisms, happen to be a
very significant description of second-order formulas.

\newpage

\section{Control hyperdoctrines}

\subsection{The second-order $\lambda\mu$-calculus}


The Curry-Howard correspondence, illustrated for intuitionistic logic by the simply typed $\lambda$-calculus, can be extended to classical logic through Michel Parigot's $\lambda\mu$-calculus~\cite{lambdamu}. It adds new operators to the $\lambda$-calculus, in
order to enable the notion of \textit{control}. Hence, the calculus allows to use the output as if it was sent to many outputs, 
which correspond to the sequents with several conclusions of classical logic. As an example, the well-known control 
command \texttt{call/cc} and its semantics can be encoded in the $\lambda\mu$-calculus.
There are two different paradigms, which differ in the reduction rules of the control operators: the call-by-name and the call-by-value
$\lambda\mu$-calculi. Peter Selinger proved in~\cite{selinger} that these two calculi are dual.

Here we consider the second-order extension of this calculus, in a
call-by-name paradigm, and with the disjunction type introduced by Selinger in~\cite{selinger}. This system will be called $\lambda\mu2$ in
the rest of the paper.

The grammar of types is the following:

$$A=\top\mid\ \bot\mid\ X\mid\ A× A\mid\ A\fl A\mid\ A\ppar A\mid\ \forall X.A$$

The grammar of terms is:

$$\begin{array}{cl}
t::=&x\mid\ \star\mid\ (t,t)\mid\ \pi_1(t)\mid\ \pi_2(t)\mid\ tt\mid\ \lambda x^A.t\mid\ [\alpha]t\mid\ \mu\alpha^A.t\\
&\mid\ [\alpha,\beta]t\mid\ 
\mu(\alpha^A,\beta^B).t\mid\ \Lambda X.t\mid\ t\{A\}\end{array}$$

The variables $\alpha$ will be called \textbf{names}. If $[\alpha]t$ appears in the scope of a $\mu\alpha^A$ it will be called a 
\textbf{bound name}; if not it is a \textbf{free name}; the set of free names of a term $t$ will be denoted by $FN(t)$.
The set of free term variables (resp. free type variables) appearing in a term is denoted $FV(t)$ (resp. $FTV(t)$). 

In order to control the free type variables appearing in a sequent, we introduce the enabling 
judgement $\XX\Vdash A$: it expresses the fact that the free type variables of a type $A$ are chosen among $X_1,\dots,X_n$, 
and it is defined by the following inference rules:

\bec
\AxiomC{$X\in\XX$}
\UnaryInfC{$\XX\Vdash X$}
\DisplayProof\qquad
\AxiomC{\phantom{$X\in\XX$}}
\UnaryInfC{$\XX\Vdash \top$}
\DisplayProof\qquad
\AxiomC{\phantom{$X\in\XX$}}
\UnaryInfC{$\XX\Vdash \bot$}
\DisplayProof


\AxiomC{$\XX\Vdash A$}
\AxiomC{$\XX\Vdash B$}
\BinaryInfC{$\XX\Vdash A\fl B$}
\DisplayProof
\qquad
\AxiomC{$\XX\Vdash A$}
\AxiomC{$\XX\Vdash B$}
\BinaryInfC{$\XX\Vdash A× B$}
\DisplayProof
\qquad
\AxiomC{$\XX\Vdash A$}
\AxiomC{$\XX\Vdash B$}
\BinaryInfC{$\XX\Vdash A\ppar B$}
\DisplayProof


\AxiomC{$\XX,X\Vdash A$}
\UnaryInfC{$\XX\Vdash \forall X.A$}
\DisplayProof
\eec

\medskip

The sequents of our calculus take the form $\XX;\Gamma\vdash t:A\barreD$ where $t$ is the term, $A$ is the type, $\Gamma$ is a context for 
variables (a sequence of typing assignments $x_i:A_i$, where $x_i$ is a variable that appears at most once in $\Gamma$), 
$\Delta$ is a context for names
(a set of typing assignments $\alpha_i:A_i$, where $\alpha_i$ is a name that appears at most once in $\Delta$) and 
$\XX$ is a set of type variables. The typing rules are:

\bec
\AxiomC{$\XX \Vdash A_1\quad\dots\quad\XX \Vdash A_n\quad\XX \Vdash B_1\quad\dots\quad\XX \Vdash B_p$}
\LeftLabel{(ax)}
\UnaryInfC{$\XX;x_1:A_1,\dots,x_n:A_n\vdash x_i:A_i\barre\alpha_1:B_1,\dots,\alpha_p:B_p$}
\DisplayProof


\AxiomC{$\XX\Vdash\Gamma\qquad \XX\Vdash\Delta$}
\LeftLabel{($\top$)}
\UnaryInfC{$\XX;\Gamma\vdash\star:\top\barreD$}
\DisplayProof


\AxiomC{$\XX;\Gamma,x:A\vdash t:B\barre\Delta$}
\LeftLabel{($\to I$)}
\UnaryInfC{$\XX;\Gamma\vdash\lambda x^A.t:A\to B\barre\Delta$}
\DisplayProof


\AxiomC{$\XX;\Gamma\vdash t:A\to B\barre\Delta$}
\AxiomC{$\XX;\Gamma\vdash u:A\barreD$}
\LeftLabel{($\to E$)}
\BinaryInfC{$\XX;\Gamma\vdash tu:B\barreD$}
\DisplayProof


\AxiomC{$\XX;\Gamma\vdash t:A\barreD$}
\AxiomC{$\XX;\Gamma\vdash u:B\barreD$}
\LeftLabel{($× I$)}
\BinaryInfC{$\XX;\Gamma\vdash (t,u):A× B\barreD$}
\DisplayProof


\AxiomC{$\XX;\Gamma\vdash t:A× B\barreD$}
\LeftLabel{($× E1$)}
\UnaryInfC{$\XX;\Gamma\vdash \pi_1(t):A\barreD$}
\DisplayProof
\quad\AxiomC{$\XX;\Gamma\vdash t:A× B\barreD$}
\LeftLabel{($× E2$)}
\UnaryInfC{$\XX;\Gamma\vdash \pi_2(t):B\barreD$}
\DisplayProof


\AxiomC{$\XX;\Gamma\vdash t:A\barreD$}
\LeftLabel{(naming rule)}
\UnaryInfC{$\XX;\Gamma\vdash [\alpha]t:\bot\barreD$}
\DisplayProof if $\alpha:A\in\Delta$


\AxiomC{$\XX;\Gamma\vdash t:\bot\barre\alpha:A,\Delta$}
\LeftLabel{($\mu$-rule)}
\UnaryInfC{$\XX;\Gamma\vdash\mu\alpha^A.t:A\barreD$}
\DisplayProof


\AxiomC{$\XX;\Gamma\vdash t:A\ppar B\barreD$}
\LeftLabel{(double naming rule)}
\UnaryInfC{$\XX;\Gamma\vdash [\alpha,\beta]t:\bot\barreD$}
\DisplayProof if $\alpha:A,\beta:B\in\Delta$


\AxiomC{$\XX;\Gamma\vdash t:\bot\barre\alpha:A,\beta:B,\Delta$}
\LeftLabel{(double $\mu$-rule)}
\UnaryInfC{$\XX;\Gamma\vdash\mu(\alpha^A,\beta^B).t:A\ppar B\barreD$}
\DisplayProof


\AxiomC{$\XX,X;\Gamma\vdash t:A\barreD$}
\LeftLabel{($\forall I$)}
\UnaryInfC{$\XX;\Gamma\vdash \Lambda X.t:\forall X.A\barreD$}
\DisplayProof if $X\notin FTV(\Gamma)\cup FTV(\Delta)$


\AxiomC{$\XX;\Gamma\vdash t:\forall X.A\barreD$}
\AxiomC{$\XX\Vdash B$}
\LeftLabel{($\forall E$)}
\BinaryInfC{$\XX;\Gamma\vdash t\{B\}:A[B/X]\barreD$}
\DisplayProof


\eec

Finally, the equational theory of $\lambda\mu2$ is defined by the sequents $\XX;\Gamma\vdash t=u:A\barreD$ (with $\XX;\Gamma\vdash t:A\barreD$ and $\XX;\Gamma\vdash u:A\barreD$) generated by congruence relations that can be classified as follows:\label{lm2eq}

\textbf{$\lambda$-calculus with products:}
\begin{align}
t=\star&\qquad\qquad:\ \top&\tag{($\top$)}\\
\pi_1((u,v))= u&\qquad\qquad:\ A&\tag{($\pi_1$)}\\
\pi_2((u,v))= v&\qquad\qquad:\ B&\tag{$\pi_2$}\\
(\pi_1(u),\pi_2(u))=u&\qquad\qquad:\ A\times B&\tag{$×$}\\
(\lambda x^A.t)u= t[u/x]&\qquad\qquad:\ B&\tag{$\beta$} \\
\lambda x^A.tx= t&\qquad\qquad:\ A\fl B&
\text{if $x\notin FV(t)$}\tag{$\eta$}
\end{align}

\textbf{$\lambda\mu$-calculus with disjunction:}
\begin{align}
(\mu\alpha^{A\fl B}.t)u=\mu\beta^B.t[[\beta](-)u/[\alpha](-)]&\qquad:\ B&\text{if $\beta\notin FN(t,u)$}\tag{$\mu^\fl$}\\
\pi_i(\mu\alpha^{A_1× A_2}.t)=\mu\beta^{A_i}.t[[\beta]\pi_i(-)/[\alpha](-)]&\qquad:\ A_i&\text{if $\beta\notin FN(t)$}\tag{$\mu^×$}\\
[\beta,\gamma](\mu\alpha^{A\ppar B}.t)=t[[\beta,\gamma](-)/[\alpha](-)]&\qquad:\ \bot&\tag{$\mu^\ppar$}\\
(\mu\alpha^{\forall X.A}.t)\{B\}=\mu\beta^{A[B/X]}.t[[\beta](-)\{B\}/[\alpha](-)]&\qquad:\ A[B/X]&\text{if $\beta\notin FN(t)$}\tag{$\mu^\forall$}
\end{align}
\begin{align}
[\alpha']\mu\alpha^A.t=t[\alpha'/\alpha]&\qquad\qquad:\ \bot&\tag{$\rho^{\mu}$}\\
[\alpha',\beta']\mu(\alpha^A,\beta^B).t=t[\alpha'/\alpha,\beta'/\beta]&\qquad\qquad:\ \bot&\tag{$\rho^{\ppar}$}\\
[\xi]t=t&\qquad\qquad:\ \bot&\text{if $\xi:\bot\in\Delta$}\tag{$\rho^{\bot}$}\\
\phantom{coucou}&&\notag\\
\mu\alpha^A.[\alpha]t=t&\qquad\qquad:\ A&\text{if $\alpha\notin FN(t)$}\tag{$\theta^{\mu}$}\\
\mu(\alpha^A,\beta^B).[\alpha,\beta]t=t&\qquad\qquad:\ A\ppar B&\text{if $\alpha,\beta\notin FN(t)$}\tag{$\theta^{\ppar}$}
\end{align}

\textbf{Second order quantification:}
\begin{align}
(\Lambda X.t)\{B\}= t[B/X]&\qquad\qquad:\ A[B/X]&\tag{$\beta2$} \\
\Lambda X.t\{X\}= t
&\qquad\qquad:\ \forall X.A&\text{if $X\notin FTV(t)$}\tag{$\eta2$}
\end{align}


In the above relations, the contextual substitution
$s_{\alpha,C}(M)=M[C(-)/[\alpha](-)]$ where $M$ is a term, $t\mapsto
C(t)$ is an operation on terms and $\alpha:A$ appears in the name
context, has to be defined by induction on $M$:
\begin{itemize}
\item $s_{\alpha,C}([\alpha]M)=C([\alpha]s_{\alpha,C}(M))$
\item
  $s_{\alpha,C}([\alpha,\beta]M)=C(\mu\alpha^A.[\alpha,\beta]s_{\alpha,C}(M))$
\item
  $s_{\alpha,C}([\beta,\alpha]M)=C(\mu\alpha^A.[\beta,\alpha]s_{\alpha,C}(M))$
\item $s_{\alpha,C}$ commutes with all other base operations on terms (with
  the requirement to avoid captures).
\end{itemize}

\medskip

Now that the system $\lambda\mu2$ is completely defined, one can give the definition of a type isomorphism:

\begin{Def}[type isomorphism]
Let $A$ and $B$ be two types of $\lambda\mu2$. We say that there is a type isomorphism between $A$ and $B$ if there exist two terms $t$
and $u$ such that:
\begin{itemize}
\item $\XX;\vdash t:A\fl B\barre$
\item $\XX;\vdash u:B\fl A\barre$
\item $\lambda x^B.t(ux)=\lambda x^B.x$
\item $\lambda y^A.u(ty)=\lambda y^A.y$
\end{itemize}
\end{Def}

\subsection{Definition of a control hyperdoctrine}\label{contr_hyp}
We wish to give a categorical model of $\lambda\mu2$. For this we use two ingredients : first, the notion of 
\textit{hyperdoctrine},
introduced by Lawvere~\cite{hyperdoc1}, with which Seely~\cite{hyperdoc2} and Pitts~\cite{hyperdoc3} have proposed a categorical 
interpretation of system F; second, the notion of \textit{control category}~\cite{selinger}, which introduces a disjunction $\ppar$ 
to characterize
models of the $\lambda\mu$-calculus. We chose to give preference to control categories rather than categories of continuations,
because using continuation categories would require to build a CPS-translation transforming the connector $\forall$
into the connector $\exists$, and to build a theory for categories of continuations with the connector $\exists$: as our model is 
based on the interpretation of the connector $\forall$, we did not choose this option.

In the following definition, $\textbf{CCC}$ is the category of
cartesian closed categories with strict morphisms of ccc's
($G:C\fl D$ is a strict morphism if the specified cartesian closed
structure of $\mathbf{C}$ is sent to the specified cartesian
closed structure of $\mathbf{D}$).

\begin{Def}[hyperdoctrine]\label{hyperdocdef}
An \textbf{hyperdoctrine} $\mathbf{H}$ is specified by:
\begin{itemize}
\item a base category $|\mathbf{H}|$ with terminal object $\top$ and binary products
\item a distinguished object $U$ in $|\mathbf{H}|$ such that for all
$I\in|\mathbf{H}|$ there exists $n\in\mathbb{N}$ such that $I=U^n$
(with the convention $U^0=\top$); we denote $\pi^i_{n}:U^n\fl U$ the projection on the $i$th component, 
and $\pi_{U^n}=\pi^1_{n+1}×\dots×\pi^n_{n+1}:U^{n+1}\fl U^n$
\item a functor $F:|\mathbf{H}|^{op}\fl \textbf{CCC}$ such that if we
compose $F$ with the forgetful functor
$\textit{fff}:\textbf{CCC}\fl \textbf{Set}$ we obtain the functor
$|\mathbf{H}|(-,U)$
\item for each $I\in|\mathbf{H}|$, a functor $\Pi_I:F(I× U)\fl
F(I)$ such that~:
\begin{itemize}
\item $\Pi_I$ is right adjoint to the functor $F(\pi_{I× U}):F(I)\fl F(I× U)$
\item $\Pi_I$ is natural in $I$: for any $\alpha:I\fl J$,
$F(\alpha)\circ\Pi_J=\Pi_I\circ F(\alpha× id_U)$
\item for any $\alpha:I\fl J$, for any object $A$ of $F(J\times U)$,
  the morphism $(F(\alpha)\circ\Pi_J)(A)\fl(\Pi_I\circ F(\alpha× id_U))(A)$
  generated by the adjunction is the identity.
\end{itemize}
\end{itemize}
The functors $F(C)$, with $C$ object
of $|\mathbf{H}|^{op}$, are called the \textbf{specialization functors}.
\end{Def}

The intuitions of such a categorical description are the following: the objects (resp. the morphisms) of $F(U^n)$ will correspond to the 
types (resp. the terms) where each free type variable that appears is chosen between $X_1,\dots,X_n$. Hence,
$F(\pi_{I\times U}):F(I)\fl F(I\times U)$ is simply the standard way to make the number of free type variables growing: thus we will
often address this functor as $-\mapsto (-)_I^{I\times U}$. $\Pi_I$ makes this
number of free type variables decreasing by quantifying one of them.

We adopt the following notations for a cartesian closed category: $\diamond_A:A\fl 1$ is the terminal arrow, $\pi_1$, $\pi_2$
are the two projections, $(f,g)$ is for pairing, $\epsilon_{A,B}:B^A× A\fl B$ is the evaluation and the curryfication
of a map $f:B× A\fl C$ is denoted $\Lambda(f):B\fl C^A$. We sometimes note $\flt{ccc}$ to denote trivial isomorphisms in
a ccc.
We also note $\kappa
:\textit{Hom}_{F(I× U)}((C)_I^{I× U},A)\fl
\textit{Hom}_{F(I)}(C,\Pi_I(A))$ the bijection associated with the adjunction $F(\pi_{I× U})\dashv\Pi_I$.

We now introduce the notion of
\textbf{control hyperdoctrine}, in order to adapt hyperdoctrines to a description of second-order \textit{classical} logic: 
the path from hyperdoctrines to control hyperdoctrines will take
the same form as the one from cartesian closed categories to
control categories:
\begin{minilist}
\item we first give the hyperdoctrine structure
\item we introduce the symmetric pretensor $\ppar$ together with the
neutral element $\bot$
\item we then require the existence of codiagonals, \ie\ for each
object $A$ two central morphisms $i_A:\bot\fl A$ and
$\Delta_A:A\ppar A\fl A$ such that $\langle
A,i_A,\Delta_A\rangle$ is a symmetric monoid compatible with the
premonoidal structure
\item we introduce a new condition, \textbf{hypermonoidality}, that asks for the 
commutation of the specialization functors with the premonoidal structure and the codiagonals, and the
preservation of centrality through the hyperdoctrine
adjunction
\item we require the distributivity of $\ppar$ over the cartesian
product
\item we introduce the \textbf{exponential strength}: the (already
existing) morphism $s_{A,B,C}:B^A\ppar C\fl (B\ppar C)^A$ is a
\emph{natural isomorphism} which respects some coherence conditions
\item  we also introduce the \textbf{quantification strength}: the (already
existing) morphism $p_{A,B}:\Pi_I(A)\ppar B\fl\Pi_I(A\ppar
(B)_I^{I\times U})$ is a \emph{natural isomorphism} which respects a
condition of centrality.
\end{minilist}

In the following definition, $[C]$ denotes the class of objects of a category $C$, regarded as a discrete subcategory.

\begin{Def}[binoidal hyperdoctrine]
A \textbf{binoidal hyperdoctrine} $\mathbf{H}$ is an hyperdoctrine together with, for each $I\in|\mathbf{H}|$, a binoidal functor 
$\ppar_I$, \ie\ a couple of two bifunctors $\ppar^1_I:F(I)\times
[F(I)]\fl F(I)$ and $\ppar^2_I:[F(I)]\times F(I)\fl F(I)$ such that 
$A\ppar^1_I B=B\ppar^2_I A$ for all pairs of objects $A,B$.
\end{Def}

We recall the definition of a central morphism: in the ccc $F(I)$, $f:A\fl A'$ is central if for every $g:B\fl B'$
one has $(f\ppar_I B')\circ (A\ppar_I g)=(A'\ppar_I g)\circ(f\ppar_I B)$ and $(B'\ppar_I f)\circ (g\ppar_I A)=(g\ppar_I A')\circ(B\ppar_I f)$.

\begin{Def}[premonoidal hyperdoctrine]
A \textbf{premonoidal hyperdoctrine} is a binoidal hyperdoctrine $\mathbf{H}$ together with, for each $I\in|\mathbf{H}|$, an object $\bot_I$ 
and central natural isomorphisms $a_{A,B,C}:(A\ppar_I B)\ppar_I C\fl A\ppar_I(B\ppar_I C)$, $l_A:A\fl A\ppar_I\bot_I$ and $r_A:A\fl\bot_I\ppar_I A$ 
making the following diagrams commute:

$$
\xymatrix{
((A\ppar_I B)\ppar_I C)\ppar_I D\ar[r]^{a}\ar[d]_{a\ppar D}&(A\ppar_I B)\ppar_I(C\ppar_I D)\ar[r]^{a}&A\ppar_I(B\ppar_I(C\ppar_I D))\\
(A\ppar_I(B\ppar_I C))\ppar_I D\ar[rr]^{a}& &A\ppar_I((B\ppar_I C)\ppar_I D)\ar[u]_{A\ppar a}}
$$

$$
\xymatrix{
&A\ppar_I B \ar[dl]_{l\ppar B}\ar[dr]^{A\ppar r} &\\
(A\ppar_I\bot_I)\ppar_I B \ar[rr]^{a} &&A\ppar_I(\bot_I\ppar_I B)}
$$

It is called a \textbf{symmetric premonoidal hyperdoctrine} if there are in addition central natural isomorphisms $c_{A,B}:A\ppar_I B\fl B\ppar_I A$ 
such that  $c_{A,B}\circ c_{B,A}=id_{A\ppar_I B}$ and:
$$
\xymatrix{
(A\ppar_I B)\ppar_I C\ar[r]^{a}\ar[d]_{c\ppar C}&A\ppar_I (B\ppar_I C)\ar[r]^{c}&(B\ppar_I C)\ppar_I A\ar[d]^{a}\\
(B\ppar_I A)\ppar_I C\ar[r]^{a}&B\ppar_I (A\ppar_I C)\ar[r]^{B\ppar c}&B\ppar_I(C\ppar_I A)}
$$
$$
\xymatrix{
&A\ar[dl]_{l}\ar[dr]^{r}&\\
A\ppar_I \bot_I\ar[rr]^{c}&&\bot_I\ppar_I A}
$$
\end{Def}

\begin{Def}[symmetric monoid, codiagonals]
Let $\hh$ be a symmetric premonoidal hyperdoctrine. A symmetric monoid in $\hh$ for an object $A\in F(I)$ ($I\in\hhh$) is a pair of central morphisms 
$i_A:\bot_I\fl A$ and $\nabla_A:A\ppar_I A\fl A$ such that:

$$
\xymatrix{
A\ppar_I\bot_I\ar[r]^{A\ppar i}\ar[dr]_{l^{-1}}&A\ppar_I A\ar[d]_{\nabla}&\bot_I\ppar_I A\ar[l]_{i\ppar A}\ar[dl]^{r^{-1}}\\
&A&}
$$

$$
\xymatrix{
(A\ppar_I A)\ppar_I A\ar[r]^{\ \ \ \nabla\ppar A}\ar[dd]_{a}&A\ppar_I A\ar[dr]^{\nabla}\\
&&A\\
A\ppar_I(A\ppar_I A)\ar[r]^{\ \ \ A\ppar\nabla}&A\ppar_I A\ar[ur]^{\nabla}}
\qquad
\xymatrix{
A\ppar_I A\ar[dr]^{\nabla}\ar[dd]_{c}&\\
&A\\
A\ppar_I A\ar[ur]^{\nabla}}
$$

We say that a a symmetric premonoidal hyperdoctrine has \textbf{codiagonals} if, for each $I\in\hhh$, there is a symmetric monoid for every $A\in F(I)$,
which is compatible with the premonoidal structure:

$$
\xymatrix{
\\
i_{\bot_I}=id_{\bot_I}\qquad\\
}
\qquad
\xymatrix{
\bot_I\ar[dd]_{l=r}\ar[dr]^{i_{A\ppar B}}&\\
&A\ppar_I B\\
\bot_I\ppar_I\bot_I\ar[ur]_{i_A\ppar i_B}&
}
\qquad
\xymatrix{
A\ppar_I B\ppar_I A\ppar_I B\ar[dd]_{A\ppar c\ppar B}\ar[dr]^{\ \ \nabla_{A\ppar B}}&\\
&A\ppar_I B\\
A\ppar_I A\ppar_I B\ppar_I B\ar[ur]_{\ \ \nabla_A\ppar\nabla_B}&
}
$$
\end{Def}

The central morphism $\nabla_A$ recovers the notion of \textit{contraction} from linear logic. One can also define the 
\textit{weakening} in a premonoidal hyperdoctrine with codiagonals: $w=A\flt{l}A\ppar_I\bot_I\flt{A\ppar i}A\ppar_I B$.

\begin{Def}[focality]
A morphism $f:A\fl B$ is \textbf{focal} if it is central and the two following diagrams commute:

$$
\xymatrix{
&\bot_I\ar[dl]_{i_A}\ar[dr]^{i_B}&\\
A\ar[rr]^{f}&&B
}
\qquad
\xymatrix{
A\ppar_I A\ar[d]_{\nabla_A}\ar[r]^{f\ppar f}&B\ppar_I B\ar[d]^{\nabla_B}\\
A\ar[r]^{f}&B}
$$
\end{Def}

A premonoidal \textit{category} with codiagonals will be called a pre-control category. A strict morphism $\mu:C\fl D$
of pre-control categories is such that it sends each element of the structure of $C$ in the corresponding element of the structure 
$D$: $\mu(A\ppar B)=\mu(A)\ppar\mu(B)$, $\mu(\bot)=\bot$, $\mu(\sigma\ppar A)=\mu(\sigma)\ppar\mu(A)$, 
$\mu(a_{A,B,C})=a_{\mu(A),\mu(B),\mu(C)}$, etc.

\begin{Def}[hypermonoidality]
Let $\hh$ be a symmetric premonoidal hyperdoctrine with codiagonals. We say that $\hh$ has \textbf{hypermonoidality} if the 
specialization functors are strict morphisms of pre-control categories and if
$\kappa$, $\kappa^{-1}$ preserve centrality of morphisms.
\end{Def}

\begin{Def}[distributivity]
Let $\hh$ be a symmetric premonoidal hyperdoctrine with codiagonals. $\hh$ is said to be \textbf{distributive} if:
\begin{itemize}
\item the projections $\pi_1$ and $\pi_2$ are focal
\item for each $I\in\hhh$ and $A\in F(I)$ the functor $- \ppar_I A$ preserves finite products: the natural morphisms 
$(\pi_1\ppar_I C,\pi_2\ppar_I C):(A× B)\ppar_I C\fl(A\ppar_I C)×(B\ppar_I C)$ and $\diamond_{1\ppar_I C}:1\ppar_I C\fl 1$ are isomorphisms, 
whose inverses are respectly denoted $d_{A,B,C}$ and $\diamond'_C$.
\end{itemize}
\end{Def}

\begin{Def}[control hyperdoctrine]
Let $\hh$ be a distributive symmetric premonoidal hyperdoctrine with codiagonals and hypercentrality. For $A,B,C\in F(I)$, 
let $s_{A,B,C}:(B^A\ppar_I C)\fl(B\ppar_I C)^A$ be the canonical morphism obtained by currying 
$$\hat{\epsilon}_{A,B,C}:(B^A\ppar_I C)× A\flt{(B^A\ppar C)× (l;A\ppar i)}(B^A\ppar_I C)×(A\ppar_I C)\flt{d}(B^A× A)\ppar_I C\flt{\epsilon\ppar C}B\ppar_I C$$
For $A\in F(I× U)$ and $B\in F(I)$, let 
$$p_{A,B}=\kappa(\kappa^{-1}(id_{\Pi_I(A)})\ppar_{I\times U}
(B)_{I}^{I\times U}):\Pi_I(A)\ppar_IB\fl\Pi_I(A\ppar_{I× U} (B)_{I}^{I\times U})$$
$\hh$ is called a \textbf{control hyperdoctrine} if
\begin{itemize}
\item $s_{A,B,C}$ is a natural isomorphism in $C$ satisfying:

$$
\xymatrix{
B^A\ppar_I C^D\ar[rr]^{s'}\ar[d]_{s}&&(B^A\ppar_I C)^D\ar[d]^{s^D}\\
(B\ppar_I C^D)^A\ar[r]^{s'^A}&((B\ppar_I C)^D)^A\ar[r]^{ccc}&((B\ppar_I C)^A)^D
}
$$
where $s'_{A,B,C}=B\ppar_I C^A\flt{c}C^A\ppar_I B\flt{s}(C\ppar_I B)^A\flt{c^A}(B\ppar_I C)^A$ and:
$$
\xymatrix{
B^A\ppar_I B^A\ar[r]^{s'}\ar[dr]_{\nabla_{B^A}}&(B^A\ppar_I B)^A\ar[r]^{s^A}&(B\ppar_I B)^{A× A}\ar[dl]^{\nabla_A^{\Delta_A}}\\
&B^A&
}
\qquad
\xymatrix{
\bot_I\ar[rr]^{ccc}\ar[dr]_{i_{B^A}}&&\bot_I^1\ar[dl]^{(i_B)^{\diamond_A}}\\
&B^A&
}
$$
where $\Delta_A=(\id_A,\id_A):A\fl A× A$.

\medskip

\item $p_{A,B}$ is a central isomorphism.
\end{itemize}

$s_{A,B,C}$ is called the \textbf{exponential strength}, whereas $p_{A,B}$ is the \textbf{quantification strength}.
\end{Def}

\textsc{Remark:} The naturality of $s_{A,B,C}$ in $A$ and $B$ follows from its definition, as well as the naturality of $p_{A,B}$ in $A$ and $B$.

\subsection{Interpretation of the calculus}

Thanks to the notion of control category, we are able to give a categorical interpretation of $\lambda\mu2$.

\textbf{Interpretation of types:}

If $I=U^n$, we note $\ppar_n$ for $\ppar_I$, $\bot_n$ for $\bot_I$ and $\Pi_n$ for $\Pi_I$.
Each type $A$ such that $\XX\Vdash A$
is interpreted as an object $A^*$ of $F(U^n)$ as follows:

\bec
$\begin{array}{c@{\qquad}c@{\qquad}c}
\bot^*=\bot_n&\top^*=1&X_i^*=\pi^i_{n}\\
(A× B)^*=A^*× B^*&(A\ppar B)^*=A^*\ppar_n B^*&(A\fl B)^*=(B^*)^{A^*}\\
(\forall X_{n+1}.A)^*=\Pi_{n}(A^*)&&
\end{array}$
\eec

Note that the interpretation of $X_i$ is a morphism $\pi^i_{n}:U^n\fl U$ in the base category $\hhh$: 
actually, we use here the fact that the composition of $F$ with the forgetful functor
$\textit{fff}:\textbf{CCC}\fl \textbf{Set}$ generates the functor $|\mathbf{H}|(-,U)$. Hence, it is equivalent to define the 
interpretation of a type as an object in $F(U^n)$ or as a morphism from $U^n$ to $U$ in $\hhh$.

\begin{Lemme}
Let $A$ and $B$ be two types such that $FTV(A),FTV(B)\in\{X_1,\dots,X_n\}$.
We note $-\mapsto(-)[U^n,B]$ for $F(id_{U^n}\times B^*)$. Then $(A[B/X_{n+1}])^*=(A^*)[U^n,B]$.
\end{Lemme}

\textsc{Proof:} We prove it by a structural induction on $A$: as $F(id_{U^n}\times B^*)$ is a strict morphism of pre-control categories, the only
cases to check are $A=\forall X_j.A'$ and $A=X_i$. The first case is ensured by
the naturality of $\Pi_n$, whereas the second one only requires a direct verification.\cqfd

\textbf{Interpretation of terms:}

A typing judgement of the form $\XX;\Gamma\vdash t:A\barreD$ will be interpreted as a morphism $\Gamma^*\fl A^*\ppar_n\Delta^*$ in the
category $F(U^n)$.

In what follows, for the sake of simplicity we use $A$ instead of $A^*$ when dealing with an object of the category $F(U^n)$.

$\begin{array}{ll}
\trad{\debinf x_i:B_i\barreD}&=\Gamma\flt{\pi_i}B_i\flt{w}B_i\ppar_n\Delta\\
\trad{\debinf \star:\top\barreD}&=\Gamma\flt{\diamond}1\flt{\cong}1\ppar_n\Delta\\
\trad{\debinf (t,u):A× B\barreD}&=\Gamma\flt{(\trad{t},\trad{u})}(A\ppar_n\Delta)×(B\ppar_n\Delta)\flt{d}(A× B)\ppar_n\Delta\\
\trad{\debinf \pi_1(t):A\barreD}&=\Gamma\flt{\trad{t}}(A× B)\ppar_n\Delta\flt{\pi_1\ppar\Delta}A\ppar_n\Delta\\
\trad{\debinf \pi_2(t):B\barreD}&=\Gamma\flt{\trad{t}}(A× B)\ppar_n\Delta\flt{\pi_2\ppar\Delta}B\ppar_n\Delta\\
\trad{\debinf tu:B\barreD}&=\Gamma\flt{(\trad{t},\trad{u})}(B^A\ppar_n\Delta)×(A\ppar_n\Delta)\flt{d}(B^A× A)\ppar_n\Delta\flt{\epsilon\ppar\Delta}B\ppar_n\Delta\\
\trad{\debinf \lambda x^A.t:A\fl B\barreD}&=\Gamma\flt{\Lambda(\trad{t})}(B\ppar_n\Delta)^A\flt{s^{-1}}B^A\ppar_n\Delta\\
\trad{\debinf[\alpha_i]t:\bot\barreD}&=\Gamma\flt{\trad{t}}A_i\ppar_n\Delta\flt{w_i\ppar\Delta}\Delta\ppar_n\Delta\flt{\nabla}\Delta\flt{\cong}\bot_n\ppar_n\Delta\\
\trad{\debinf\mu\alpha^A.t:A\barreD}&=\Gamma\flt{\trad{t}}\bot_n\ppar_nA\ppar_n\Delta\flt{\cong}A\ppar_n\Delta\\
\trad{\debinf[\alpha_i,\alpha_j]t:\bot\barreD}&=\Gamma\flt{\trad{t}}A_i\ppar_nA_j\ppar_n\Delta\flt{w_i\ppar w_j\ppar\Delta}\Delta\ppar_n\Delta\ppar_n\Delta\flt{\nabla\ppar\Delta;\nabla}\Delta\flt{\cong}\bot_n\ppar_n\Delta
\end{array}$

$\begin{array}{ll}
\trad{\debinf\mu(\alpha^A,\beta^B).t:A\ppar B\barreD}&=\Gamma\flt{\trad{t}}\bot_n\ppar_nA\ppar_nB\ppar_n\Delta\flt{\cong}(A\ppar_nB)\ppar_n\Delta\\
\trad{\debinf\Lambda X.t:\forall X.A\barreD}&=\Gamma\flt{\kappa(\trad{t})}\Pi_n(A\ppar_{n+1} (\Delta)_{U^n}^{U^{n+1}})\flt{p^{-1}}\Pi_n(A)\ppar_{n}\Delta\\
\trad{\debinf t\{B\}:A[B/X]\barreD}&=\Gamma\flt{\kappa^{-1}(\trad{t};p)[U^n,B]}A[U^n,B]\ppar_n\Delta
\end{array}$

\medskip


\begin{Th}[soundness]
The interpretation of second-order $\lambda\mu$-terms in a control hyperdoctrine is sound: for any couple of terms $t,u$ such that 
$\debinf t=u:A\barreD$,
we have $\trad{\debinf t:A\barreD}=\trad{\debinf u:A\barreD}$. Thus, every control hyperdoctrine is a model of $\lambda\mu2$.
\end{Th}

The main steps of the proof of this theorem can be found in the appendix.

\section{The Game Model}

Game models have originally been introduced by Hyland-Ong and Nickau~\cite{ho,gamesn}, and 
Abramsky-Jagadeesan-Malacaria~\cite{ajm}, giving rise to two different paradigms. In this section, we introduce a game model for $\lambda\mu2$ by choosing an HON-style of games. This requires to introduce a notion of \textit{arena}, on which we have to define the notion of \textit{play}.

\subsection{Polymorphic arenas}

In this section we will describe polymorphic arenas, \ie\ the arborescent structure by which we are going to interpret types. We need to
be very precise in defining this structure, because the control hyperdoctrine structure we wish to obtain forces us to have equalities
like $(A\ppar B)[C/X]=A[C/X]\ppar B[C/X]$, which are non-trivial in a purely geometrical structure.

Hence, the name of nodes will carry an information about how the arena has been built: for example, the arena $A\fl(B× C)$ will
be very similar to $(A\fl B)×(A\fl C)$, except that the nodes will carry the information that the product has been made ``before'' 
the arrow. Actually, our arenas will be so near to formulas that we are allowed to use this correspondence to define substitution: 
instead of
introducing it as an operation on forests, we define the operation $A\mapsto A[C/X]$ as the transformation of an arena $A$ described
by a formula $F$ into the arena $A[C/X]$ described by the formula $F[C/X]$. This trivial definition  is not the original goal
of our presentation of arenas, but it is one of its advantages.

This arborescent structure of arenas hides a structure of hyperforests (\ie\ a forest with additional structure), as introduced by 
Dominic Hughes~\cite{Hug97}. This structure is more convenient for dealing with plays on arenas, but we cannot introduce it from the
beginning because of the precision we want for the objects of our model.

Polymorphic arenas are built with the constructors $\ppar$, $×$ and $¬$; the constructor $\fl$ is introduced at the end.

\textbf{Construction of arenas:}

We consider the set of type variables $X, Y, \dots$ to be in bijection with $\mathbb{N}\backslash\{0\}$, and we will further write this
set $\mc{X} = \{X_j \mid\ j > 0\}$.

We define the set $\mc{N}$ of \textbf{nodes}, based on the following grammar:
$$c::=\star\mid\ x_i\mid\ x^{(j,c)}\mid\ \forall(c)\mid\ (c,1)\mid\ (c,2)\mid\ ¬(c)\mid\ (c\ppar c)\mid\ (c,c,1)\mid\ (c,c,2)$$
for $i,j\in\mathbb{N}$. The nodes $x_i$ (resp. $x^q$) will be called \textbf{free variables} (resp. \textbf{bound variables}).

For a given node $c$, we note $V(c)$ the sets of nodes of the form $x_i$, $x^{(j,c')}$ or $\star$ occurring in $c$ 
(we call them the \textbf{variables} of $c$).
We define on the grammar the operation $c\mapsto c[c'/x]$ (where $x$ can be either $\star$, $x_i$ or $x^{j,c''}$), which simply consists 
in replacing each occurrence of $x$ in $c$ by $c'$, and the operation $c\mapsto c[\star]$ which consists 
in replacing each occurrence of a variable of $V(c)$ by $\star$.

A forest $A$ is a set of nodes $E_A$, together with an order relation $\leq_A$ on $E_A$ such that for every $c$ in $E_A$, the set
$\{c'\mid\ c'\leq c\}$ is finite and totally ordered by $\leq_A$. We note $a<_Ab$ if $a\leq_Ab$ and $a\neq b$, and we say that $c$ is a
root of $A$ if there is no $c'$ in $E_A$ such that $c'<_Ac$.
Finally, for two nodes $a$ and $a'$ in a forest $A$ such that $a\leq a'$, we note $d(a,a')$ the number of nodes $x$ such that $a<x\leq a'$

If $A$ and $B$ are two forests, we define the following forests:
\begin{itemize}
\item $\top$, $\bot$ and $X_i$ are defined by:
\begin{itemize}
\item $E_{\top}=\emptyset\qquad E_{\bot}=\{\star\}\qquad E_{X_i}=x_i$
\item $<_{\top}$, $<_{\bot}$ and $<_{X_i}$ are the empty relations
\end{itemize}
\item $A× B$ is given by:
\begin{itemize}
\item $E_{A× B}=\{(a,1)\mid\ a\in A\}\cup\{(b,2)\mid\ b\in B\}$
\item $c<_{A× B}c'$ iff ($c=(a,1)$, $c'=(a',1)$ and $a<_A a'$) or ($c=(b,1)$, $c'=(b',1)$ and $b<_B b'$)
\end{itemize}
\item $¬ A$ is given by:
\begin{itemize}
\item $E_{¬ A}=\{\star\}\cup\{¬(a)\mid\ a\in A\}$
\item $c<_{¬ A}c'$ iff ($c=\star$ and $c'\neq\star$) or ($c=¬(a)$, $c'=¬(a')$ and $a<_A a'$)
\end{itemize}
\item $A\ppar B$ is given by:
\begin{itemize}
\item $E_{A\ppar B}=\{(a\ppar b)\mid\ a\textrm{ root of }A\land b\textrm{ root of }B\}\cup
\{(a,b_0,1)\mid\ b_0\textrm{ root of }B\land \exists a_0\in E_A,a_0<_Aa\}\cup
\{(b,a_0,2)\mid\ a_0\textrm{ root of }A\land \exists b_0\in E_B,b_0<_Bb\}$
\item $c<_{A\ppar B}c'$ iff ($c=(a\ppar b)$, $c'=(a',b,1)$ and $a<_Aa'$) or  ($c=(a\ppar b)$, $c'=(b',a,2)$ and $b<_Bb'$) or
($c=(a,b,1)$, $c'=(a',b,1)$ and $a<_A a'$) or ($c=(b,a,2)$, $c'=(b',a,2)$ and $b<_B b'$)
\end{itemize}
\item $\forall X_i.A$ is given by:
\begin{itemize}
\item $E_{\forall X_i.A}=\{\forall(a[x^{(0,a[\star])}/x_i])\mid\ a\textrm{ root of }A\}\cup
\{a[x^{(d(a_0,a),a_0[\star])}/x_i]\mid\ a_0\textrm{ root of }A\land a_0<_Aa\}$
\item $c<_{\forall X_i.A}c'$ iff ($c=a[x^{(d(a_0,a),a_0[\star])}/x_i]$, $c'=a'[x^{(d(a_0,a'),a_0[\star])}/x_i]$ and $a_0<_Aa<_Aa'$) or
($c=\forall(a[x^{(0,a[\star])}/x_i])$, $c'=a'[x^{(d(a,a'),a[\star])}/x_i]$ and $a<a'$)
\end{itemize}
\end{itemize}

\textsc{Remark:} The variables $x^{(j,c)}$ correspond to bound type variables, and hence are related to an occurrence of $\forall$;
but the challenge is to be able to characterize
which one ! For this reason, they carry two pieces of information:
first the distance (in the forest) of the node where this occurrence appears; second, the name of the node corresponding to this 
occurrence. Note that it is still normally not enough to say which occurrence of $\forall$ they are related to (think to $A\ppar A$ for 
example), but thanks to the uniqueness of the construction of an arena, that we establish further, it becomes a sufficient information.

\textbf{Example~1:} Let us consider the arena $A=\forall X_3.(¬\bot\ppar X_3)\ppar \forall X_3.(¬ X_2\ppar X_3)$. It can be represented graphically as
follows:
\bec\epsfig{file=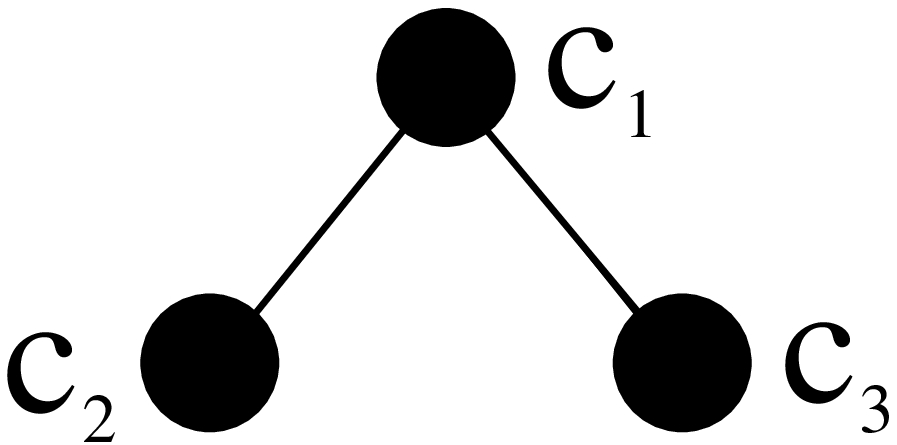,height=1.5cm}\eec

with:

$\begin{array}{l}
c_1=\forall(\star\ppar x^{(0,\star\ppar\star)})\ppar\forall(\star\ppar x^{(0,\star\ppar\star)})\\
c_2=((¬(\star),x^{(0,\star\ppar\star)},1),\forall(\star\ppar x^{(0,\star\ppar\star)})),1)\\
c_3=((¬(x_2),x^{(0,\star\ppar\star)},1),\forall(\star\ppar x^{(0,\star\ppar\star)})),2)
\end{array}$

There are many things to note: first, the occurrence of $x^{(0,\star\ppar\star)}$ in $c_2$ and $c_3$ does not imply any dependency on the occurrences of $\forall$ in $c_1$. Moreover, one can remark that the two occurrences of $x^{(0,\star\ppar\star)}$
in $c_1$ are a priori difficult to bind with a specific occurrence of $\forall$. However, there is no ambiguity if one can retrieve the way the arena
has been built up.

All of this will be explicited by the following definitions.\finex

\medskip

The set $\mc{A}$ of \textbf{polymorphic arenas} is the smallest set of forests containing $\top$, $\bot$, $X_i$ for $i\in\mathbb{N}$, and which
is closed under the constructions product, lift, par and quantification.
We note $FTV(A)=\{X_i\mid\ \exists c\in E_A,x_i\textrm{ appears in }c\}$. If $FTV(A)=\emptyset$, then $A$ is called a \textbf{closed arena}. 
The set of closed arenas is denoted $\mc{H}$.

As a consequence of the definition, a polymorphic arena is described by a second-order formula built over $¬$, $\ppar$, $×$.
Actually, this description is essentially unique:
let us define the congruence rule $=_\rho$ by:
\begin{itemize}
\item $\top\ppar A=_\rho\top$
\item $A\ppar\top=_\rho\top$
\item $\top×\top=_\rho\top$
\item $¬\top=_\rho\bot$
\item $\forall X_i.\top=_\rho\top$
\end{itemize}
The $\mathbf{\alpha\rho-}$\textbf{equivalence} on formulas is the equivalence relation built over the congruence rule $=_\rho$ and the $\alpha-$equivalence.

\begin{Lemme}\label{unique}
Let $A$ be a polymorphic arena, there exists a formula describing $A$. Furthermore, let $F$ and $F'$ 
be two such formulas, then $F$ and $F'$ are equal up to $\alpha\rho-$equivalence.
\end{Lemme}

\textsc{Proof:} This can be proved by induction on $A$. If $A$ is empty, then either $F=F_1\ppar F_2$ with $F_1$ or $F_2$ describing an empty 
arena, or $F=\forall X.F'$ with $F'$ describing an empty arena, or  $F=F_1× F_2$ with $F_1$ and $F_2$ describing an empty 
arena
(indeed, the other cases lead to non-empty arenas).

If $A$ is not empty, let $F$ be a formula describing $A$. Consider a root $c$ of $A$:
\begin{itemize}
\item if $c=x_i$ then $F=X_i$
\item if $c=\star$, then we have two possibilities: either $E_A=\{c\}$, and necessarily $F=\bot$ or $F=¬ F'$ with $F'$ describing an empty arena, 
or $E_A$ is not reduced to $c$,
and then $A=¬ A'$ for some non-empty $A'\in\mc{A}$, and we necessarily have $F=¬ F'$
\item if $c=(c_1\ppar c_2)$, then we necessarily have $F=F_1\ppar F_2$, $A=A_1\ppar A_2$, and the names of nodes allows 
to recognize the arenas $A_1$ and $A_2$
\item if $c=(c',1)$ or $c=(c',2)$, then we necessarily have $F=F_1× F_2$, $A=A_1× A_2$, and the names of nodes 
allows to recognize the arenas $A_1$ and $A_2$
\item if $c=\forall(c')$ then each root $c_n$ can be written $c_n=\forall(c'_n)$, and we necessarily have $F=\forall X_i.F(i)$, $A=\forall X_i.A(i)$ 
for some $X_i\notin FTV(A)$,
where $A(i)$ is built by replacing each $c_n$ by $c'_n$, and by replacing each occurrence of 
$x^{(j,c_n'[\star])}$ in a node $c_0\geq_A c_n$ (with $d(c'_n,c_0)=j$)\footnote{This is the reason why we specify the couple $q=(j,c)$ for a variable 
$x^q$: this allows us to determine exactly which are the variables $x^q$ related to a specific occurrence of $\forall$.} by $x_i$. These occurrences $x^{(j,c'_n[\star])}$
are called the \textbf{bound variables} of the node $c_n$, and the nodes of $A$ where they appear are called its \textbf{bound nodes}. By induction hypothesis, $F$ is unique up to $\alpha\rho$-equivalence.
\end{itemize}\cqfd

This result means that there is a one-to-one correspondence between
arenas and $\alpha\rho$-equivalence
classes of formulas.


\begin{Def}[variable substitution]
Let $A$ and $B$ are two polymorphic arenas, respectively described by $F_1$ and $F_2$. We define the substitution $A[B/X_i]$ as the arena described by the
formula $F_1[F_2/X_i]$.
\end{Def}

Note that this definition makes sense only because each arena corresponds to exactly one formula up to $\alpha\rho$-equivalence, and because the class
of $\alpha\rho$-equivalence of $F_1[F_2/X_i]$ does not depend on the representatives we choose for $F_1$ and $F_2$.

\textbf{Hyperforest:}

Now that we have defined a  notion of substitution on our arena, which trivially respects the required equalities like
$(A\ppar B)[C/X]=A[C/X]\ppar B[C/X]$, $(A× B)[C/X]=A[C/X]× B[C/X]$, etc, we can express the structure of 
\textbf{hyperforests} that is hidden in these arenas: the notion presented here is directly inspired by Hughes' hyperforests which he
introduced for his game model of system F. Actually, as we shall see further, this structure completely carries the equivalence relation
corresponding to type isomorphisms.

For a given set $E$, $\mathbf{P}(E)$ is the set of \textit{finite multisets} of elements of $E$. The multiset containing the elements $a_1,a_2,\dots,a_n$ will be denoted $\{\{a_1,a_2,\dots,a_n\}\}$.

\begin{Def}[hyperforest]
An hyperforest $H=(E,\leq,\RRR,\DD)$ is a finite forest
$(E,\leq)$, together with a multiset of \textbf{hyperedges}
$\mc{R}\in\mathbf{P}(E\times\mathbf{P}(E))$ such that,
for each $b=(S,t)$ in $\mc{R}$, we have $t\leq s$ whenever $s\in
S$, and a function $\DD:E\mapsto\mathbf{P}(\mc{X})$ which associates to each node its multiset of \textbf{decorations}.
\end{Def}

\begin{Def}[quantifiers]\label{quant}
Let $A$ be a polymorphic arena. For every $c\in E_A$, we define the multiset $\var(c)$ by induction on $c$:
\begin{itemize}
\item $\var(x_i)=\var(x^q)=\var(\star)=\emptyset$
\item $\var((a\ppar b))=\var(a)+\var(b)$
\item $\var((a,a',1))=\var((a,a',2))=\var((a,1))=\var((a,2))=\var(¬(a))=\var(a)$
\item $\var(\forall(a))=\var(a)+\{S\}$ where $S$ is the multiset\footnote{A node appears $n$ times in $S$ if it contains $n$ bound variables of $\forall(a)$.} of bound nodes of $\forall(a)$.
\end{itemize}

The multiset $\RRR_A$ of \textbf{quantifiers} of an arena $A$ is defined by $\RRR_A=\{\{(t,S)\mid\ t\in E_A\land S\in\var(t)\}\}$.
For any quantifier $b=(t,S)$, we note $\mc{T}(b)=t$ (the \textbf{target} of $b$) and $\mc{S}(b)=S$ (the \textbf{source} of $b$). 

Finally, for 
$c\in E_A$, we note $\quant(c)=\{(c,S)\in\RRR_A\}$.
\end{Def}

\begin{Def}[free variable publisher]
Let $A$ be a polymorphic arena. For every $c\in E_A$ and $i\in\mathbb{N}$, we define the multiset $\DD_A(c)$, called the \textbf{free variable publisher}, by induction on $c$:
\begin{itemize}
\item $\DD_A(x_i)=\{X_i\}$
\item $\DD_A(x^q)=\DD_A(\star)=\emptyset$
\item $\DD_A((a\ppar b))=\DD_A(a)+\DD_A(b)$
\item $\DD_A((a,a',1))=\DD_A((a,a',2))=\DD_A((a,1))=\DD_A((a,2))=\DD_A(\forall(a))=\DD_A(¬(a))=\DD_A(a)$
\end{itemize}
\end{Def}

\begin{Lemme}
For any polymorphic arena $A$, $(E_A,\leq_A,\RRR_A,\DD_A)$ is an hyperforest.
\end{Lemme}

\textbf{Example~2:}  For the arena $A$ defined in the preceding
 example, $\var(c_1)=\{\{S_1,S_2\}\}$ where $S_i$ contains only the $i$th occurrence of 
$x^{(0,\star\ppar\star)}$. 
So, $\RRR_A=\{\{b_1,b_2\}\}$ with $b_1=(c_1,\{\{c_1\}\})$ and $b_2=(c_1,\{\{c_1\}\})$.

Besides, $\DD_A(c_1)=\DD_A(c_2)=\emptyset$ and $\DD_A(c_3)=\{\{X_2\}\}$.

Hence, the hyperforest associated to the arena $A$ can be represented graphically by:
\bec\epsfig{file=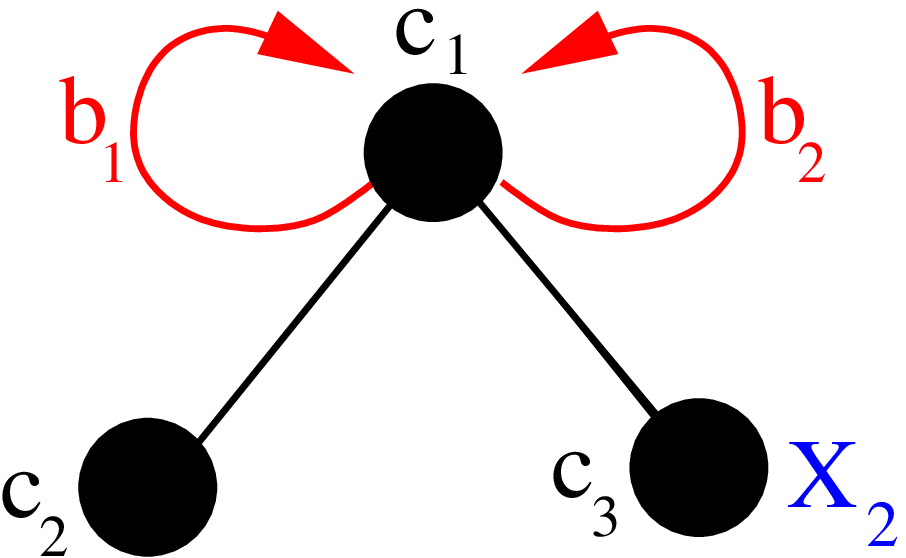,height=2cm}\eec
where straight lines stand for the relation $\leq_A$
(the smallest element is at the top), arrows stand for the
hyperedges, and decorations are attached to nodes.\finex
\medskip

\textbf{Substitution for a quantifier:}

In order to define moves in an arena, we wish to give a definition of substitution for a quantifier, \ie\ to define $A[B/b]$ for 
$A,B\in\mc{A}$ and $b\in\RRR_A$. According to the definition~\ref{quant}, the quantifier $b$ is necessarily related to a specific occurrence 
of $\forall$ in the node $\mc{T}(b)$. Moreover, we know from the proof of lemma~\ref{unique} that this occurrence of $\forall$ is itself 
related to a subformula $C=\forall X_i.C(i)$ in
the formula describing $A$. Then we define the arena $A'$, which is described by the formula $A$ where $C$ has been substituted by $C(i)$ 
(with $X_i\notin FTV(A)$), and 
we set $A[B/b]=A'[B/X_i]$ (note that this definition does not depend on the choice of $i$).

\textbf{Origin:}

As our arenas are nearly equivalent to formulas, we introduced the
substitution through formulas. However, we could have given an
explicit formulation of the arena $A[B/X_i]$, starting from $A$ and
$B$.  But this formulation would have been very technical, whereas for
our model we just need to know that each node of $A[B/X_i]$ is related
to a particular node of $A$. This idea is expressed in the following
lemma. The notation $[\alpha(x)/x]_{x\in V(a)}$ with
$\alpha:V(a)\fl\mc{N}$ indicates successive substitutions
$[\alpha(x)/x]$ for $x$ varying in $V(a)$.

\begin{Lemme}
Let $A$, $B$ be two polymorphic arenas, and $c$ a node of
$E_{A[B/X_i]}$.  Then there exists a unique node $a\in E_A$ and a
function $\alpha:V(a)\fl\mc{N}$ such that $c=a[\alpha(x)/x]_{x\in
  V(a)}$.


The node $a$ is called the \textbf{origin} of $c$ in $A$, and denoted $\textit{origin}(c)$.
\end{Lemme}

\textsc{Proof:} First we prove the uniqueness of the node $a$: suppose
$a,b\in E_A$ with $a[\alpha(x)/x]_{x\in V(a)}=b[\beta(x)/x]_{x\in
  V(a)}$. Then an induction on $A$ ensures that $a=b$:
\begin{itemize}
\item if $A=X_i$ or $A=\bot$ it is obvious
\item if $A=\forall X_i.A_0$ we have $a=\forall(a_0)$ and
  $b=\forall(b_0)$ with $a_0[\alpha(x)/x]_{x\in
  V(a_0)}=b_0[\beta(x)/x]_{x\in V(b_0)}$
\item if $A=A_1\ppar A_2$ then $a=a_1\ppar a_2$ or $a=(a_1,a_2,1)$ or
  $a=(a_1,a_2,2)$. Then we have respectively $b=b_1\ppar b_2$ or
  $a=(b_1,b_2,1)$ or $a=(b_1,b_2,2)$, with $a_1[\alpha(x)/x]_{x\in
  V(a_1)}=b_1[\beta(x)/x]_{x\in V(b_1)}$ and $a_2[\alpha(x)/x]_{x\in
  V(a_2)}=b_2[\beta(x)/x]_{x\in V(b_2)}$
\item the other cases are similar.
\end{itemize}



The existence of $a$ can be proved by a structural induction on $c$:

\begin{itemize}
\item if $c=x_i$ or $c=\star$ or $c=x^{(j,q)}$ the proof is trivial
\item if $c=(c',1)$ or $c=(c',2)$ then $A[B/X_i]=G_1× G_2$, and from
  the definition of substitution we deduce that either $A=X_i$ or
  $A=A_1× A_2$ with $G_1=A_1[B/X_i]$ and $G_2=A_2[B/X_i]$; in this
  case $c'$ is a node of $G_1$ or $G_2$, hence
  $c'=a_0[\alpha'(x)/x]_{x\in V(a_0)}$ with $a_0$ node of
  $A_1$ or $A_2$
\item if $c=¬(c')$ then $A[B/X_i]=¬ G$, and from the definition of
  substitution we deduce that either $A=X_i$ or $A=¬ A'$ with
  $G=A'[B/X_i]$; in this case $c'$ is a node of $G$, hence
  $c'=a_0[\alpha'(x)/x]_{x\in V(a_0)}$ with $a_0$ node of $A'$
\item if $c=c_1\ppar c_2$ then $A[B/X_i]=G_1\ppar G_2$, and from the
  definition of substitution we deduce that either $A=X_i$ or
  $A=A_1\ppar A_2$ with $G_1=A_1[B/X_i]$ and $G_2=A_2[B/X_i]$; in this
  case $c_1$ is a root of $G_1$ and $c_2$ is a root of $G_2$, hence
  $c_1=a_1[\alpha_1(x)/x]_{x\in V(a_1)}$ with $a_1$ root of
  $A_1$, and $c_2=a_2[\alpha_2(x)/x]_{x\in V(a_2)}$ with $a_2$
  root of $A_2$
\item if $c=(c_1,c_2,1)$ then $A[B/X_i]=G_1\ppar G_2$, and from the
  definition of substitution we deduce that either $A=X_i$ or
  $A=A_1\ppar A_2$ with $G_1=A_1[B/X_i]$ and $G_2=A_2[B/X_i]$; in this
  case $c_1$ is a node of $G_1$ and $c_2$ is a root of $G_2$, hence by
  induction hypothesis $c_1=a_1[\alpha_1(x)/x]_{x\in V(a_1)}$
  with $a_1$ node of $A_1$, and $c_2=a_2[\alpha_2(x)/x]_{x\in
    V(a_2)}$ with $a_2$ root of $A_2$
\item if $c=(c_1,c_2,2)$ then $A[B/X_i]=G_1\ppar G_2$, and from the
  definition of substitution we deduce that either $A=X_i$ or
  $A=A_1\ppar A_2$ with $G_1=A_1[B/X_i]$ and $G_2=A_2[B/X_i]$; in this
  case $c_1$ is a node of $G_2$, $c_2$ is a root of $G_1$ and
  $n_{c_2}<n_c$, hence by induction hypothesis
  $c_1=a_1[\alpha_1(x)/x]_{x\in V(a_1)}$ with $a_1$ node of
  $A_2$, and $c_2=a_2[\alpha_2(x)/x]_{x\in V(a_2)}$ with $a_2$
  root of $A_1$
\item if $c=\forall(c')$ then $A[B/X_i]=\forall X_k.G$ for some
  $k\in\mathbb{N}$, and from the definition of substitution we deduce
  that either $A=X_i$ or $A=\forall X_{k'}.A'$.  In this case, thanks
  to $\alpha$-equivalence, one can choose $k=k'\neq i$, and so $G=A[B/X_i]$;
  $c'$ is a node of $G$, hence $c'=a_0[\alpha'(x)/x]_{x\in
    V(a_0)}$ with $a_0$ node of $A'$.
\end{itemize}
\cqfd

As a consequence, one can also establish a notion of origin for the
substitution $A\mapsto A[B/b]$, defined similarly: for any node $c$ of
$E_{A[B/X_i]}$, there exists a unique node $a\in E_A$ and a function
$\alpha:V(a)\fl\mc{N}$ such that $c=a[\alpha(x)/x]_{x\in V(a)}$, or
$c=a'[\alpha(x)/x]_{x\in V(a)}$ where $a'$ is obtained from $a$ by
erasing one occurrence of $\forall$.


\begin{Def}[offspring]
Let $A$, $B$ be two polymorphic arenas, $c$ node of $E_{A[B/X_i]}$
(resp. of $E_{A[B/b]}$) and $a=\textit{origin}(c)$. We say that $c$ is
an \textbf{offspring} of $a$ in $A[B/X_i]$ (resp. $A[B/b]$) if $c$ is
minimal among the $c'$ such that $a=\textit{origin}(c')$.
\end{Def}

\textbf{Example~3:} Let us consider the arena $A=\forall
X_3.(¬\bot\ppar X_3)\ppar \forall X_3.(¬ X_2\ppar X_3)$ from the
preceding examples, and $B=(\neg X_1)\ppar X_1$. Then
$A[B/X_2]=\forall X_3.(¬\bot\ppar X_3)\ppar \forall X_3.(¬ ((\neg
X_1)\ppar X_1)\ppar X_3)$ can be represented as follows:
\bec\epsfig{file=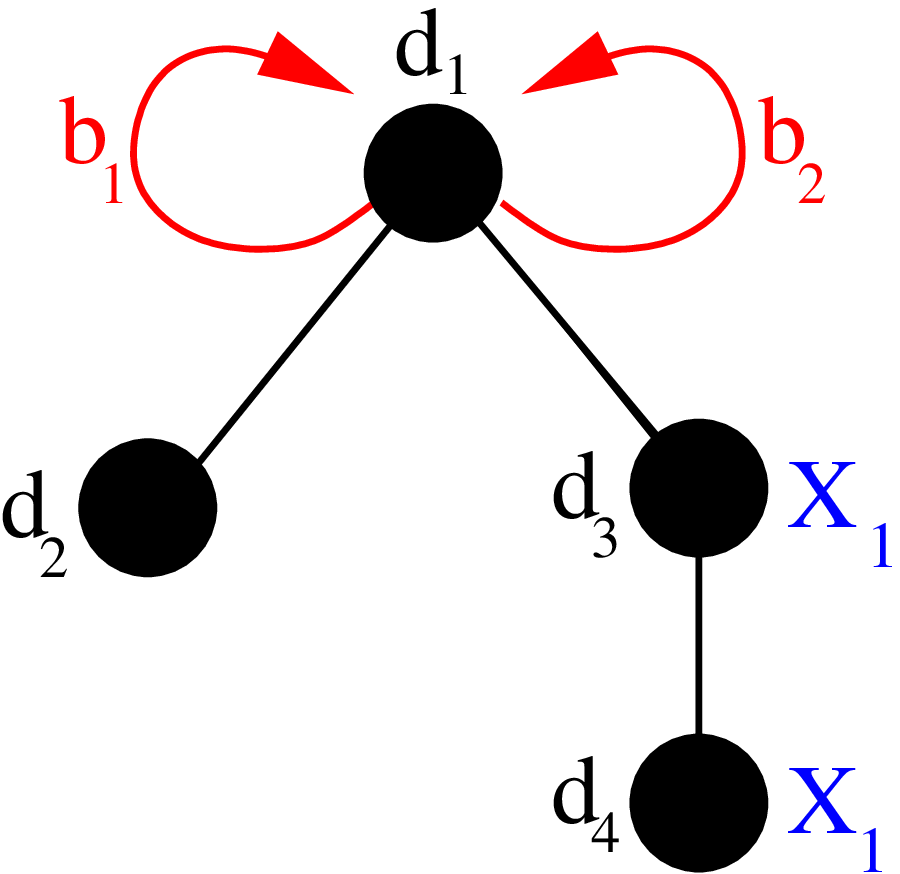,height=3cm}\eec
with:

$\begin{array}{l}
d_1=\forall(\star\ppar x^{(0,\star\ppar\star)})\ppar\forall(\star\ppar x^{(0,\star\ppar\star)})\\
d_2=((¬(\star),x^{(0,\star\ppar\star)},1),\forall(\star\ppar x^{(0,\star\ppar\star)})),1)\\
d_3=((¬(\star\ppar x_1),x^{(0,\star\ppar\star)},1),\forall(\star\ppar
x^{(0,\star\ppar\star)})),2)\\
d_4=((¬(\neq x_1,x_1,1),x^{(0,\star\ppar\star)},1),\forall(\star\ppar
x^{(0,\star\ppar\star)})),2)
\end{array}$

Consider for example the node $d_3$: we can write $d_3=a[b/x_2]$ with
$a=((¬(x_2),x^{(0,\star\ppar\star)},1),\forall(\star\ppar
x^{(0,\star\ppar\star)})),2)\in E_A$ and $b=\star\ppar x_1\in
E_B$. Hence, $a$ is the origin of $d_3$, \ie\ the part of the move
played in $A$ and $b$ is the part played in $B$. Similarily,
$d_4=a[c/x_2]$ with $c=(\neq x_1,x_1,1)\in E_B$, so the origin of $d_4$ is also
$a$. But $d_3$ is an offspring of $a$ whereas $d_4$ is not: indeed,
$d_3$ is the minimal node in $A[B/X]$ whose origin is $a$.  \finex


\textsc{Remark:} For the rest of this article, we introduce the constructor $A\fl B=(¬ A)\ppar B$.
We will generally identify $B$ (resp. $A$) to its trivially isomorphic part (resp. to its copies) in $A\fl B$.

\subsection{The notion of game in a polymorphic arena}

We now informally describe a \textbf{play} in a polymorphic
arena $A$, with parameters
$\vec{X}=(X_1,\dots,X_n)$ (this parameters will correspond
to the free type variables appearing in a term).

As in propositional HON-games, a play is a sequence of \textbf{moves}, played alternately by two players:
$\mathbf{P}$ (the Player) and $\mathbf{O}$ (the Opponent). But this time, because of second-order, playing a move does not simply consist in choosing a node in the arena: it is a more complicated process. Each
move follows different steps :
\begin{minilist}
\item one choose a node in the forest given by the moves justifying
this move
\item one \textbf{instantiates} all the quantifiers whose target are the
chosen node
\item one substitutes the arena where we are playing
\item if necessary, one chooses a new node in the arena we
obtained, and one iterates the process.
\end{minilist}

\medskip

Let us take a look further at how the process is actually working.

A \textbf{move} $m$ is played in a closed arena $H$ by
$\mathbf{P}$ or $\mathbf{O}$, who
begins by choosing a node $c_1$ in $B^0=H$, and then
\textbf{instantiates} by a closed arena each quantifier $b$ such
that $\mc{T}(b)=c_1$. This modifies the arena consequently:
each node $d\in\mc{S}(b)$ from $B^0$ is replaced by the
closed arena instantiating $b$, this gives us the closed arena
$B^1$.

But what happens if $c_1\in\mc{S}(b)$? Once the node has been
replaced by its instantiation $G$, one may have a doubt on the
node the player really chose: for example, if $G$ contains two
trees, one must say which of the two roots has been chosen. Moreover, 
if the chosen node contains quantifiers, they have to be 
instantiated as well. So,
the player has to choose another node $c_2$ in $B^1$, and possibly
to instantiate the corresponding quantifiers to obtain a new
closed arena $B^2$, etc. This process is an "horizontal" enlarging
\footnote{In $\lambda\mu2$, this horizontal enlarging would
correspond to a term with successive type instantiations, like
$t\{\forall X.X\}\{\forall Y.\bot\fl Y\}\{\forall Z.Z\fl Z\}...$.},
in the sense that one does not go deeper in the closed arena, but
one makes it evolve until there is no ambiguity on the chosen
node. The player finally stops on a node $c_n$ such that there is
no $b$ for which $c_n\in\mc{S}(b)$. $c_n$ is called the
\textbf{resulting node} of $m$, and the closed arena $B^n$ is its
\textbf{resulting arena}.

As in a propositional setting, a play in a polymorphic arena $A$ is a
sequence of moves with a relation of \textbf{justification}, but this
time if $m_i$ justifies $m_j$ we require $m_j$ to be played in the
resulting arena of $m_i$: consequently, a move in a play can be chosen
only when the quantifiers above it has been instantiated. Moreover,
each initial move begins with a function $\theta:\{X_1,\dots,X_n\}\fl
\mc{H}$ which instantiates every variable $X_i$ by a closed arena; the
player replaces in the arena $A$ each node decorated by $X_i$ by the
closed arena $\theta(X_i)$, and then plays a move beginning with a
root in the closed arena finally obtained: this allows to play in any
polymorphic arena instead of just in a closed one.

Explicit examples will come after the definition to make all these
intuitions clearer.

\subsection{Moves, plays and strategies}

\begin{Def}[move]
A \textbf{move} in a closed arena $H$ takes the form
$$\begin{array}{cl}
m=&[c_1:A_1^1/b_1^1;A_2^1/b_2^1;\dots;A_{k_1}^1/b_{k_1}^1][c_2:A_1^2/b_1^2;\dots;
A_{k_2}^2/b_{k_2}^2]\dots\dots\\
&[c_n:A_1^n/b_1^n;\dots;
A_{k_n}^n/b_{k_n}^{n}]\end{array}$$
with the following conditions: \begin{minilist}
\item $c_1$ is a node of
$B^1=H$ (called the \textbf{first node} of $m$) such that
$quant(c_1)=\{b_1^1,\dots,b_{k_1}^1\}$; we note
$B^2=H[A_1^1/b_1^1,\dots,A_{k_1}^1/b_{k_1}^1]$ 
\item for all $r\in[2,n]$, $c_r$ is a node of
$B^r$ such that $c_r$ offspring of $c_{r-1}$ in $B^r$ and $quant(c_r)=\{b_1^r,\dots,b_{k_r}^r\}$;
we note $B^{r+1}=B^r[A_1^r/b_1^r,\dots,A_{k_r}^r/b_{k_r}^r]$
\item for all $r\in[1,n-1]$, there exists $b_l$ such that
$c_r\in\mc{S}(b_l)$ in $B^r$; for $c_n$, there is no $b_l$ such
that $c_n\in\mc{S}(b_l)$ in $B^n$
\end{minilist}
We call \textbf{resulting arena} of $m$ the closed arena $B^{n+1}$,
and
\textbf{resulting move} the move $c_n$, which is a root of $B^n$.
\end{Def}

\textsc{Remark: }In order to avoid any confusion, take care to the fact that, in this model, the notions of \textit{node} 
and \textit{move} do not coincide as they did in the propositional model.

\begin{Def}[initial move]
An \textbf{initial move} in a polymorphic arena $A$ with
parameters $\vec{X}=(X_1,\dots,X_n)$ ($FTV(A)\subseteq \vec{X}$) takes the form $m=\theta m'$ where
$\theta:\{X_1,\dots,X_n\}\fl\mc{H}$ and $m'$ is a move in the closed arena $H=A[\theta(X_1)/X_1]\dots[\theta(X_n)/X_n]$
such that the first node of $m'$ is a root of $H$.
\end{Def}

\textbf{Example~4:} Consider the arena $A=\forall X_3.(¬\bot\ppar X_3)\ppar \forall X_3.(¬ X_2\ppar X_3)$ described in the example~1, 
and the arenas $H_1=\forall X.¬ X$, $H_2=\bot$, $H_3=\bot$.

Then $m=\theta[c_1:H_1/b_1,H_2/b_2][c':H_3/b_3]$, where $\theta(X_1)=\theta(X_2)=\bot$, $c'$ is the root of 
$A'=((¬\bot)\ppar(\forall X_4.¬ X_4))\ppar((¬\bot)\ppar\bot)$ and $b_3$ is the unique quantifier of $A'$,
is an initial move in $A$ with parameters $\vec{X}=(X_1,X_2)$.

The resulting arena of this move is $A''=((¬\bot)\ppar(¬\bot))\ppar((¬\bot)\ppar\bot)$, which can be represented graphically as:
\bec\epsfig{file=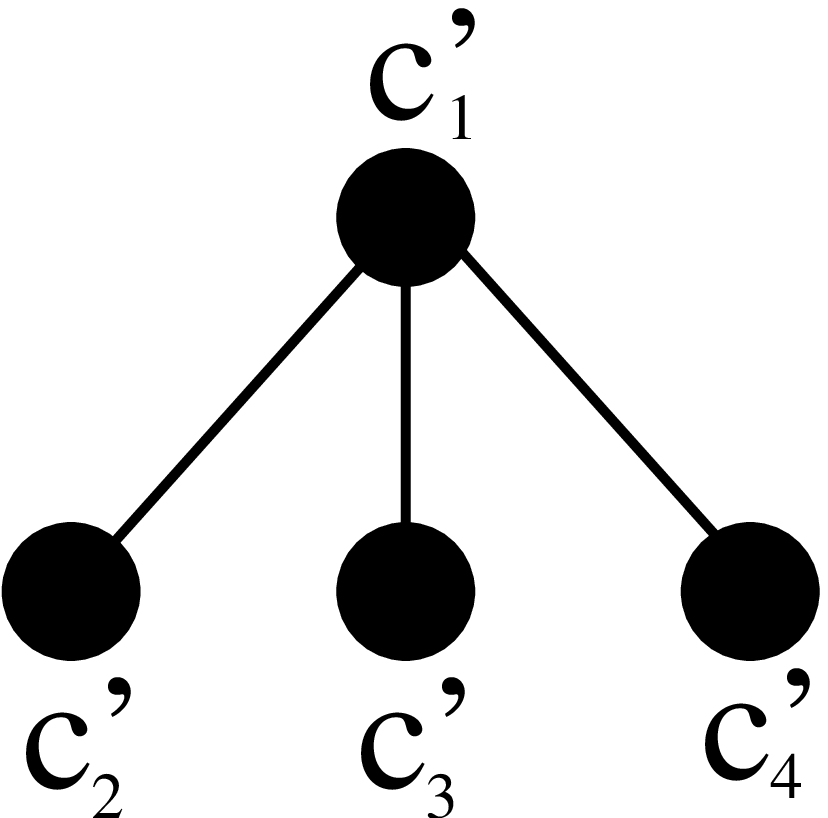,height=2.2cm}\eec

with:

$
\begin{array}{l}
c'_1=(\star\ppar\star)\ppar(\star\ppar\star)\\
c'_2=((¬(\star),\star,1),\star\ppar\star,1)\\
c'_3=((¬(\star),\star,2),\star\ppar\star,1)\\
c'_4=((¬(\star),\star,1),\star\ppar\star,2)\\
\end{array}$

\finex

With these definitions, the internal structure of moves carries all
the second-order complexity, so that the external structure will
now take the same form as in a propositional setting: the
definitions of plays, strategies, etc, will be the standard ones.

\begin{Def}[justified sequence, play]
A \textbf{justified sequence} in a polymorphic arena $A$ with
parameters $\vec{X}$ is a finite sequence of (initial and
non-initial) moves $s=m_1\dots m_n$, together with a partial
function $\fref:\{m_1,\dots,m_n\}\rightharpoonup\{m_1,\dots,m_n\}$
such that\footnote{Note that $\fref$ is actually a partial
function from occurrences of moves to occurrences of moves.}:
\begin{minilist}
\item if $\fref(m_i)$ is not defined, then $m_i$ is an initial
move in $A$ with parameters $\vec{X}$
\item if $\fref(m_j)=m_i$ then $i<j$ and $m_j$ non-initial move
in the closed arena $H^i$ (resulting arena of $m_i$) such
that its first node is a son of the resulting node of $m_i$.
\end{minilist}

In a justified sequence $s$, we say that a move $m$ is played by
$\mathbf{P}$ (resp. by $\mathbf{O}$) if the greatest natural
number $n$ such that $\text{ref}^n(m)$ is defined is odd (resp.
even); then we note $\lambda(m)=\mathbf{P}$ (resp.
$\lambda(m)=\mathbf{O}$). 

A \textbf{play} is a justified sequence
$s=m_1\dots m_n$ such that, for all $1\leq j\leq n-1$, we have
$\lambda(m_{j+1})\neq\lambda(m_j)$. The set of plays on $A$ with
parameters $\XX$ is denoted $\mc{P}_{\XX}(A)$.
A \textbf{thread} is a play $s=m_1\dots m_n$ such that $m_1$ is
the only initial move in $s$. A $P$-view (resp. an $O$-view) is a play 
$s=m_1\dots m_n$ such that $\fref(m_j)=m_{j-1}$ for each $j$ odd (resp. even).
A \textbf{bi-view} is both a $P$-view and an $O$-view.
\end{Def}

\textbf{Example~5:}\\ 
Consider the term
$t=\Lambda Z.\Lambda X.\lambda x^{\forall Y.Y}.\lambda u^X.\lambda
v^X.(x\{\forall U.U\}\{X\fl\bot\})u$ of type $A=\forall Z\forall X.(\forall Y.Y)\fl X\fl X\fl \bot$.
As we see further, this term will be interpreted in our model as a set of plays: we are going to describe one of these plays.
For the sake of simplicity, we do not pay attention to the name of nodes, we only explicit the hyperforest structure associated to an arena.

The starting arena is $G_0$, interpretation of $A$:

\btab{m{3cm}@{}m{3cm}}
$G_0\quad =$&\epsfig{file=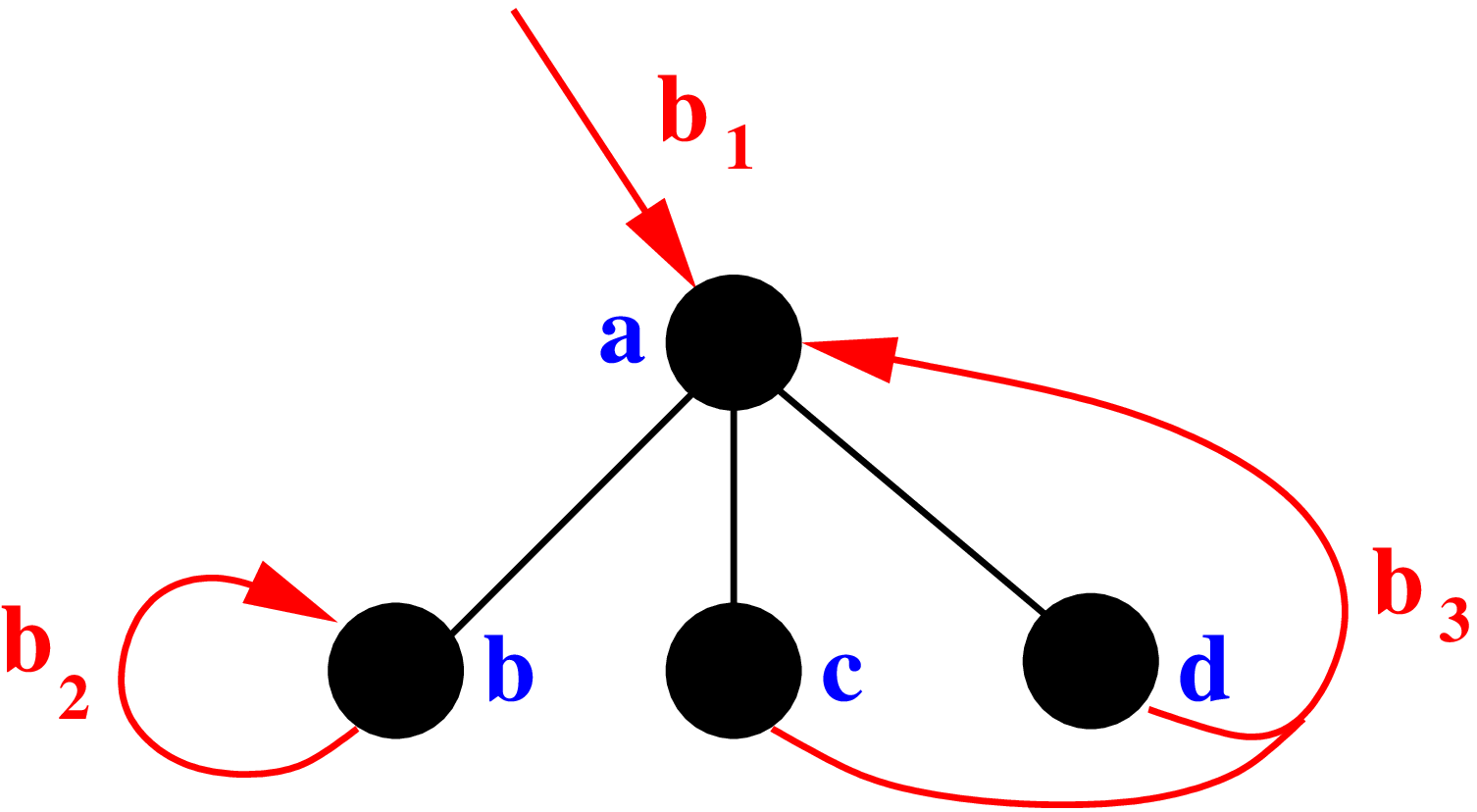,height=2.5cm}
\etab

We define three other arenas:

\bec\btab{m{1cm}@{}m{1cm}llllm{1cm}@{}m{0.7cm}}
$H=$&\epsfig{file=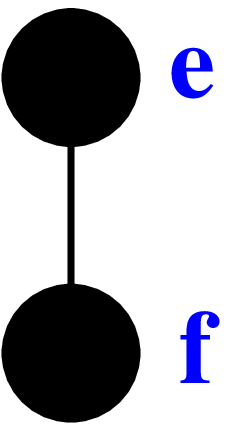,height=1cm}&$H'
=$&\epsfig{file=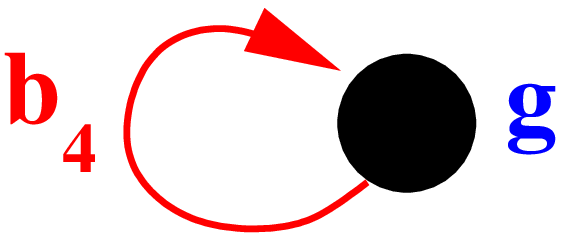,height=0.7cm}
\etab\eec

\bec\btab{m{1cm}@{}m{2cm}}
$H''=$&\epsfig{file=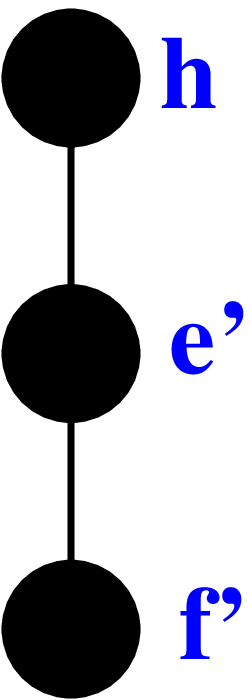,height=2cm}\etab\eec

The first move of Opponent is $m_1=\theta[a:H'/b_1;H/b_3]$ ($\theta$ does not play any role here) and it transforms $G_0$ into:

\btab{m{3cm}@{}m{3cm}}
$G_1\quad =$&\epsfig{file=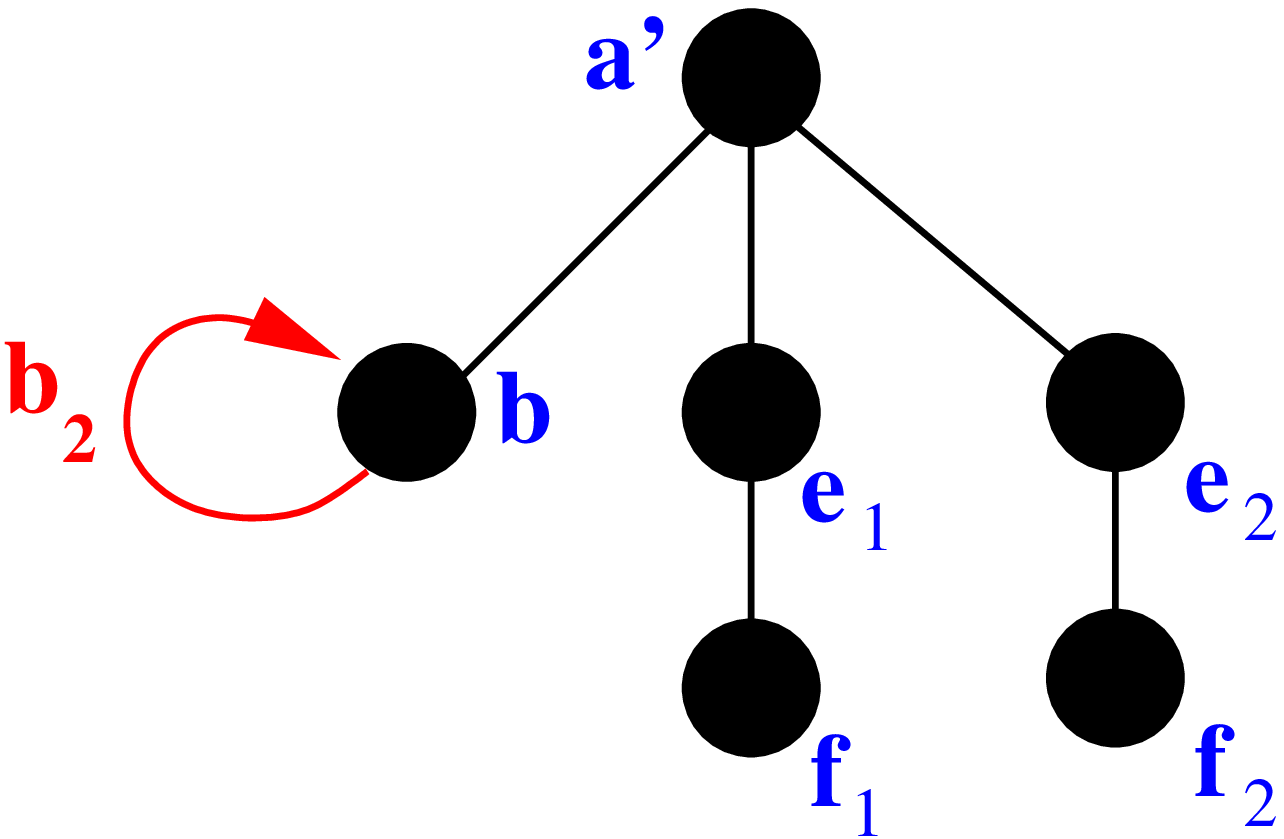,height=2.5cm}
\etab

The second move is played by Player: it is written
$m_2=[b:H'/b_2][g':H''/b_4][h':]$ and it transforms
$G_1$ successively in:

\btab{m{3cm}@{}m{3cm}}
$G_2\quad =$&\epsfig{file=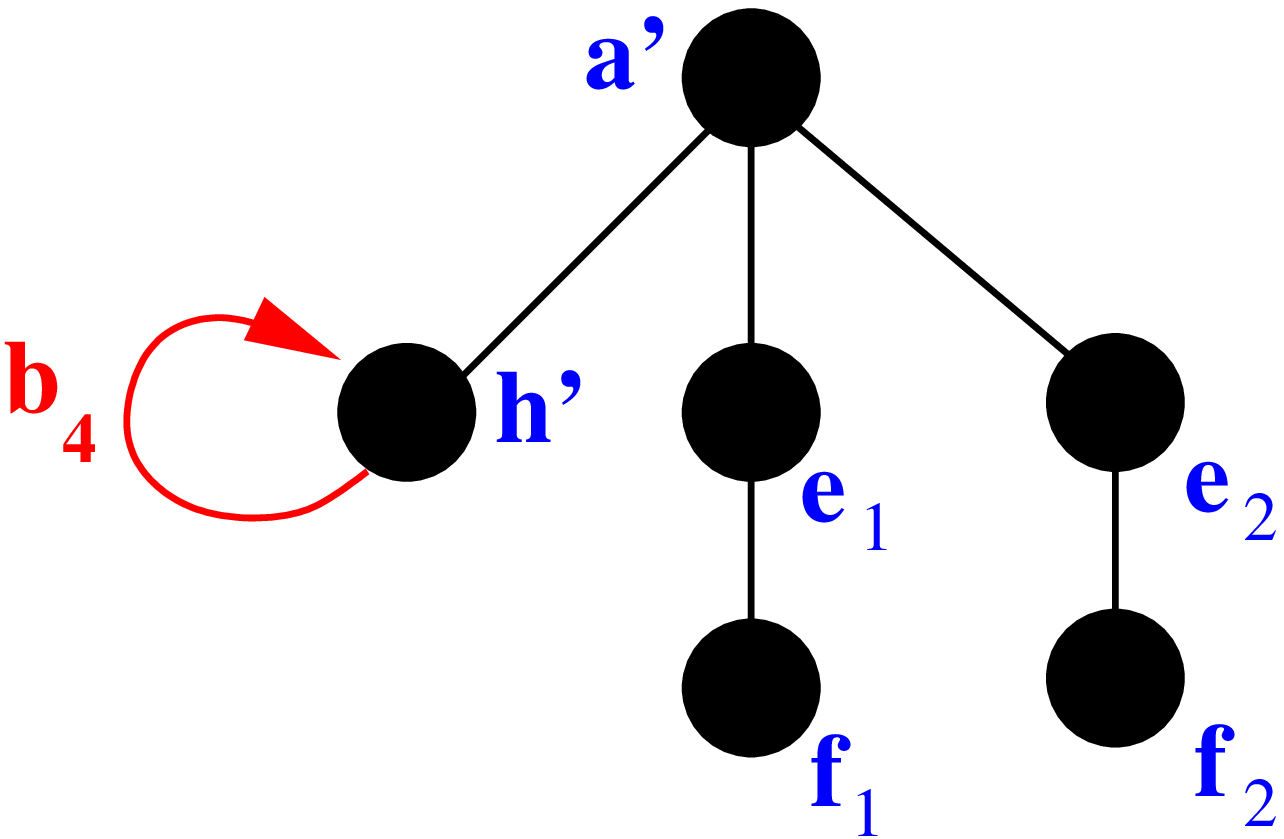,height=2.5cm}
\etab

and in:

\btab{m{3cm}@{}m{3cm}}
$G_3\quad =$&\epsfig{file=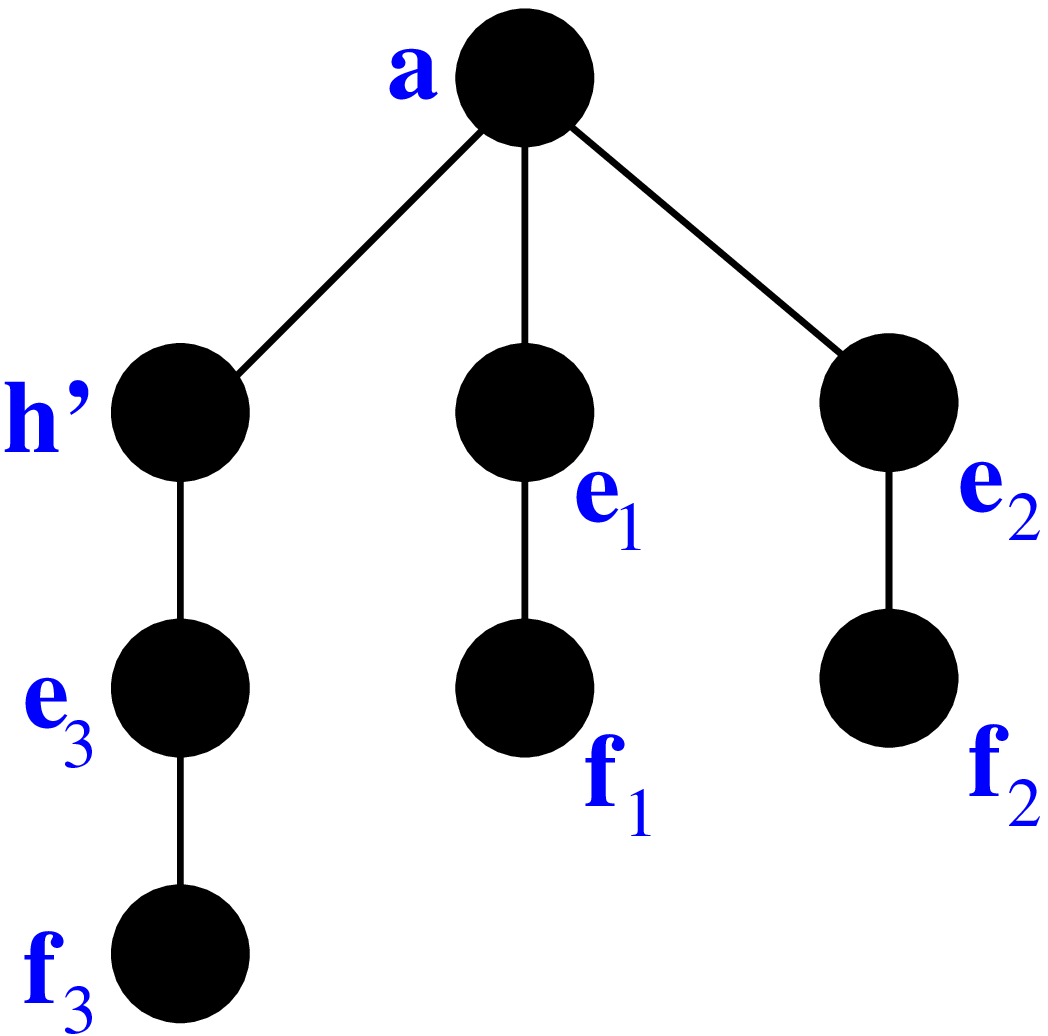,height=2.5cm}
\etab

We now understand why it is important to have a sequence of nodes with their associated closed arenas:
when we play $[b:H'/b_2]$, this corresponds to $x\{\forall U.U\}$, and when we play $[g':H''/b_4]$ this corresponds to $\{X\fl\bot\}$.
Finally, $[h':]$ simply indicates the last node we chose (there could be several choices if $b_4$ were instantiated by a product for example).

The rest of the play is a simple dialog between Opponent and Player in the arena $G_3$:
\begin{align*}
m_3&=[e_{3}:]\\
m_4&=[e_{1}:]\\
m_5&=[f_{1}:]\\
m_6&=[f_{3}:]
\end{align*}
\finex
\medskip

\begin{Def}[oldest ancestor]
The \textbf{oldest ancestor} of $m$ in a play $s$, denoted $\text{ref}^\infty(m)$,
is the move $m'$ such that $\textit{ref}^n(m)=m'$ for some $n$ and
$\textit{ref}(m')$ is not defined: it is an initial move. We note $\theta_m$ the $\theta$ function appearing at the beginning of the move
$\text{ref}^\infty(m)$.
\end{Def}

Let us consider the arena $A\fl B=\neg A\ppar B$, and
let $s\in\mc{P}_{\XX}(A\fl B)$ and $m$ be a move of $s$. Let $d$ be the origin of the first node of $m$.
There can be two cases: either $d\in B$, in which case each node appearing in $m$ is written
$(c_1,\star,2)$; then we define the move $\tilde{m}$ by replacing each node $(c_1,\star,2)$ by $c_1$. Or $d\in A$, in which case each node 
appearing in $m$ is written
$(\neg(c_1),c_2,1)$; then we define the move $\tilde{m}$ by replacing each node $(\neg(c_1),c_2,1)$ by $c_1$ and, if $d$ is a root of $A$, by
adding the function $\theta_m$ at the beginning of the move. These notations allow us to define the notion of restriction of a play:

\begin{Def}[restriction]
Let $s\in\mc{P}_{\XX}(A\fl B)$. The \textbf{restriction} of $s$ to $A$
(resp. to $B$), denoted $s\restr_A$ (resp. $s\restr_B$), is the
sequence of moves $\tilde{m_1},\dots,\tilde{m_n}$ (with the same
pointers as in $s$, wherever it is possible), where $m_1,\dots,m_n$
are the moves such that the origin of their first node is a node of
$A$ (resp. of $B$) : we say that these moves are \textbf{played} in
$A$ (resp. in $B$).
\end{Def}

With this definition, $s\restr_A$ (resp. $s\restr_B$) is a justified sequence in $A$ (resp. in $B$) with parameters $\XX$.


\begin{Def}[strategy]
A \textbf{strategy} $\sigma$ in an arena $A$ with parameters
$\vec{X}$, denoted $\sigma:A;\vec{X}$, is a non-empty set of
even-length plays of $\mc{P}_{\XX}(A)$, which is closed by
even-length prefix and deterministic: if $sa$ and $sb$ are two plays
of $\sigma$ then $sa=sb$.
\end{Def}

\begin{Def}[central strategy]
Let $\sigma:A\fl B;\XX$. $\sigma$ is \textbf{central} if
\begin{itemize}
\item in each play of $\sigma$, 
for each initial move $m$ there is exactly one move $m'$ played in $A$ and justified by $m$
\item for each initial move $m$, there is a play $mm'\in\sigma$ with $m'$ played in $A$.
\end{itemize}
\end{Def}

\subsection{Cartesian closed structure}\label{trans}

We now have the ingredients for our model: polymorphic arenas and strategies. Let us
give some basic categorical structure on these objects.

\begin{Def}[identity]
The \textbf{identity strategy} on $A$,
$\id_A:A\fl A;\XX$, is defined by $id_{A;\vec{X}}=\{s\in\mc{P}_{\XX}(A_1\fl A_2)\mid \forall t \textit{ even prefix of
}s,\ t\restr_{A_1}=t\restr_{A_2}\}$ (where $A_1$ and $A_2$ stand
for the two occurrences of $A$ in $A\fl A$).
\end{Def}

We let the reader check that this indeed defines a (central) strategy.

\begin{Def}[composition]
Let $A$, $B$, $C$ be three polymorphic arenas. An
\textbf{interaction} on $A,B,C$ with parameters $\vec{X}$ is a
justified sequence on $(A\fl B)\fl C$ with parameters $\vec{X}$
such that $u\restr_{A,B}\in\mc{P}_{\XX}(A\fl B)$,
$u\restr_{B,C}\in\mc{P}_{\XX}(B\fl C)$ and
$u\restr_{A,C}\in\mc{P}_{\XX}(A\fl C)$. We note
$\textit{int}_{\XX}(A,B,C)$ the set of such interactions.
Let $\sigma:A\fl B;\vec{X}$ and
$\tau:B\fl C;\vec{X}$, we call \textbf{composition} of $\sigma$
and $\tau$ the set of plays
$\sigma;\tau=\{u\restr_{A,C} \mid u\in int_{\XX}(A,B,C),\
u\restr_{A,B}\in\sigma\ \textit{and}\ u\restr_{B,C}\in\tau\}$
\end{Def}

We shall now recover many properties which have already been proved in
game semantics for the propositional case (see for example~\cite{phdharmer}). As we said, the inner
structure of moves is very different in second-order games, but
not the structure of plays and strategies, so that all reasonings
in the propositional case will still be valid in this case. In order
to reuse them directly instead of rewriting them, we establish a
translation of second-order objects (polymorphic arenas, moves,
...) into a propositional setting with infinite forests: basically, the idea is to associate, to 
each occurrence of a move in a play, its non-empty bi-view, and to consider it as a ``propositional'' move
(the prefix order on bi-views will give us the order relation in the arena).

Pay attention to the fact that this translation is only a tool, not a
necessary construction, and that in particular it has nothing to do
with the interpretation of $\lambda\mu2$ in a control hyperdoctrine
(although we chose the same notation).

\begin{Def}[translation] We consider one set of parameters $\vec{X}=(X_1,\dots,X_n)$.
Let $A$ be a polymorphic arena such that $FTV(A)\subset\XX$, and $\mc{BV}_A$ be the set of
non-empty bi-views on $A$. If we note $\leq$ the prefix order on bi-views,
then $\trad{A}=(\mc{BV}_A,\leq)$ is a (generally infinite) forest,
called the \textbf{translation} of the arena $A$.

Let $s\in\mc{P}_{\XX}(A)$ with $s=m_1\dots m_n$. To each occurrence $m_i$ in
$s$ one can associate the unique bi-view $v_i$ it belongs to, and
this bi-view is a move in $\trad{A}$. Then $\trad{s}=v_1\dots v_n$
(with the same pointers as $s$) is a play in $\trad{A}$, called
the \textbf{translation} of $s$.
The \textbf{translation} of a set $\sigma$ of plays on $A$ is
$\trad{\sigma}=\{\trad{s}\mid s\in\sigma\}$.
\end{Def}

\begin{Lemme}\label{trad1}
\begin{enumerate}[(i)]
\item
$\LL\sigma\RR$ is a strategy on $\LL A\RR$ if, and only if, $\sigma$ is a
strategy on $A$.
\item If $\LL \sigma\RR=\LL \tau\RR$ then $\sigma=\tau$.
\end{enumerate}
\end{Lemme}

\textsc{Proof:} 
\begin{enumerate}[(i)]
\item From the definition of the translation we immediately deduce that $\trad{\sigma}$ is a non-empty set of even-length plays, closed 
by even-length prefix if, and 
only if, it is the case for $\sigma$. 

Besides, suppose $\sigma$ is deterministic and $\trad{sa},\trad{sb}\in\trad{\sigma}$. Then $sa,sb\in\sigma$, so 
$sa=sb$ and $\trad{a}=\trad{b}$. Reciprocally, if $\trad{\sigma}$ is deterministic and $sa,sb\in \sigma$, then $\trad{sa},\trad{sb}\in\trad{\sigma}$, so
$\trad{sa}=\trad{sb}$ and $sa=sb$

\item It suffices to remark that the translation $s\mapsto\trad{s}$ on plays is injective. 
\end{enumerate}
\cqfd

But we have to take caution to the fact that we do not have $\LL
A\fl B\RR=\LL A\RR\fl\LL B\RR$. However, there is an isomorphism between $\LL
A\fl B\RR$ and a subforest of $\trad{A}\fl\trad{B}$:
to each move $\mu$ of $\trad{A\fl B}$ (which is in fact a non-empty bi-view $\mu=m_1\dots m_n$ played on $A\fl B$), one associates the move $\mu'$ defined in the
following way:
\begin{itemize}
\item if $\origin(m_n)\in B$, $\mu'=\mu$
\item if $\origin(m_n)\in A$, $\mu'=(\theta_{m_1}m_2)m_3\dots m_n$ (which is indeed a bi-view in $A$).
\end{itemize}
In what follows, we will call this operation an \textbf{adaptation}.

We can extend this definition to a play: if $s=\mu_1\dots mu_n$ then $s'=\mu'_1\dots mu'_n$, and to a set of plays:
$\sigma'=\{s'\mid\ s\in\sigma\}$, and we have:

\begin{Lemme}\label{trad2}
$\trad{\sigma;\tau}'=\trad{\sigma}';\trad{\tau}'$
\end{Lemme}

\textsc{Proof:} Note first that the first composition takes place in a second-order setting, whereas the second one is the usual propositional 
composition. Remark also that the part of $\trad{A}\fl\trad{B}$ which is isomorphic to $\trad{A\fl B}$ is composed of the nodes of 
$\trad{B}$ and of the copies of nodes of $\trad{A}$ which begin with the same $\theta$ function as their ancestor in $\trad{B}$.

Suppose $\sigma:A\fl B$ and $\tau:B\fl C$. Then
$\alpha_1=\trad{\sigma;\tau}'$ is a set of plays on $\trad{A}\fl\trad{C}$, as well as $\alpha_2=\trad{\sigma}';\trad{\tau}'$.

For each $s\in\alpha_1$, $s=t'$, where $t$ is such that there exists $u$ played on $(A\fl B)\fl C$ verifying $t=\trad{u\restr{A,C}}$, 
$u\restr{A,B}\in\sigma$ and $u\restr{B,C}\in\tau$. By translating and adapting the moves of $u$, one obtain a justified sequence
$v$ played on $(\trad{A}\fl\trad{B})\fl\trad{C}$, such that $v\restr{\trad{A},\trad{C}}=s$, $v\restr{\trad{A},\trad{B}}=(\trad{u\restr{A,B}})'$ and
$v\restr{\trad{B},\trad{C}}=(\trad{u\restr{B,C}})'$. Hence $v\restr{\trad{A},\trad{B}}\in(\trad{\sigma})'$, 
$v\restr{\trad{B},\trad{C}}\in(\trad{\sigma})'$ and $s\in\alpha_2$.

Conversely, if $s\in\alpha_2$, there exists $u$ played on $(\trad{A}\fl\trad{B})\fl\trad{C}$ such that $s=u\restr{\trad{A},\trad{C}}$,
$u\restr{\trad{A},\trad{B}}\in(\trad{\sigma})'$ and $u\restr{\trad{B},\trad{C}}\in(\trad{\tau})'$. The two last conditions
forces $u$ to be in the part of $(\trad{A}\fl\trad{B})\fl\trad{C}$ which is isomorphic to $\trad{(A\fl B)\fl C}$, so that $u$ can be 
adapted to obtain a sequence of $\trad{(A\fl B)\fl C}$.
By taking the inverse translation of this adaptation
of $u$, one obtain a justified sequence
$v$ played on $(A\fl B)\fl C$ such that $v\restr{A,B}\in\sigma$, $v\restr{B,C}\in\tau$ and $v\restr_{A,B}=t$ with $s=\trad{t}'$. 
Hence $s\in\alpha_2$.\cqfd

This result allows us to manipulate the translation of composition easily, and this yields to the following results:

\begin{Lemme}
\begin{enumerate}[(i)]
\item If $\sigma:A\fl B;\XX$ and $\tau:B\fl C;\XX$, then $\sigma;\tau$ is a strategy on $A\fl C$ with parameters $\XX$.

\item If $\sigma:A\fl B;\XX$, $\sigma;\id_B=\id_A;\sigma=\sigma$.

\item If $\sigma:A\fl B;\XX$, $\tau:B\fl C;\XX$ and $\rho:C\fl D$, we have $(\sigma;\tau);\rho=\sigma;(\tau;\rho)$.
\end{enumerate}
\end{Lemme}

\textsc{Proof:} As an example, we prove the last assertion:

$\trad{(\sigma;\tau);\rho}'=\trad{\sigma;\tau}';\trad{\rho}'=\trad{\sigma}';\trad{\tau}';\trad{\rho}'=\trad{\sigma}';\trad{\tau;\rho}'$. As it is trivial
that $\alpha'=\beta'$ iff $\alpha=\beta$, we have $\trad{\alpha}'=\trad{\beta}'$ iff $\alpha=\beta$, so $(\sigma;\tau);\rho=\sigma;(\tau;\rho)$.
\cqfd

Finally, for each sequence of variables $\XX=X_1,\dots, X_n$, we obtain a category of games : objects are polymorphic
arenas whose variables are chosen between $X_1,\dots,X_n$,
and morphisms are strategies on these arenas with parameters
$\XX$. We note $\GG_0(X_1,\dots,X_n)$
this category.

To obtain a cartesian closed structure, we add innocence:

\begin{Def}
Let $s$ be a play on an arena $A$, we define the view of $s$ (which is indeed a $P$-view), denoted $\ulcorner s\urcorner$, by:
\begin{itemize}
\item $\ulcorner\epsilon\urcorner=\epsilon$
\item $\ulcorner sm\urcorner=m$ if $m$ is an initial move
\item $\ulcorner sm\urcorner=\ulcorner s\urcorner m$ if $m$ is a
$\mathbf{P}$-move
\item $\ulcorner smtn\urcorner=\ulcorner sm\urcorner n$ if $n$ is an
$\mathbf{O}$-move justified by $m$.
\end{itemize}

A strategy $\sigma:A$ is called innocent if, for every play $sn$
of $\sigma$, the justifier of $n$ is in $\ulcorner s\urcorner$,
and if we have: if $smn\in\sigma$, $t\in\sigma$, $tm$ play in $A$
and $\ulcorner sm\urcorner=\ulcorner tm\urcorner$ then
$tmn\in\sigma$.
\end{Def}

Note that the game isomorphisms between isomorphic arenas we have built in proposition~\ref{gameariso} are innocent.

The structure of plays is preserved by translation, so that we
have : $\LL\sigma\RR$ is innocent if and only if $\sigma$ is
innocent. So, $\id_A$ is innocent and if $\sigma$ and $\tau$ are
innocent then $\sigma;\tau$ is innocent.

Thanks to these properties, we obtain a subcategory of
$\GG_0(X_1,\dots,X_n)$ by considering only innocent
strategies. We denote this subcategory
$\GG(X_1,\dots,X_n)$.

\begin{Prop}
$\GG(X_1,\dots,X_n)$ is a cartesian closed category.
\end{Prop}

\textsc{Proof:} Let $A$ and $B$ be two arenas in propositional game semantics, we note $A\cong B$ when 
there is an isomorphism between the forests $A$ and $B$.

We have isomorphisms $\trad{A× B}\cong\trad{A}×\trad{B}$ and 
$\trad{A\fl B}\cong\trad{A}\fl\trad{B}$, that we treat as equalities.

We introduce the projections $\pi_1=\{s\in\mc{P}_{\XX}(A× B\fl A)\mid\ s\restr_{A\fl A}\in\id_A\}$,
$\pi_2=\{s\in\mc{P}_{\XX}(A× B\fl A)\mid\ s\restr_{B\fl B}\in\id_B\}$ and the product of strategies 
$\sigma×\tau=\{s\in\mc{P}_{\XX}((A× C)\fl (B× D))\mid\ s_{A\fl B}\in\sigma\land s_{C\fl D}\in\tau\}$,
and we check that 
$\trad{\pi_1}'=\pi_1:\trad{A}×\trad{B}\fl\trad{A}$,
$\trad{\pi_2}'=\pi_2:\trad{A}×\trad{B}\fl\trad{B}$ and $\trad{\sigma×\tau}'=\trad{\sigma}'×\trad{\tau}':(\trad{A}×\trad{C})\fl(\trad{B}×\trad{C})$.
Similarly, we introduce 
$A\fl\sigma=\{s\in\mc{P}_{\XX}((A\fl B)\fl(A\fl C))\mid\ s\restr_{A\fl A}\in\id_A\land s\restr_{B\fl C}\in\sigma\}$
and we can check that $\trad{A\fl\sigma}=\trad{A}\fl\trad{\sigma}$.

All the commutative diagrams and unicity properties we need to check are guaranteed by lemmas~\ref{trad1} and~\ref{trad2}.
\cqfd

\subsection{Construction of an hyperdoctrine}

In this section, we set up the operations that will give rise to an hyperdoctrine.

The first ingredient of this construction is a base category $\mathbb{B}$: here it has as objects natural numbers and
as morphism $n\fl m$ the $m$-tuples $\langle
A_1,\dots,A_m\rangle$, where $A_i\in\mc{G}(X_1,\dots,X_n)$ for
$1\leq i\leq m$.
The composition in this category is substitution:
if $\vec{A}=\langle A_1,\dots,A_m\rangle:n\fl m$ and $\vec{B}=\langle
B_1,\dots,B_n\rangle:k\fl n$ then $\vec{A}\circ\vec{B}=\langle
A_1[\vec{B}/\vec{X}],\dots,A_m[\vec{B}/\vec{X}]\rangle:k\fl m$.

We then have to define a functor $\mc{G}:\mathbb{B}^{op}\fl \mathbf{CCC}$
(where $\textbf{CCC}$ is the category of
cartesian closed categories with strict morphisms of ccc's).
We choose $\mc{G}(k)=\mc{G}(X_1,\dots,X_k)$, and for each
$\vec{C}:n\fl m$ we define $\GG(\vec{C}):\mc{G}(m)\fl \mc{G}(n)$ (the specialization functor):
\begin{minilist}
\item for every $A\in\mc{G}(m)$, we set $\GG(\vec{C})(A)=A[\vec{C}]$
\item for every $\sigma:A\fl B;X_1,\dots,X_m$, we set $\GG(\vec{C})(\sigma)=\sigma[\vec{C}]$ where $\sigma[\vec{C}]:A[\vec{C}];X_1,\dots,X_n$ is defined by:
$s\in\sigma[\vec{C}]$
if and only if $\tilde{s}\in\sigma$, where $\tilde{s}$ is obtained
by replacing each initial move $m_{in}=\theta m'_{in}$ in $s$ by
$m''_{in}=\theta'm'_{in}$ with
$\theta'(X_k)=C_k[\theta(X_1)/X_1,\dots,\theta(X_m)/X_m]$.
\end{minilist}

We let the reader ascertain that if $\sigma$ is an innocent strategy then $\sigma[\vec{C}]$ is an innocent strategy. 

We check that this actually gives us a functor $\mc{G}:\mathbb{B}^{op}\fl \mathbf{CCC}$: indeed, $\GG(\vec{C}):\mc{G}(m)\fl \mc{G}(n)$
is a strict morphism of ccc's (we know that $(A× B)[\vec{C}]=A[\vec{C}]× B[\vec{C}]$, $(A\fl B)[\vec{C}]=A[\vec{C}]\fl B[\vec{C}]$,
we have to check that $(\sigma×\tau)[\vec{C}]=\sigma[\vec{C}]×\tau[\vec{C}]$, $(\sigma;\tau)[\vec{C}]=\sigma[\vec{C}];\tau[\vec{C}]$, etc...).
And the composition coincides with substitution: $\GG(\vec{C})\circ\GG(\vec{C}')=\GG(\vec{C}'[\vec{C}])$.

One can also check easily that composing this functor with the forgetful functor $\textit{ob}:\mathbf{CCC}\fl \mathbf{Set}$ gives us the representable 
functor $\mathbb{B}(-,1)$.

For what follows, we need an additional definition:
\begin{Lemme}
Let $m$ be a move in an arena $H$, such that $H=(A\fl B)[C/X]$ for some arenas $A$, $B$ and $C$. Suppose that the origin of $m$
is a root of $A$ (resp. of $B$). 
Then there is a unique move $m'$ in the arena $(\forall X.A)\fl
B[C/X]$ (resp. $A[C/X]\fl(\forall X.B)$) such that:
\begin{itemize}
\item the arenas occurring in $m'$ are the same as the ones occurring in $m$, plus the arena $C$
\item the resulting arena and the resulting nodes of $m$ and $m'$ are the same.
\end{itemize}
The move $m'$ is called the \textbf{abstraction} of $m$ along $A\fl B$.
\end{Lemme}

\textsc{Proof:} Consider the case where $m$ is played in $B$ and suppose
$m=\theta[c_1:A_1^1/b_1^1;A_2^1/b_2^1;\dots;A_{k_1}^1/b_{k_1}^1]\dots[c_n:A_1^n/b_1^n;\dots;
A_{k_n}^n/b_{k_n}^{n}]$. We define $c_0=\origin(c_1)$ and $b_0$ the quantifier of target $c_0$ corresponding to $\forall X$ 
in $\forall X.B$.

If $X_{n+1}\notin\DD(c_0)$, then
$m'=\theta'[c_1:C/b_0;A_1^1/b_1^1;\dots;A_{k_1}^1/b_{k_1}^1]\dots[c_n:A_1^n/b_1^n;\dots;\newline
A_{k_n}^n/b_{k_n}^{n}]$. Otherwise, let $c'_1$ be the origin of $c_1$ in $B^1=(A\fl B)[\theta(X_k)/X_k]_k$ 
(\ie\ \textit{before} the substitution $C/X$). The quantifers among $b_1^1,\dots,b_{k_1}^1$ which come from $C$ are named
$b_{i^1_1}^1,\dots,b_{i^1_{p1}}^1$, the others are named $b_{j^1_1}^1,\dots,b_{j^1_{q1}}^1$. Let $c'_2$ be the origin of $c_2$ in 
$B^2=B^1[A_{i^1_1}^1/b_{i^1_1}^1,\dots,A_{i^1_{p1}}^1/b_{i^1_{p1}}^1]$ 
(\ie \textit{before} the substitutions $A_{j^1_1}^1/b_{j^1_1}^1,\dots,A_{j^1_{q1}}^1/b_{j^1_{q1}}^1$). The quantifers among 
$b_1^2,\dots,b_{k_2}^2$ which come from $A_{i^1_1}^1,\dots,A_{i^1_{p1}}^1$ are named
$b_{i^2_1}^2,\dots,b_{i^2_{p2}}^2$, the others are named $b_{j^2_1}^2,\dots,b_{j^2_{q2}}^2$, etc. By iterating this process, 
we finally get to a node $c'_n$ which is the origin of 
$c_n$ before some substitutions $A_{j^{n-1}_1}^{n-1}/b_{j^{n-1}_1}^{n-1},\dots,A_{j^{n-1}_{q(n-1)}}^{n-1}/b_{j^{n-1}_{q(n-1)}}^{n-1}$. 
The quantifers among 
$b_1^n,\dots,b_{k_n}^n$ which come from $A_{i^{n-1}_1}^{n-1},\dots,A_{i^{n-1}_{p(n-1)}}^{n-1}$ are named
$b_{i^n_1}^n,\dots,b_{i^n_{pn}}^n$, the others are named $b_{j^n_1}^n,\dots,b_{j^n_{qn}}^n$. If $pn=0$ then
$m'=\theta[c'_1:C/b_0;A_{j^1_1}^1/b_{j^1_1}^1;\dots;A_{j^1_{q1}}^1/b_{j^1_{q1}}^1]\dots[c'_n:A_{j^n_1}^n/b_{j^n_1}^n;\dots;A_{j^n_{qn}}^1/b_{j^n_{qn}}^n]$; if $pn\neq 0$ then $m'=\theta[c'_1:C/b_0;A_{j^1_1}^1/b_{j^1_1}^1;\dots;A_{j^1_{q1}}^1/b_{j^1_{q1}}^1]\dots[c_n:A_{i^n_1}^n/b_{i^n_1}^n;\dots;A_{i^n_{pn}}^1/b_{i^n_{pn}}^n]$.

We let the reader check that the move defined above is in $(A[C/X]\fl(\forall X.B)$. Note that this technical definition is just the natural way to define a node of $A[B/X]\fl(\forall X.B)$, starting from a move in $A[B/X]\fl B[C/X]$.

The case where $m$ is played in $A$ is similar.
\cqfd

In the category $\mathbb{B}$, the projection is $\vec{X}=\langle
X_1,\dots,X_n\rangle:n+1\fl n$. It gives us a functor
$\vec{X}^*=\GG(\XX):\GG(n)\fl \GG(n+1)$. We have to find a right adjoint for $\vec{X}$, and for this we introduce the notion of morphism quantification:
\begin{Def}[morphism quantification]
Let $\sigma:A\fl B;X_1,\dots,X_n,X_{n+1}$. We define the strategy
$\forall\sigma:(\forall X_{n+1}.A)\fl(\forall X_{n+1}.B);X_1,\dots,X_n$ as the set of plays $\forall
s$ for $s\in\sigma$, where $\forall s$ is defined from $s$ via the following operations:
\begin{minilist}
\item each initial move
$m=\theta m_B$ is replaced by $m'=\theta' m'_B$, where $\theta'(X_k)=\theta(X_k)$ for $1\leq k\leq n$, and 
$m'_B$ is the abstraction of $m_B$ along $(A\fl
  B)[\theta(X_1)/X_1,\dots,\\ \theta(X_n)/X_n]$
\item each move $m_A$ whose origin is a root of $A$ is replaced by its abstraction along $(A\fl B)[\theta(X_1)/X_1,\dots,\theta(X_n)/X_n]$.
\end{minilist}
\end{Def}

Once again, we let the reader verify that if $\sigma$ is an innocent strategy, then $\forall\sigma$ is an innocent strategy.

We can now define the
functor $\Pi_n:\GG(n+1)\fl \GG(n)$ by $\Pi_n(A)=\forall X_{n+1}.A$
and $\Pi_n(\sigma)=\forall \sigma$.

\begin{Prop}
$\Pi_n$ is a right adjoint of $\vec{X}^*$.
\end{Prop}

\textsc{Proof:} We first have to establish for each
$C\in\mc{G}(n)$ a bijection $\kappa
:\GG(n+1)(\vec{X}^*(C),A)\fl
\GG(n)(C,\forall X_{n+1}A)$. We notice that
$\vec{X}^*(C)=C[\vec{X}]=C$.

If $\sigma:C\fl A;X_1,\dots,X_n,X_{n+1}$, $\kappa(\sigma)=\{\kappa(s)\mid\ s\in\sigma\}$, where $\kappa(s)$ is obtained from $s$ by
replacing each initial move of the form $m=\theta m_0$ by an initial move 
$m'=\theta'm'_0$ in $C\fl\forall X_{n+1}A$ such that $\theta'(X_k)=\theta(X_k)=$ for $1\leq k\leq n$, and $m'_0$ is the abstraction of $m_0$
along $C\fl A[\theta(X_1)/X_1,\dots,\theta(X_n)/X_n]$.

We finally just need to check the naturality of this bijection, namely
that $\tau;\kappa(\sigma)=\kappa(\vec{X}^*(\tau);\sigma)$ and
$\kappa(\tau;\sigma)=\kappa(\tau);\Pi_n(\sigma)$.
This comes directly from the action of $\vec{X}^*$ and $\Pi_n$ on
strategies.\cqfd

\begin{Lemme}
$\Pi_n$ is natural in $n$: $\GG(\vec{C})\circ\Pi_n=\Pi_m\circ\GG(\vec{C},X_{m+1})$.
\end{Lemme}

\textsc{Proof:} This is easy to check for objects: for $A\in\GG(n+1)$, the formulas representing respectively $(\forall X_{n+1}.A)[\vec{C}]$ and 
$\forall X_{m+1}.A[\vec{C},X_{m+1}]$ are $\alpha$-equivalent, hence the arenas are equal.

On morphisms, this requires to check that $(\forall\sigma)[\vec{C}]=\forall(\sigma[\vec{C},X_{m+1}])$, which is easy because the substitution 
$\sigma\mapsto\sigma[\vec{C}]$ does only modify the $\theta$ function of the initial moves.
\cqfd

One can now conclude, using the results of~\cite{hyperdoc2,hyperdoc3}:

\begin{Th} The structure $\mc{M}$ defined by the base category $\mathbb{B}$ and the functor 
$\mc{G}:\mathbb{B}^{op}\fl \mathbf{CCC}$
is an hyperdoctrine, and therefore a model of system F.
\end{Th}

The interpretation of a type $A$ in this model is a polymorphic arena $A^*$, whereas the interpretation
of a typing derivation ending with the judgement $\XX;x_1:A_1,\dots,x_n:A_n\vdash t:A$ is a strategy $\sigma_t:A_1^*×\dots× A_n^*\fl A^*;\XX$.

\subsection{Arena isomorphisms}

Before defining our control hyperdoctrine, we introduce the notions of arena isomorphisms and game isomorphisms that will be useful to define some structural morphisms. Note that these notions will become really important when dealing with type isomorphisms.

There are two ways to define an isomorphism between arenas: it can be either 
an isomorphism using strategies, or a (trivial) geometrical equality between hyperforests. We prove here that the first notion of isomorphism is implied by the latter.

\begin{Def}[arena isomorphism]
Let $A$ and $B$ two polymorphic arenas. We say that there is an \textbf{arena isomorphism} between $A$ and $B$ if there is a bijection $g:E_A\fl E_B$
preserving the hyperforest structure:
$g(\RRR_A)=\RRR_B$ and $\DD_B\circ g=\DD_A$. We note this $g:A\simeq_aB$, or simply $A\simeq_aB$.
\end{Def}

\textbf{Example~6:} The arenas $A=\forall X.\forall Y.((\forall Z.(\bot\times Z)\fl X)\times(\forall U.U))\fl\bot$
and $B=(\forall X.X)\fl(\forall Y.(\forall Z.\bot\fl Z\fl Y)\fl(\forall U.\bot))$
are isomorphic: 
indeed, they can be described by the following hyperforests:

\bec\epsfig{file=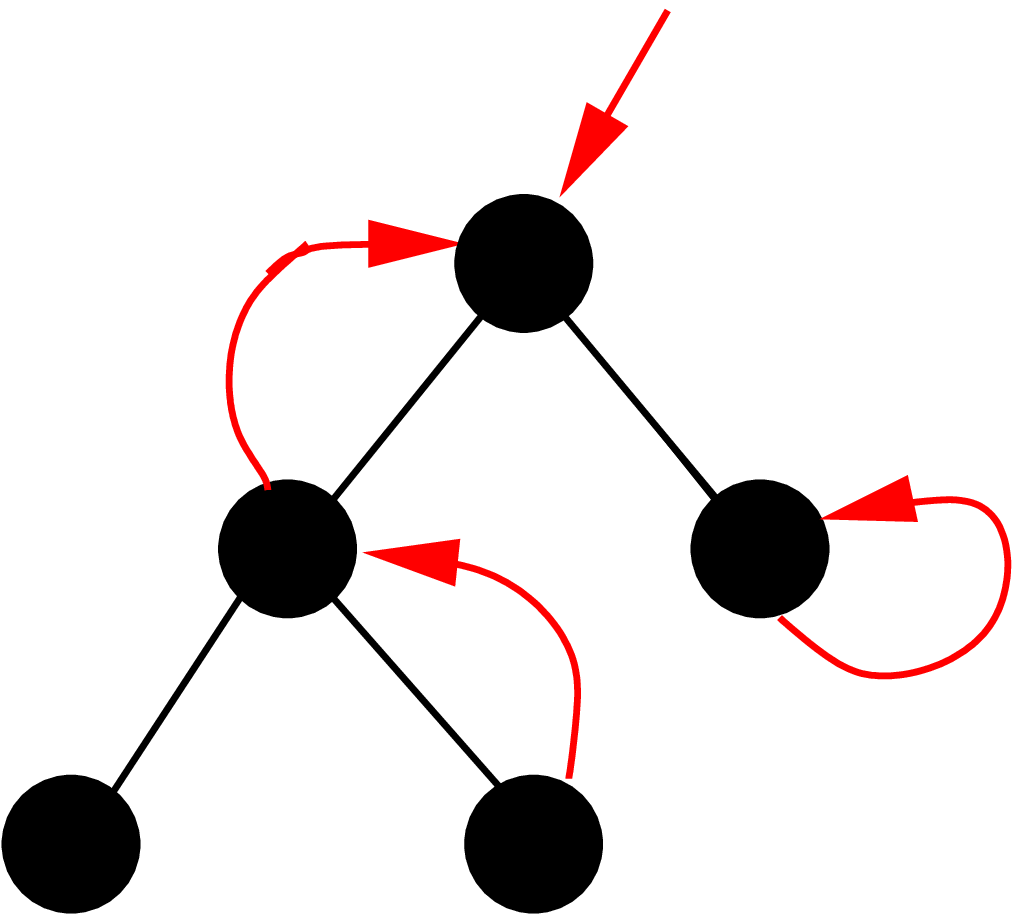,height=2.5cm}\qquad\text{ and }\qquad\epsfig{file=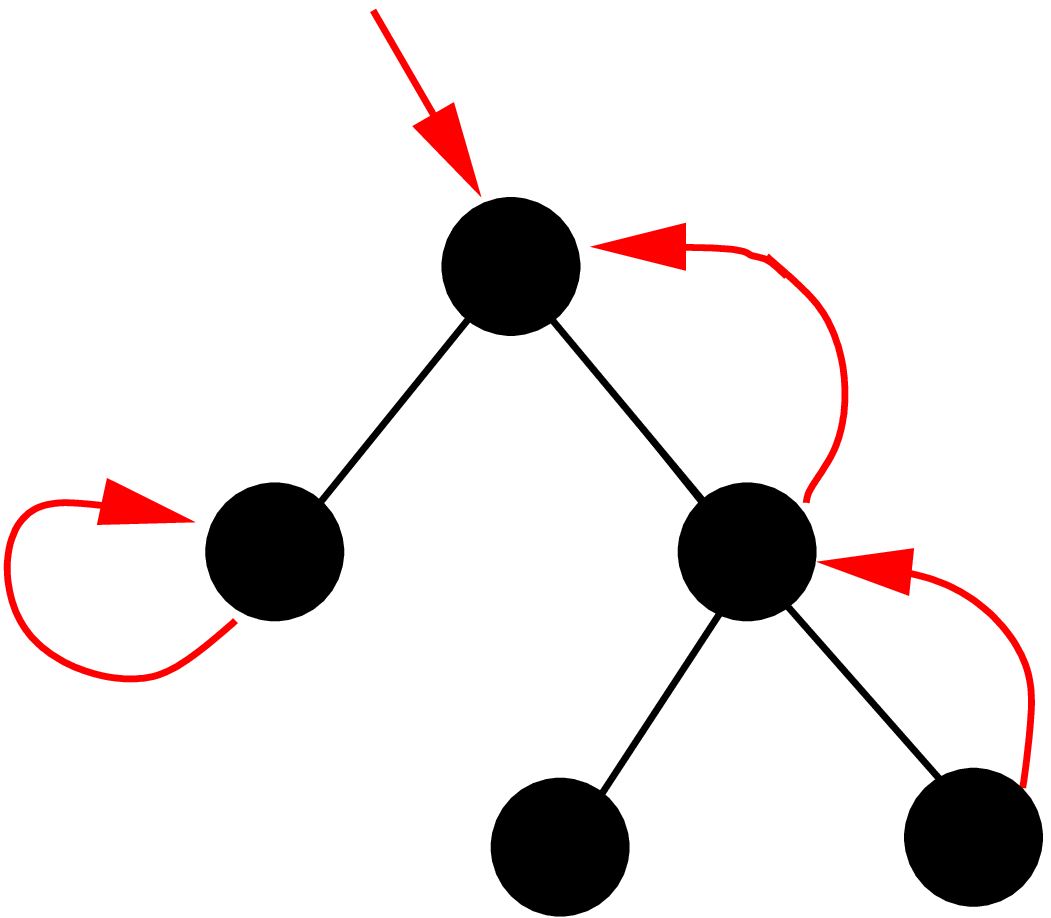,height=2.5cm}\eec
\finex
\medskip

\begin{Def}[game isomorphism]
Let $A$ and $B$ two polymorphic arenas. We say that there is a \textbf{game isomorphism} $(\sigma,\tau)$ between $A$ and $B$ 
($A\simeq_gB$) if there are two strategies $\sigma:A\fl B;\XX$ and $\tau:B\fl A;\XX$ such that $\sigma;\tau=\id_A$ and $\tau;\sigma=\id_B$. We note this 
$(\sigma,\tau):A\simeq_gB$, or simply $A\simeq_gB$.
\end{Def}

\begin{Prop}\label{gameariso}
Let $A$ and $B$ two polymorphic arenas such that $A\simeq_aB$. Then there is a game isomorphism $(\sigma,\tau)$ between $A$ and $B$;
moreover, $\sigma$ and $\tau$ are central strategies.
\end{Prop}

\textsc{Proof:} We wish to extend the function $g$ into a function on plays.

First note that if $g:A\simeq_aB$ then $A[C/X_i]\simeq_aB[C/X_i]$ and, if $b\in\RRR_A$, $A[C/b]\simeq_aB[C/g(b)]$. We note
$g[C/X_i]$ (or $g[C/b]$) the function realizing this isomorphism. For a given move 
$m=(\theta)[c_1:A_1^1/b_1^1;\dots;A_{k_1}^1/b_{k_1}^1]\dots[c_p:A_1^p/b_1^p;\dots;A_{k_p}^n/b_{k_p}^{p}]$ played in $A$ 
(the notation $(\theta)$ indicates that a $\theta$ function may appear or not), we define
$\tilde{g}(m)=(\theta)[g^1(c_1):A_1^1/g^1(b_1^1);\dots;A_{k_1}^1/g^1(b_{k_1}^1)]\\
\dots[g^p(c_p):A_1^p/g^p(b_1^p);\dots;A_{k_p}^n/g^p(b_{k_p}^{p})]$
with $g^1=g[\theta(X_1)/X_1]\dots[\theta(X_n)/X_n]$ and $g^{i+1}=g^{i}[A_1^i/g^i(b_1^i)][A_{k_i}^i/g^i(b_{k_i}^i)]$.

We note $\tilde{g}/m$ the function realizing the isomorphism between the resulting arenas of $m$ and $\tilde{g}(m)$ (\ie\ $\tilde{g}=g^{p+1}$).
For a given play $s=m_1,\dots m_n$, we define the functions $g_j$ by:
$g_1=g$ and $g_{j+1}=\tilde{g_j}/m_j$. Finally, we set $\bar{g}(s)=\tilde{g_1}(m_1)\dots \tilde{g_n}(m_n)$.

Consider $\sigma=\{s\in\mc{P}_{\XX}(A\fl B)\mid \forall t \textit{ even prefix of
}s,\ t\restr_{A}=\bar{g}(t\restr_{B})\}$ and $\tau=\{s\in\mc{P}_{\XX}(B\fl A)\mid \forall t \textit{ even prefix of
}s,\ t\restr_{B}=\bar{g}(t\restr_{A})\}$. Then $\sigma$ and $\tau$ are indeed central strategies from on $A\fl B$ and $B\fl A$
respectively, and they verify $\sigma;\tau=\id_A$ and $\tau;\sigma=\id_B$.

The game isomorphisms we have constructed here will sometimes be called the \textbf{trivial isomorphisms} between $A$ and $B$.
\cqfd

The fundamental result of the third section of this article will be to prove that, in a certain submodel of this one, the converse of this
proposition is also true.

\subsection{Construction of a control hyperdoctrine}

As we wish to establish that we have a model of the $\lambda\mu2$-calculus and we already have a structure of hyperdoctrine, we need to recover the additional requirements of the 
section~\ref{contr_hyp}. The important part of the job will in fact consist in characterizing $\ppar$ as a binoidal functor, 
and distinguishing central morphisms.

Let $s$ be a justified sequence on $A\ppar B$; we want to define a justified sequence $s\restr_A$ on $A$. The basic idea is the following: each node $c$ in $A\ppar B$ ``comes from'' a 
node of $A$ or $B$. $s\restr_A$ will consist of the part of $s$ which comes from $A$.

Formally, if $s=\epsilon$ then $s\restr_A=\epsilon$; otherwise, let $s=s'm$ with
$m=(\theta)[c_1:A_1^1/b_1^1;\dots;A_{k_1}^1/b_{k_1}^1]\dots[c_n:A_1^n/b_1^n;\dots;
A_{k_n}^n/b_{k_n}^{n}]$. All the $c_i$ are necessarily of the same form: $c_i=(a_i\ppar c'_i)$ or $c_i=(a_i,c'_i,1)$ or 
$c_i=(c'_i,a_i,2)$.
\begin{itemize}
\item If $c_i=(a_i\ppar c'_i)$, let $p$ be the least $i$ such that $c'_{i+1}=c'_i$ ($i=n$ if this equality in never true),
$b_{i_m}^1$ be the quantifiers appearing in $A[(\theta(X_j)/X_j)_j]$, and $b_{i_m}^l$ (for $2\leq l\leq p$ and $1\leq m\leq km$) 
be the quantifiers appearing in $A_{i_1}^{l-1}/b_{i_1}^{l-1}$, \ldots, $A_{i_{k(l-1)}}^{l_1}$; then
$s\restr_A=s'\restr_Am'$ with 
$m'=\theta[a_1:A_{i_1}^1/b_{i_1}^1;\dots;A_{i_{k1}}^1/b_{i{k1}}^1]\dots[a_n:A_{i_p}^p/b_{i_p}^p;\dots;A_{i_{kp}}^p/b_{i_{kp}}^{n}]$
\item If $c_i=(a_i,c'_i,1)$, $s\restr_A=s'\restr_Am'$ with $m'=[a_1:A_1^1/b_1^1;\dots;A_{k_1}^1/b_{k_1}^1]\dots[a_n:A_1^n/b_1^n;\dots;
A_{k_n}^n/b_{k_n}^{n}]$
\item If $c_i=(c'_i,a_i,2)$, $s\restr_A=s'\restr_A$.
\end{itemize}

$t\restr_B$ is defined similarly.

\textbf{Example~7:} Let us come back to the arena $A$ and the initial move $m$ of example~4. We have $A=B\ppar C$, with 
$B=\forall X_3.(¬\bot\ppar X_3)$ and $C=\forall X_3.(¬ X_2\ppar X_3)$,
$m=\theta[c_1:H_1/b_1,H_2/b_2][c':H_3/b_3]$ with $c_1=\forall(\star\ppar x^{(0,\star\ppar\star)})\ppar\forall(\star\ppar x^{(0,\star\ppar\star)})$ and 
$c'=c'_1=(\star\ppar\star)\ppar(\star\ppar\star)$.

If we consider the play $s=m$, one has $s\restr_B=\theta[\forall(\star\ppar x^{(0,\star\ppar\star)}):H_1/b_1][(\star\ppar\star):
H_3/b_3]$ and
$s\restr_B=[\forall(\star\ppar x^{(0,\star\ppar\star)}):H_2/b_2][(\star\ppar\star):]$. 
The reader can check that one has kept in $s\restr_B$ the ``left part'' of the moves $a\ppar a'$, and every instantiation of a 
quantifier which is related to the ``left part'' of the arena $A$. \finex

\medskip

This definition is such that, for $s$ justified sequence on $A\ppar B$, one has $s\restr_A$ justified sequence on $A$. If
$s$ is a justified sequence on $(A\ppar C)\fl(B\ppar D)$, one can define as well $s\restr_{A\fl B}$, composed of the moves of
$(s\restr_{A\ppar C})\restr_A$ and the moves of $(s\restr_{B\ppar D})\restr_B$. 

\begin{Prop}
Let $\tau:C\fl D;\XX$ be a strategy, and $\sigma:A\fl B;\XX$ a central strategy. Let $\sigma\ppar\tau=\{s\in\mc{P}_{\XX}((A\ppar C)\fl
(B\ppar D))\mid\ s\restr_{A\fl B}\in\sigma\land s\restr_{C\fl D}\in\tau\}$. Then $\sigma\ppar\tau$ is a strategy. If $\sigma$ and $\tau$ 
are innocent then $\sigma\ppar\tau$ is innocent. Moreover, we have 
$(\sigma\ppar\id_C);(\id_B\ppar\tau)=(\id_A\ppar\tau);(\sigma\ppar\id_D)$.
\end{Prop}

\textsc{Proof:} Once again we make use of the translation into propositional game semantics. We recall that, in 
propositional game semantics, the nodes of an arena $A\ppar B$ are denoted $(a_0,b_0)$ for roots, $(a,b_0,1)$ for copies of nodes of $A$
and $(b,a_0,2)$ for copies of nodes of $B$.

First, one can verify that $\trad{\sigma}$ is a central strategy (in the propositional game semantics) iff $\sigma$ is a central 
strategy.

The correspondence $\trad{A\ppar B}\cong\trad{A}\ppar\trad{B}$ is true, but not completely trivial: actually, this is precisely given by
the preceding definition: to each non-empty bi-view $sm$ of $A\ppar B$ (which is a node of $\trad{A\ppar B}$) ending with 
the move $m$, one associates a node $\mu$ in $\trad{A}\ppar\trad{B}$ by proceeding by cases:
\begin{itemize}
\item if the nodes of $m$ take the form $(a_i\ppar b_i)$, then $\mu=(sm\restr_A,sm\restr_B)$
\item if the nodes of $m$ take the form $(a_i,b_i,1)$, then $\mu=(sm\restr_A,b_i,1)$
\item if the nodes of $m$ take the form $(b_i,a_i,2)$, then $\mu=(sm\restr_B,a_i,2)$.
\end{itemize}
This defines a bijection from the nodes of $\trad{A\ppar B}$ to the nodes of $\trad{A}\ppar\trad{B}$.

By extending this correspondence to plays and sets of plays, one obtain, for each set $\sigma$ of plays on $A\ppar B$, a set 
$\trad{\sigma}''$ on $\trad{A}\ppar\trad{B}$, and we have: $\trad{\sigma\restr_A}=\trad{\sigma}''\restr_{\trad{A}}$.
For a strategy $\tau:C\fl D;\XX$ and a central strategy $\sigma:A\fl B;\XX$, one has 
\begin{align*}
\trad{\sigma\ppar\tau}''&=\{\trad{s}''\mid\ s\in\mc{P}_{\XX}((A\ppar C)\fl(B\ppar D)\land s\restr_{A\fl B}\in\sigma\land s\restr_{C\fl D}\in\tau\}\\
&=\{s\in\mc{P}_{\XX}((\trad{A}\ppar\trad{C})\fl(\trad{B}\ppar\trad{D})\mid\
s\restr_{\trad{A}\fl\trad{B}}\in\trad{\sigma}\\
&\hspace{0.7cm}\land s\restr_{\trad{C}\fl\trad{D}}\in\trad{\tau}\}\\
& =\trad{\sigma};\trad{\tau}\end{align*}

Now we can refer to what has been done for propositional game semantics~\cite{phdlaurent} and retrieve the expected results (thanks to lemmas~\ref{trad1} and~\ref{trad2}):
$\trad{\sigma;\tau}''$ is a strategy, so $\sigma;\tau$ is a strategy. If and $\tau$ are innocent, then $\trad{\sigma;\tau}''$ is innocent, so $\sigma;\tau$ is innocent. Finally, $\trad{(\sigma\ppar\id_C);(\id_B\ppar\tau)}''=\trad{(A\ppar\tau);(\sigma\ppar D)}$, 
so $(\sigma\ppar\id_C);(\id_B\ppar\tau)=(A\ppar\tau);(\sigma\ppar D)$.
\cqfd

Of course, $\tau\ppar\sigma$ for any $\tau$ and a central $\sigma$ is defined similarly.
This gives us the following result:

\begin{Prop}
In every category $\GG(n)$, $\ppar$ is a binoidal functor. Central morphisms for this structure are exactly central strategies.
\end{Prop}

\textsc{Proof:} We define $\sigma\ppar C=\sigma\ppar\id_C$ and $C\ppar\sigma=\id_C\ppar\sigma$. As 
$\trad{(\sigma;\tau)\ppar C}=\trad{(\sigma\ppar C);(\tau\ppar C)}$ and $\trad{C\ppar(\sigma;\tau)}=\trad{(C\ppar\sigma);(C\ppar\tau)}$,
$\ppar$ is a binoidal functor.

Moreover,\\
\begin{tabular}{lll}
$\sigma$ is central for the binoidal structure &$\Leftrightarrow$&$\trad{\sigma}''$ is central for the\\
&&(propositional) binoidal structure\\
&$\Leftrightarrow$&$\trad{\sigma}''$ is a central strategy\\
&$\Leftrightarrow$&$\sigma$ is a central strategy\end{tabular}\\
\phantom{coucou}\cqfd

\begin{Th}
$\mc{M}$ is a control hyperdoctrine.
\end{Th}

\textsc{Proof:} For each $n\in\mathbb{N}$, we define $\ppar_n=\ppar$ and $\bot_n=\bot$.

To construct the central isomorphisms $a_{A,B,C}$, $l_A$, $r_A$, $c_{A,B}$, we use the fact that there are trivial arena isomorphisms
between the arenas they bind together: for example, there is an arena isomorphism between $(A\ppar B)\ppar C$ and $A\ppar(B\ppar C)$, and $a_{A,B,C}$
is defined to be the corresponding game isomorphism. The reader can check the naturality of these isomorphisms and the commutation of 
associated diagrams.

The symmetric monoid of an object $A$ is defined in the following way:
\begin{itemize}
\item $i_A=\{\epsilon\}\cup\{mm'\mid\ m\textrm{ initial move of }A\land m'=(¬(\star),c_n,1)\land c_n\textrm{ last node of }m\}$
\item $\nabla_A=\{s\in\mc{P}_{\XX}(A_1\ppar A_2\fl A_0)\mid\ \forall t\leq s\textrm{ with $|t|$ even },t\restr_{A_1\fl A_0}\in\id_A\land
t\restr_{A_2\fl A_0}\in\id_A\}$ 
\end{itemize}
The reader can check the innocence of these strategies and the commutativity of the required diagrams.

At this point, we have proved that $\mc{M}$ is a symmetric premonoidal hyperdoctrine with codiagonals. $\mc{M}$ is also distributive:
$\pi_1$ and $\pi_2$ are focal, and $(\pi\ppar C,\pi_2\ppar C)$ is the trivial isomorphism between $(A× B)\ppar C$ and 
$(A\ppar C)×(B\ppar C)$, and the empty strategy is 
the trivial isomorphism between $\top\ppar C=\top$ and $\top$.
Checking naturality is left to the reader.

Hypercentrality is easy to check: because of their definitions, $\kappa$ and $\kappa^{-1}$ preserve centrality. Besides, thanks to the way
we have defined the substitution for an arena (through substitution of the formula) and for a strategy (through a simple operation on 
the $\theta$ function), the specialization functors commute with $\ppar$; and the operation
of substitution for strategies of course transforms trivial isomorphisms into trivial isomorphisms.

Finally, $s_{A,B,C}$ is the trivial isomorphism between $(A\fl B)\ppar C$ and $A\fl(B\ppar C)$ (because $\hat{\epsilon}_{A,B,C}:(B^A\ppar C)× A\flt{(B^A\ppar C)× (l;A\ppar i)}(B^A\ppar C)×(A\ppar C)\flt{\epsilon\ppar C}B\ppar C$ is trivially equivalent to $ev\ppar C$), and $p_{A,B}$ is the trivial isomorphism between $(\forall X_n.(A\ppar B)$ and $\forall X_n(A)\ppar B$ (if $X_n\notin FTV(B)$).
One again, the naturality of $s_{A,B,C}$ in $C$ and the commutativity of additional diagrams are left to the reader.
\cqfd

\section{Characterization of isomorphisms}

Having defined our model, we would like to use it to characterize
second-order type isomorphisms. Unfortunately, there are too many
isomorphisms in our model: for example, there exists an
isomorphism ($\sigma$,$\tau$) between $\forall X.\bot$ and $\forall X\forall Y.\bot$.

Indeed, the set of closed arenas $\mc{H}$ is countable, hence there exists a bijection $k:\mc{H}\fl\mc{H}\times\mc{H}$ ($k=(k_1,k_2)$).
Then the innocent strategy $\sigma:(\forall X.\bot)\fl(\forall X\forall Y.\bot)$ can be defined by its views, which take the form 
$s_{H,H'}=[\forall(\forall(\star)):H/b_1,H'/b_2][(\neg(\forall(\star)),\forall(\forall(\star)),1):k^{-1}(H,H')/b_0]$ for $H,H'\in\mc{H}$. Similarly, the innocent strategy $\tau:(\forall X\forall Y.\bot)\fl(\forall X.\bot)$ can be defined by its views, which take the form 
$t_{H}=[\forall(\star):H/b_0][(\neg(\forall(\forall(\star))),\forall(\star)),1):k_1(H)/b_1,k_2(H)/b_2]$
for $H\in\mc{H}$.

But the problem is, that this isomorphism does not exist in our language $\lambda\mu2$\footnote{It exists in Curry-style system F, but our language is defined in the Church-style.} !

In order to characterize type isomorphisms more precisely, we will
introduce a new property, called \textbf{uniformity}, which tends
to move the model nearer to the behavior of $\lambda\mu2$ itself. In
particular, this property will break down the high level of
symmetry between $\mathbf{P}$ and $\mathbf{O}$ (as it is the case for innocence).

\subsection{Uniformity}

\begin{Def}[rank] Let $s\in\mc{P}_{\XX}(A)$ and $m$ a move of $s$ which takes the form
\begin{align*}m=&(\theta)[c_1:A_1^1/b_1^1;\dots;A_{k_1}^1/b_{k_1}^1],\dots\dots,[c_n:A_1^n/b_1^n;\\&\dots;
A_{k_n}^n/b_{k_n}^{n}]\end{align*}
For each occurrence $H$ of a closed arena in $m$ ($H=A_i^j$ for some $(i,j)$ or $H=\theta(X_i)$ for some $X_i$),
we define the rank of $H$, denoted $\rank_m(H)$, by\footnote{Note that if $m$ is not an initial move, then $\rank_m(A_i^j)=j$.}:
\begin{minilist}
\item if $H=\theta(X_i)$ for some $X_i\in\XX$, or $H=A_i^j$ with $b_i^j$ quantifier of the arena where $m$ is played,then $\rank_m(H)=1$ 
\item if $H=A_i^j$ where $b_i^j$ is an quantifier of a closed arena $H'$ occurring in $m$ before $H$, then $\rank_m(H)=\rank_m(H')+1$.
\end{minilist}
\end{Def}

\begin{Def}[paths and instantiation traces]
Let $s\in\mc{P}_{\XX}(A)$. For every
move $m$ of $s$, we define:
\begin{minilist}
\item the \textbf{path} of $m$: $\underline{m}=c_1\dots c_n$
\item the \textbf{instantiation traces} of $m$:
$\overline{m}^i=B_1\dots B_p$ is the sequence of occurrences of arenas $B_j$ appearing in $m$ such that $\rank_m(B_i)\leq i$.
\end{minilist}

We note $\mathsf{C}$ the set of paths and $\mc{I}$ the set of
instantiation traces.
If $s=m_1\dots m_r$, we note
$\underline{s}=\underline{m_1}\dots\underline{m_r}$ ($\fref$ is
implicit in $\underline{s}$) and
$\overline{s}^j=\overline{m_1}^j\dots\overline{m_r}^j$ for
$j\in\mathbb{N}$.
\end{Def}

The path of $m$ is the description of the move without looking at
the instantiated arenas.
Instantiations traces are the sequences of arenas
effectively instantiated. The existence of many instantiation
traces depending on the node is required by the proof of theorem \ref{isos}.

Beside the set $\mc{X}$ of variable names that can be used as
parameters, we put another set $\mc{Y}=\{C_i\mid i\in\mathbb{N}\}$
where the $C_i$'s will represent \textbf{holes}, whose destiny is
to be replaced by a closed arena.

\begin{Def}[arenas with holes]
An \textbf{arena with holes} is a polymorphic arena built on the set of free variables $\mc{X}\cup\mc{Y}$.
The set of arenas with holes will be denoted $\mc{K}$.
\end{Def}

\textsc{Remark:} This definition means that, to define arenas with holes, we extend the grammar of nodes 
with the variables $c_i$ for $i\in\mathbb{N}$.

\begin{Def}[uniform strategy]
A strategy $\sigma:A;\XX$ is called \textbf{uniform} if there
exists a partial function
$f:\mathsf{C}^*\rightharpoonup\mathsf{C}^*$, and a sequence of
functions $F_1,\dots,F_n,\dots:\mathsf{C}^*\fl\mc{K}^*$ such that,
if $s\in\sigma$ and $sm$ play in $A$, then: $smm'\in\sigma$ if
and only if
$\underline{smm'}=f(\underline{sm})$ and
$\overline{m'}^i=F_i(\underline{sm})[\ov{sm}^i]$ for
every $i\in\mathbb{N}$.
\end{Def}

The notation $F_i(\underline{sm})[\ov{sm}^i]$ means that
the sequence of closed arenas $\overline{m'}^i$ is obtained first
by building the sequence $F_i(\underline{sm})$ of arenas with
holes, then by applying the substitution
$[\overline{sm}^i(1)/C_1,\dots,\overline{sm}^i(p)/C_p]$
in these arenas, where $\overline{sm}^i(k)$ stands for the $k$th arena occurring in 
$\overline{sm}^i$ (and $p$ is the length of this sequence).

This way, the arenas with holes cannot depend on already
instantiated arenas: these ones can only fill the holes to
generate the arenas for the following moves. This
corresponds to the fact that, in a term of $\lambda\mu2$, $\textbf{P}$
does not have a direct access to the instantiation of types by
$\textbf{O}$, he can only reuse them (think about the term
$\Lambda X.\lambda x^{\forall Y.Y}.x\{X\fl\bot\}$ for example).
Likewise, via the function $f$ we see that the paths of the moves
of $\mathbf{P}$ in a uniform strategy do not depend on already
instantiated arenas, but only on the names of already played
moves.

Note also that the functions $f,F_1,\dots,F_n,\dots$ suffice to
recover the uniform strategy $\sigma$. This definition of uniform strategies is inspired by the work of Murawski and Ong~\cite{muraw}.

\textbf{Example~8:} Let us go back to the play described in example~5.
This play belongs to a strategy $\sigma$ which is the interpretation of a $\lambda\mu2$ term. As we shall prove further, 
this implies that $\sigma$ is uniform. 

We are interested by the two first moves of this play: 
$m_1=\theta[a:H'/b_1;H/b_3]$ and $m_2=[b:H'/b_2][g':H''/b_4][h':]$.
The property of uniformity implies that:
\begin{itemize}
\item the choice of the nodes $b,g',h'$ only depends on the choice of $a$ by Opponent: $f(a)=a\cdot b g'h'$
\item the closed arenas $H'$ and $H''$ are determined by two arenas with holes $G_1=F_1(a)$ and $G_2=F_2(a)$, and we have 
$G_1[H'/C_1,H/C_2]=H'$ and $G_2[H'/C_1,H/C_2]=H''$; actually, $G_1$ is simply $H'$ and $G_2=C_2\fl\bot$.
\end{itemize}

The interest of this example lies in the comparison with the term $t=\Lambda Z.\Lambda X.\lambda x^{\forall Y.Y}.\lambda u^X.\lambda
v^X.(x\{\forall U.U\}\{X\fl\bot\})u$ interpreted by $\sigma$. Indeed, the arenas with holes $G_1$ and
$G_2$ can in fact be read directly off this term: they correspond to the instantiations by $\forall U.U$ and $X\fl\bot$.
This case is actually a bit more simple than the definition, because each arena played by $m_1$ is of rank $1$.\finex

\medskip


The above definition of uniformity is not very convenient for its
use in our proofs. That is why we introduce a lemma giving
a very useful and widely used consequence of uniformity. 


\begin{Lemme}\label{unif} Let $s\in\sigma$ with $\sigma$ uniform.
Let $m\in s$ be such that $\lambda(m)=\mathbf{O}$ and
$H/b_i$ be one of its instantiations. Suppose that each node of
$H$ played during $s$ is a root of $H$. Then, if $H'$ is an arena
whose roots have the same names as the roots of $H$, we have $s'\in\sigma$, where $s'$
is the play $s$ modified in the following way: first, the node
$m$ is replaced by $m'$, which is identical to $m$ except that it
instantiates $b_i$ by $H'$; second, each time an arena with holes
does a reference to $H$, we give $H'$ instead. We have, in
particular, $\und{s}=\und{s'}$
\end{Lemme}


\textsc{Proof:} We set $s=s_1ms_2$. By induction on the length of $s_2$:
\begin{minilist}
\item if $s_2=m_1$, we have $\und{s_1m}=\und{s_1m'}$ and we can
conclude thanks to uniformity
\item if $s_2=m_1\dots m_n$ with $n\geq3$, we have by induction
hypothesis:\newline $\und{s_1 mm_1\dots m_{n-2}}=\und{s_1 mm'_1\dots
m'_{n-2}}$ (where $m'_i$ is the move corresponding to $m_i$ in
$s'$); besides $m'_{n-1}=m_{n-1}$, so $\und{s_1 mm_1\dots
m_{n-1}}=\und{s_1 mm'_1\dots m'_{n-1}}$, and we can conclude
thanks to uniformity.
\end{minilist}\cqfd

\subsection{The uniform model}

Our goal in this section is to prove that, by restricting the model to uniform strategies, we still obtain a control hyperdoctrine,
whose structural objects are actually the same as in the original model.
We first check that the basic operations on strategies preserve uniformity:

\begin{Prop}
 If $\sigma$ and $\tau$ are uniform and $A$ is an arena, then $\sigma×
\tau$, $\sigma\ppar A$, $A\ppar\sigma$ and $\sigma;\tau$ are uniform.
\end{Prop}

\textsc{Proof:}
Preservation of uniformity through $×$ is trivial. For $\ppar$, we recall that
$\sigma\ppar A=\{s\in\mc{P}_{\XX}((B\ppar A)\fl(C\ppar A))\mid\ s\restr_{B\fl C}\in\sigma\land s\restr_{A\fl A}\in\id_A\}$: 
hence, if one can rebuild $\und{sm\restr_{B\fl C}}$ from  $\und{s\restr_{B\fl C}}$ and $\ov{sm\restr_{B\fl C}}^i$ from  
$\ov{s\restr_{B\fl C}}^i$, one can rebuild $\und{sm}$ from $\und{s}$ and $\ov{sm}^i$ from  $\ov{s}^i$.

Let us focus our attention on composition.

Let $f,F_1,\dots,F_{n},\dots$ and
$f',F'_1,\dots,F'_{n},\dots$ be the functions associated respectively
with $\sigma$ and $\tau$. Let $smn\in\sigma;\tau$, we know that
$smn=u\restr_{A,C}$ with $u\in int(A,B,C)$,
$u\restr_{A,B}\in\sigma$ and $u\restr_{B,C}\in\tau$. Besides, as
the strategy is innocent, we can ask $smn$ to be a thread.
Finally, we note $s'=smn$.

Starting from a uniform strategy $\rho$, if we define
$\und{\rho}=\{\und{s}\mid s\in\rho\}$, we can see that, thanks to
uniformity, $\und{\rho}$ is a strategy in a propositional setting.
Thus we obtain two strategies $\und{\sigma}$ and $\und{\tau}$,
which interact to give the play $s'$ in particular. So we can use
the Zipping lemma in the propositional setting (cf.
\cite{phdharmer}) to conclude that we can reconstruct $\und{u}$
starting from $\und{s'}$.

This gives us a partial function $g_0$ such that
$\und{u}=g_0(\und{s'})$ (this function will be defined on $s'$ if
and only if we can reconstruct a finite sequence $\und{u}$
starting from $\und{s'}$). It is then easy to construct another
partial function $g$ so that $\und{s'n}=g(\und{s'})$.

Suppose now that $n$ is in $A$ (the case of $n$ being in $C$ can
be treated exactly the same way), and note $u'$ the prefix of $u$
whose last move is $m$. We know that
$\ov{n}^i=F_i(\und{s_0})[\ov{s_0}^i]$, where
$s_0=u'\restr_{A,B}$. As we can calculate $\und{s_0}$ starting from
$\und{s'}$, this gives us
$\ov{n}^i=H_i(\und{s'})[\ov{s_0}^i]$. As $s_0$ is
contained in $u'$, we have
$\ov{n}^i=H'_i(\und{s'})[\ov{u'}^i]$. We will now prove
in the following that we can keep this relation while suppressing
from $u'$ the moves played in $B$, one after the other.

Let $m_0$ be the last move in $u'$ played in $B$ and let us write
$u'=s'_0m_0u_2$. $m_0$ is played by $\mathbf{P}$ for $\sigma$ or
for $\tau$. Suppose it is for $\tau$ (the other case is
equivalent): then we have
$\ov{m_0}^i=F'_i(\und{s'_0})[\ov{s'_0}^i]$, so
$\ov{m_0}^i=F''_i(\und{s'})[\ov{s'_0}^i]$ for some
function $F''_i$. Note that formally $\ov{m_0}^i$ may differ if we consider $\sigma$ or $\tau$ (because of the $\theta$ function), but it has no practical effect thanks to the definition of restriction (the $\theta$ functions are the same in plays of $\sigma$ as in plays of $\tau$).
We now
replace the sequence of arenas with holes $H_i(\und{s'})$ by the
adequate sequence of arenas with holes (i.e. where the arenas of
$F''_i(\und{s'})$ are placed in the corresponding holes), and we
get $\ov{n}^i=H''_i(\und{s'})[\ov{s''_0}^i]$, where
$s''_0$ is obtained by suppressing $m_0$ from $u'$.

We have succeeded in suppressing the last move of $u'$ which is
played in $B$, we can proceed iteratively until we only have moves
played in $A$ or $C$. This will give us a function $G_i$ such that
$\ov{n}^i=G_i(\und{s'})[\ov{s'}^i]$.\cqfd

\begin{Prop}
Let $\XX=(X_1,\dots,X_n)$ and $\vec{C}=(C_1,\dots,C_n)$ where the
free variables in $C_1,\dots,C_n$ are chosen between
$Y_1,\dots,Y_m$. If $\sigma:A;\XX$ is uniform then
$\sigma[\vec{C}]:A[\vec{C/\vec{X}}];\vec{Y}$ is uniform.

Let $A,C\in\mc{G}(n+1)$, $D\in\mc{G}(n)$ and $\XX=X_1,\dots,X_n$.
If $\sigma:F(\XX)(C)\fl A;\XX,X_{n+1}$ is uniform then
$\kappa(\sigma):C\fl\forall X_{n+1}.A;\XX$ is uniform.
If $\tau:D\fl\forall X_{n+1}.A;\XX$ is uniform then
$\kappa^{-1}(\tau):D\fl A;\XX,X_{n+1}$ is uniform.

If $\sigma:A\fl B;\XX,X_{n+1}$ is uniform then $\forall\sigma:(\forall X_{n+1}.
A)\fl(\forall X_{n+1}.B);\XX$ is uniform.
\end{Prop}

\textsc{Proof:}
In each case, it suffices to do a slight update of the
functions related to the first uniform strategy to obtain the new
one. Note that this would not work for $\kappa$, $\kappa^{-1}$ and $\sigma\mapsto\forall\sigma$ 
if we had not introduced the notion of rank: indeed, the rank is preserved through abstraction.\cqfd

It is easy to see that each base strategy is uniform:

\begin{Lemme}
If $A,B,C\in\GG(k)$, then the following strategies are uniform: $\epsilon:\top$, $id_A:A\fl A;\XX$,
$\Delta_A:A\fl A× A;\XX$ and $\pi_A:A× B\fl A;\XX$, $a_{A,B,C}:(A\ppar B)\ppar C\fl A\ppar(B\ppar C)$,
$l_A:A\fl A\ppar\bot$, $r_A:A\fl\bot\ppar A$, $i_A:\bot\fl A$, $\nabla_A:A\ppar A\fl A$, 
$d_{A,B,C}:(A\ppar C)×(B\ppar C)\fl(A× B)\ppar C$, $s_{A,B,C}:(B\fl A)\ppar C\fl ((B\ppar C)\fl A)$, 
as well as $a_{A,B,C}^{-1}$, $l_A^{-1}$, $r_A^{-1}$, $d_{A,B,C}^{-1}$ and $s_{A,B,C}^{-1}$.
\end{Lemme}

The control hyperdoctrine $\mc{M}$ was characterized by its base strategies, which are uniform, and its fundamental operations,
which preserve uniformity. This leads us to the following:

\begin{Th}
If we restrain each category $\GG(k)$ to the subcategory $\GG_{unif}(k)$ where every strategy is uniform, we obtain a new
control hyperdoctrine $\mc{M}_{unif}$.
\end{Th}

This model is much less symmetric and comfortable than the first
one we introduced, that is why we introduced it only in a second
time. But the symmetric model $\mc{M}$ is too liberal to allow us
to deal properly with type isomorphisms; so, uniformity is an ad
hoc property to constrain the model in such a way that we do not
have more type isomorphisms in the model $\mc{M}_{unif}$ than in
$\lambda\mu2$, as we will see right now.

\subsection{Isomorphisms in the model}

We are now ready to prove the fundamental result of our work on type isomorphisms: this is essentially the converse of the 
proposition~\ref{gameariso}. All this section is dedicated to the proof of this theorem.

\begin{Th}\label{isos}
If there exists a game isomorphism $(\sigma,\tau)$ between two
polymorphic arenas $A$ and $B$, with $\sigma,\tau$ uniform and innocent, then $A$ and $B$ are isomorphic.
\end{Th}

\textsc{Proof:} The same result has been proved in a propositional
setting in~\cite{classisos}; the tricky part in our case is that
moves are not only nodes of a forest, but contain all the
second-order structure. However, we may reuse the results of
Olivier Laurent to do one part of the work. 

\textbf{Zig-zag plays:}

\begin{Def}[zig-zag play]
A play $s$ of $A\fl B$ is said to be \textbf{zig-zag} if
\begin{minilist}
\item each Player move following an Opponent move played in $A$ (resp. in
$B$) is played in $B$ (resp. in $A$)
\item each Player move played in $A$ which follows an Opponent initial move played
in $B$ is justified by it
\item $s\restr_A$ and $s\restr_B$ have the same pointers.
\end{minilist}

If $s$ is a zig-zag even-length play on $A\fl B$, we note
$\breve{s}$ the unique zig-zag play on $B\fl A$ such that
$\breve{s}\restr_A=s\restr_A$ and $\breve{s}\restr_B=s\restr_B$.
\end{Def}

We also give the definition of totality, which coincides with the usual notion:

\begin{Def}[totality]
Let $\sigma:A;\XX$. We say that $\sigma$ is total if, whenever $s\in\sigma$ and $sm\in\mc{P}_{\XX}(A)$, there exists a move $m'$ such that
$smm'\in\sigma$.
\end{Def}

\begin{Lemme}
If there is a game isomorphism $(\sigma,\tau)$ between $A$ and $B$ then:
\begin{minilist}
\item every play of $\sigma$ or $\tau$ is zig-zag
\item $\tau=\{\breve{s}\mid s\in\sigma\}$
\item $\sigma$ and $\tau$ are total.
\end{minilist}
\end{Lemme}

This lemma can be proved by using forth and back translation to the propositional setting, as it has been defined in section~\ref{trans}.
Indeed, as the lemma has been proved for the propositional setting in ~\cite{classisos}, all we need is to check the following:
\begin{itemize}
\item if $(\sigma,\tau)$ defines an isomorphism between $A$ and $B$ then $(\trad{\sigma}',\trad{\tau}')$ defines an isomorphism between 
$\trad{A}$ and $\trad{B}$
\item if $\trad{s}'$ is zig-zag then $s$ is zig-zag
\item if $\trad{\tau}=\{\breve{s}\mid s\in\trad{\sigma}\}$ then $\tau=\{\breve{s}\mid s\in\sigma\}$
\item if $\trad{\sigma}$ is total then $\sigma$ is total
\end{itemize}

Given these preliminary results, we are ready to build the bijection $g:E_A\fl E_B$ as a morphism for 
the whole structure of arenas: in order to do that, we will build $g$ together with a bijection 
$\Psi:\mc{R}_A\fl\mc{R}_B$ such that $g(\mc{T}(b))=\mc{T}(\Psi(b))$ for all $b\in\mc{R}_A$, and then 
prove that $g(\mc{S}(b))=\mc{S}(\Psi(b))$ and finally that $\DD_A(c)=\DD_B(g(c))$ for all $c\in E_A$.

\textbf{Construction of the bijection $g$:}

Suppose that the uniform strategies $\sigma$ and $\tau$ are
determined respectively by the functions $f,F_1,\dots,F_n,\dots$
and $f',F'_1,\dots,F'_n,\dots$. 
Consider an arena
$Q$ and a play (in a propositional
setting) $s=m_1\dots m_n$ on the forest $\mc{F}_Q=(E_Q,\leq_Q)$;
we can associate to this play a second-order play
$[s]=[m_1]\dots[m_n]$ on $Q$ where each move $[m_i]$ consists in
playing the node $m_i$ (or the node replacing $m_i$) and
instantiating each quantifier by $\bot$ (for an initial move, this
also means that $\theta(X_j)=\bot$ for each $X_j$). This means that
this move will be written $[m_i]=[m'_i:\bot/b_{i_1};\dots
\bot/b_{i_n}]$ or $[m_i]=[m'_i:\bot/b_{i_1};\dots \bot/b_{i_n}][m''_i:]$ where $m'_i$ and $m''_i$
can be either $m_i$ or a node of the form $m_i[\alpha'(x)/x]_{x\in V(m_i)}$, whose origin is $m_i$.


Let $a$ be a node of $A$ and $a_1\dots a_p$ be the sequence of
nodes of $A$ such that $a_1$ is a root, $a_{i+1}$ son of $a_i$ for
$1\leq i\leq p-1$ and $a_p=a$. By induction on $p$, we will define
a function $g$ from the nodes of $A$ to the nodes of $B$ and prove
that\footnote{For a better understanding of this assertion, we recall that the arena $C_i$ is the arena reduced to a node $c_i$.}
:
\begin{minilist}
\item if
$s_\sigma=[g(a_1)][a_1][a_2][g(a_2)][g(a_3)][a_3]\dots$ then
$s_\sigma\in\sigma$
\item there exists a bijection $\psi$ between the quantifiers $b_i$
such that $\mc{T}(b_i)=a_p$ and the quantifiers $b'_i$ such that
$\mc{T}(b'_i)=g(a_p)$, and we have : if $s'_\sigma$ and $s'_\tau$
are respectively the plays $s_\sigma$ and $s_\tau$ where the last
move has been suppressed, $F_1(\und{s'_\sigma})=(C_{r(i)})_{i\in[1,n]}$ where $r(i)$ is the index associated with
$b'_{\psi(i)}$ (or $b_{\psi^{-1}(i)}$), i.e. the index of the
arena which instantiates this quantifier, and
$F'_1(\und{s'_\tau})=(C_{l(i)})_{i\in[1,n]}$ where
$l(i)$ is the index associated with $b_{\psi^{-1}(i)}$ (or
$b'_{\psi(i)}$).
\end{minilist}

Note that the functions $g$ and $\psi$ should depend on $a$ and be
written $g_a$ and $\psi_a$. But actually, by construction, we see
that if $a'$ is an ancestor of $a$, then $g_a{a'}=g_{a'}(a')=g(a')$, and
the same holds for $\psi$.

\bigskip

If $p=0$ it suffices to say that $\epsilon\in\sigma$ and
$\epsilon\in\tau$.

\bigskip

If $p=p'+1$ we note
$s_1=[a_1][g(a_1)][g(a_2)][a_2]...[g(a_{p'})][a_{p'}]\in\tau$\\ and
$
s_2=[g(a_1)][a_1][a_2][g(a_2)]...[a_{p'}][g(a_{p'})]\in\sigma$. We
choose the unique move $m$ played in $B$ such that
$s_1[a_p]m\in\tau$ (it exists by totality of $\tau$) and we set
$g(a_p)=\textit{origin}(m)$. Let $b_1,\dots,b_n$ be the quantifiers
such that $\mc{T}(b_i)=a$ and $b'_1,\dots,b'_{n'}$ the quantifiers
such that $\mc{T}(b'_i)=g(a_p)$.

By uniformity of $\tau$, we have
$\ov{m}^1=F_1'(\und{s_1[a_p]})[S_1]$ where $S_1$ stands for some
substitution of holes and variables. As we have
$s_2m[a_p]\in\sigma$, we also get, from the uniformity of
$\sigma$,
\begin{equation}\label{eq2}\ov{[a_p]}^1=F_1(\und{s_2m})[S_2]\tag{$\star$}
\end{equation} where $S_2$ stands for some
substitution of holes and variables. We note
$F_1(\und{s_2m})=G_1\dots G_n$ and $F'_1(\und{s_1[a_p]})=G'_1\dots
G'_n$, and we are interested in $G_i$ for a given $i$. We know
that $G_i[S_2]=\bot$ from (\ref{eq2}).

Let $a'$ be the move obtained from $[a_p]$ by instantiating $b_i$
by $H_1=¬ H$ instead of $\bot$, where $H$ is a non-empty closed arena. Then 
$H_1$ has the same root as $\bot$. We
have $\und{a'}=\und{[a_p]}$, so $s_1a'm'\in\tau$ and $s_2
m'a'\in\sigma$, with $\und{m'}=\und{m}$ thanks to lemma
\ref{unif}. This gives
$\ov{a'}^1=F_1(\und{s_2m'})[S'_2]=F_1(\und{s_2m})[S'_2]$ ($S'_1$
and $S'_2$ are the new substitutions obtained with the new
instantiation). In particular, $H_1=G_i[S'_2]$.

At this point, we have proved on one side $\bot=G_i[S_2]$ and on the
other side $H_1=G_i[S'_2]$. As $H_1\neq\bot$ and $S'_2$ can only
use\footnote{This is where it is important to use
functions $F_i$ indexed by the rank: if we could use some arenas from $\ov{s_2m'}^2$
for example, then we could not conclude anything concerning the
arenas appearing in $\ov{m'}^1$, and we could not prove that
$n=n'$.} closed arenas from $\ov{s_2m'}^1$ (and not from
$\ov{s_2m'}^2$). This means that
$G_i$ must contain a hole $C_l$ referring to an arena from
$\ov{m}^1$ (or $\ov{m'}^1$), because otherwise we should have
$G_i[S'_2]=G_i[S_2]$ (indeed, the instantiation of $b_i$ by $H_1$
instead of $\bot$ does not affect the closed arenas from $\ov{s_2}^1$,
it affects only $\ov{m}^1$). But if, for each value $l$ satisfying
this property, $G'_l$ does not contain the hole $C_{l(i)}$ (where
$l(i)$ is the index of the arena which instantiates $b_i$ in
$s_1[a_p]m$), then $G'_l[S_1]=G'_l[S'_1]$, and so
$G_i[S'_2]=G_i[S_2]$ ($S_2$ and $S_2$ can only be differentiated
by the arenas $G'_l[S_1]$ and $G'_l[S'_1]$), which is absurd. We
then have at least one value of $l$ such that $G_i$ contains the
hole $C_l$ and $G'_l$ contains the hole $C_{l(i)}$.

It is then easy to see that $G'_l=C_{l(i)}$ : indeed,
if $G'_l$ contained anything else than the node $c_{l(i)}$, $G'_l[S_1]$
would strictly contain $\bot$ (because none of the arenas it can
refer to is empty), and $G_i[S_2]=\bot$ would also strictly contain
$\bot$.

This proves that $n'\geq n$, and that we have an injection from
the $b_i$ quantifiers to the $b'_j$ quantifiers. To show that $n\geq
n'$, we will use nearly the same reasoning but in the other
direction.

Let $j\in[1,n']$, we call $K$ the closed arena by which $m$
instantiates $b'_j$, i.e. $K=G'_j[S_1]$. Let $m'$ be the move
obtained from $m$ by instantiating $b'_j$ by $K_1$ instead of $K$,
where $K_1$ must have the same roots as $K$ but must be distinct from $K$ (note that if $K$ is
empty we can choose $K_1=\bot$). Then $\und{m''}=\und{m}$, so that we
have $s_2 m''a''\in\sigma$ and $s_1 a''m''\in\tau$ with
$\und{a''}=\und{[a_p]}$ thanks to lemma \ref{unif}. So,
$\ov{m''}^1=F'_1(\und{s_1a''})[S''_1]=F'_1(\und{s_1[a]})[S''_1]$
($S''_1$ and $S''_2$ are the new substitutions obtained with the
new instantiation). In particular, $K_1=G'_j[S''_1]$.

At this point, we have on one side $K=G'_j[S_1]$ and on the other
side $K_1=G'_j[S''_1]$. We also know that $K\neq K_1$, and $S''_1$
can only use arenas from $\ov{s_1a''}^1$, so
$G'_j$ contains at least one hole $C_r$ which refers to an arena
from $\ov{[a]}^1$ (or $\ov{a''}^1$) : indeed, the instantiation of
$b'_j$ by $K_1$ instead of $K$ does not affect the closed arenas
from $\ov{s_1}^1$, it only affects $\ov{[a]}^1$. But if, for any
value of $r$ satisfying this property, $G_r$ does not contain the
hole $C_{r(j)}$ (where $r(j)$ is the index of the arena which
instantiates $b'_j$ in $s_2m[a_p]$), then $G_r[S''_2]=G_r[S_2]$,
and so $G'_j[S''_1]=G'_j[S_1]$ which is absurd. We finally have at
least one value of $r$ such that $G'_j$ contains the hole $C_r$
and $G_r$ contains the hole $C_{r(j)}$.

To see that $G_r=C_{r(j)}$, one must first define a
move $m_0$ which is identical to $m$ except that it does not
instantiate any of its quantifiers by the empty arena (it
puts $\bot$ instead, for example). This can modify $K$, but it does
not modify the paths, so that, thanks to uniformity, we still have
$s_2 m_0a_0\in\sigma$ and $s_1 a_0m_0\in\tau$, with
$\und{a_0}=\und{[a]}$. Let $S^0_1$ and $S^0_2$ be the new
substitutions obtained with the new instantiation. If $G_r$
contained anything else than the hole $C_{r(j)}$, then
$G_r[S^0_2]$ would strictly contain $K$ (because none of the other
arenas it can refer to is empty), and $G'_j[S^0_1]=K$ would
strictly contain $K$.

Finally, we obtain a bijection $\psi$ between the $b_i$'s and the
$b'_j$'s ; $m$ instantiates each of its quantifiers to $\bot$, so
$m=[g(a_p)]$, and we have $F_1(\und{s_2m})=(C_{r(i)})_{i\in[1,n]}$ where $r(i)$ is the index associated with
$b'_{\psi(i)}$, and $F'_1(\und{s_1[a_p]})=(C_{l(i)})_{i\in[1,n]}$ where $l(i)$ is the index associated with
$b_{\psi^{-1}(i)}$.

\bigskip

The case $p=p'+1$ with $p'$ odd can be treated exactly the same
way, switching the roles of $\sigma$ and $\tau$.

\bigskip

It is now easy to associate a function $g'$ to $\sigma$ as we have
associated a function $g$ to $\tau$. By construction, $g$ and $g'$
respect filiation (because of the property on zig-zag plays
concerning pointers). We can easily check that $g\circ g'$ is the
identity on the nodes of $A$, and $g'\circ g$ is the identity on
the nodes of $B$ : this directly comes from the fact that
$[g(a_1)][a_1][a_2][g(a_2)][g(a_3)][a_3]\dots\in\sigma$ and
$[a_1][g(a_1)][g(a_2)][a_2][a_3][g(a_3)]\dots\in\tau$.

\bigskip

\textbf{Proof of $g(\RRR_A)=\RRR_B$ and $\DD_B\circ g=\DD_A$:}

The construction of the bijection $\Psi:\mc{R}_A\fl\mc{R}_B$ such that 
$g(\mc{T}(b))=\mc{T}(\Psi(b))$ for all $b\in\mc{R}_A$ is directly given by the function $\psi$.
What remains to be proved is the following : if $a_p$ appears $n$
times in $\mc{S}(b_j)$ then $g(a_p)$ appears $n$ times in
$\mc{S}(\psi(b_j))$, and if $a_p$ is decorated $n$ times by $X_k$
then $g(a_p)$ is also decorated $n$ times by $X_k$.

Suppose that $a_p$ appears $n$ times in $\mc{S}(b_j)$ whereas
$g(a_p)$ appears $n'$ times in $\mc{S}(\psi(b_j))$ with for
example $n'<n$. Let us consider the plays
$s_1=[a_1][g(a_1)]\newline[g(a_2)][a_2]...[g(a_p)][a_p]\in\tau$ and
$s_2=[g(a_1)][a_1][a_2][g(a_2)]...[a_p][g(a_p)]\in\sigma$ (here,
$p$ has been chosen even, but of course all this still holds for
$p$ odd), and take a play $s'$ nearly identical to $s_2$ but for
which $b_j$ has been instantiated by the closed arena $H=\neg\neg\dots\neg\bot$, which is
a string of length $N$ 
(\ie\ a tree of depth $N$ where
each node has at most one son). Thanks to lemma
\ref{unif}, this play is still in $\sigma$. Besides, we note that,
if $g(a_p)\in\mc{S}(b_r)$ (with $b_r\neq\psi(b_j)$) or if $g(a_p)$
is decorated by $X_r$, then $b_r$ or $X_r$ is still instantiated
by $\bot$ in $s'$ : indeed, the instantiation of $b_r$ is given by
the arena with holes $C_{\psi^{-1}(b_r)}$ (and
$\psi^{-1}(b_r)$  is still instantiated by $\bot$), and the
instantiation of $X_r$ is given by a $\theta$ function which
instantiates every variable by $\bot$. So, in the play $s'$ the last
move played is still $[g(a_p)]$, or $[g(a_p)]'$ ($[g(a_p)]'$ is
just the move $[g(a_p)]$ where $\psi(b_j)$ has been instantiated
by $H$: it is the case $\mc{T}(b_j)=a$) : we note this move
$[g(a_p)]_0$. For simplicity of notations, we will consider from
now that $a_p$ does not appear in any other set of the form
$\mc{S}(b_k)$, with $k\neq j$, and is not decorated by any
variable $X_k$. The reader can easily check that this point does
not change anything in the following, because each $b_k$, $k\neq
j$, and each variable $X_k$, are instantiated by $\bot$.

Let us note $c_1,\dots,c_N$ the successive nodes of the string
$H$ ($c_1=c$). The play $s'$ can be written
$s'=t[a_p]_0[g(a_p)]_0$ ($[a_p]_0$ is either $[a_p]$, either the
move which consists in playing $[a_p]$ by instantiating $b_j$ by
$H$, in the case $\mc{T}(b_j)=a$), and we have
$\breve{t}[g(a_p)]_0[a_p]_0\in\tau$. The instantiations by $H$
generate $n$ strings on one side and $n'$ strings on the other
side: let us simply call the nodes of these strings $(a_p-c_2,i)$, $(a_p-c_3,i)$, $\dots$, $(a_p-c_N,i)$, for
$1\leq i\leq n$, on one side, and $(g(a_p)-c_2,j)$,
$(g(a_p)-c_3,j)$, $\dots$, $(g(a_p)-c_N,j)$, for
$1\leq j\leq n'$, on the other side. Actually, these nodes correspond to ``copies'' of the nodes $c_1,\dots,c_n$,
and they are such that $|\quant((a_p-c_j,i))|=|\quant(c_j)|=0$ (this can be proved for the substitution $D\mapsto D[H/b]$ by an induction on $D$).

For each $1\leq i\leq n$, $\breve{t}[g(a_p)]_0[a_p]_0(a_p-c_2,i)\in\mc{P}_{\XX}(B\fl A)$, so that we
have $\breve{t}[g(a_p)]_0[a_p]_0(a_p-c_2,i)m_i\in\tau$ for
some move $m_i$, and $m_i$ is justified by $[g(a_p)]_0$ (this is
the property of zig-zag plays concerning pointers). Besides, the
moves $m_i$ must be pairwise distinct because
$t[a_p]_0[g(a_p)]_0m_i(a_p-c_2,i)\in\sigma$ ; but the node
$g(a_p)$ has been substituted by $\ppar_{j=1}^nH$, so there is
one value for $i$ such that the origin of $m$ is a son of $g(a_p)$
\textit{in the initial arena}. Indeed, there are only $n'$ other
alternatives, namely the moves $(g(a_p)-c_2,j)$ for $1\leq
j\leq n'$.

This implies that $g(a_p)$ has at least one son in the initial
arena : the idea of the proof is to show that there is behind
$g(a_p)$ a branch of length at least equal to $N$ : this will lead
us to a contradiction if we have chosen $N$ big enough at the
beginning (choose for example $n=h(A)+1$, where $h(A)$ is the
maximal depth of $A$).

As we have $\breve{t}[g(a_p)]_0[a_p]_0(a_p-c_2,i)m\in\tau$,
we also get $t[a_p]_0[g(a_p)]_0m(a_p-c_2,i)\in\sigma$ and,
as $t[a_p]_0[g(a_p)]_0m(a_pc_2,i)(a_pc_3,i)\in\mc{P}_{\XX}(A\fl B)$, we have
$t[a_p]_0[g(a_p)]_0\newline m(a_pc_2,i)(a_p-c_3,i)m'\in\sigma$
for some move $m'$, justified by $m$ and played in $B$. But be
careful : this time, the origin of $m$, say $d$, may have been
substituted by $\bot$ or by $H$, so the move $m'$ is not necessarily
played in the initial arena : it can be a move $(d-c_2,r)$,
played in the arena substituted for $d$ and corresponding to the node $c_2$ in $H$. In fact, we will show
further that this case, which we call a \textbf{trapped
substitution}, leads to a contradiction. Finally, one could think
that an quantifier may have $d$ as target, and this would lead to
another substitution, by something else than $\bot$ or $H$ ; but
actually it suffices to reuse the arguments explained many times
before to show, thanks to uniformity, that there cannot exist more
quantifiers having $d$ for target than having $(a_p-c_2,i)$
for target. As $|\quant((a_p-c_2,i))|=0$ this case is impossible.

We can thus go on with our proof : $\textit{origin}(m')$ is a son
of $\textit{origin}(m)$. We have $\\
\breve{t}[g(a_p)]_0[a_p]_0(a_p-c_2,i)mm'(a_pc_3,i)\in\tau$, and $\breve{t}[g(a_p)]_0[a_p]_0(a_p-c_2,i)
mm'(a_pc_3,i)(a_p-c_4,i)\in\mc{P}_{\XX}(B\fl A)$, so $\breve{t}[g(a_p)]_0[a_p]_0(a_p-c_2,i)mm'(a_p-c_3,i)
(a_p-c_4,i)m''\in\tau$ for some
move $m''$ justified by $m'$, etc. So, by systematically rejecting
trapped substitutions, we show that we can construct a branch of
length $N$ descending from $g(a_p)$. This leads to a
contradiction, if we have chosen $N$ big enough at the beginning.

We still have to show the impossibility of trapped substitutions :
suppose that $t[a_p]_0[g(a_p)]_0m_0(a_p-c_2,i)(a_p,c_3-i)m_1m_2\dots(a_p-c_k,i)(a_p-c_{k+1},i)m_l\in\sigma$ (or, equivalently,
$\breve{t}[g(a_p)]_0[a_p]_0(a_p-c_2,i)m_0m_1(a_p-c_3,i)(a_p-c_4,i)\dots(a_p-c_k,i)(a_p-c_{k+1},i)m_l\in\tau$) 
with $k\geq2$ and $m_l=(d-c_2,r)$ 
where $d=\text{origin}(m_{l-1})$. This would
correspond to the case where one of the descendants of $g(a_p)$
belongs to $\mc{S}(\psi(b_j))$, and we have then played in the
substituted arena. In this case, we have
$t[a_p]_0[g(a_p)]_0m_0(a_p-c_2,i)(a_p-c_3,i)m_1m_2\dots(a_p-c_k,i)(a_p-c_{k+1},i)m_l
(d-c_3,r)\in\mc{P}_{\XX}(A\fl B)$, so
$t[a_p]_0[g(a_p)]_0m_0(a_p-c_2,i)(a_p-c_3,i)m_1m_2\dots(a_p-c_k,i)(a_p-c_{k+1},i)m_l
(d-c_3,r)(a_p-c_{k+2},i)\in\sigma$ by totality
(because $(a_p-c_{k+2},i)$, if it exists, is the only son of
$(a_p-c_{k+1},i)$). So $\breve{t}[g(a_p)]_0[a_p]_0(a_p-c_2,i)m_0m_1
(a_p-c_3,i)(a_p-c_4,i)\dots(a_p-c_{k+2},i)(d-c_3,r)\in\tau$, and, by the same arguments,
$\breve{t}[g(a_p)]_0[a_p]_0(a_p-c_2,i)m_0m_1(a_p-c_3,i)(a_p-c_4,i)\dots(a_p-c_{k+2},i)
(d-c_3,r)(d-c_4,r)(a_p-c_{k+3},i)\in\tau$, etc.
Finally, as $k\geq2$, we necessarily come to a contradiction,
namely looking for a son of $c_N$, which does not have any by
construction.

\bigskip

We have proved that if $a_p$ appears $n$ times in $\mc{S}(b_j)$
then $g(a_p)$ appears $n$ times in $\in\mc{S}(\psi(b_j))$. We
still should show that if $a_p$ is decorated $n$ times by $X_j$
then $g(a_p)$ is decorated $n$ times by $X_j$. But actually we see
immediately that the arguments are exactly the same : here, it
suffices to consider the plays $s_1$ and $s_2$ with a new function
$\theta'$ which instantiates each variable by $\bot$, except $X_j$
which is instantiated by $H$. By lemma \ref{unif} we still obtain
a play from $\sigma$ or $\tau$, and we can follow the preceding
proof without any trouble.\cqfd

\subsection{Isomorphisms of types}

Thanks to the preceding result, and to the fact that we have a model of $\lambda\mu2$, we are able to characterize precisely 
type isomorphisms in this system.
The equational system we want to establish for type isomorphisms in $\lambda\mu2$ has been presented on figure~\ref{isof}.

On the grammar of $\lambda\mu2$ types, we consider: 
\begin{itemize}
\item products of arity $n$:
$\overset{n}{\underset{i=1}{\prod}} M_i=((M_1×
M_2)×\dots)× M_n$ ($\overset{n}{\underset{i=1}{\prod}} M_i=\top$ if $n=0$)
\item disjunctions of arity $n$:
$\overset{n}{\underset{i=1}{\ppar}} M_i=((M_1\ppar
M_2)\ppar\dots)\ppar M_n$ ($\overset{n}{\underset{i=1}{\ppar}} M_i=\bot$ if $n=0$)
\item quantifications of arity
$n$: $\overrightarrow{\forall X}_M=\forall X_{i_1}\dots\forall
X_{i_n}$ if $M=\{i_1,\dots,i_n\}$.
\end{itemize}

Inspired by the work of Roberto Di Cosmo on system F types\cite{isotypes}, we define normal forms:

\begin{Def}[canonical form]
A second order type $N$ is called a \textbf{canonical form} if it
is written
$N=\overset{n}{\underset{i=1}{\prod}} \overrightarrow{\forall
X}_{M_i}.N_i\fl\alpha_i$ with $\alpha_i=\overset{m}{\underset{j=1}{\ppar}}X_{k_j}$ and $N_i$ canonical form.
\end{Def}

\begin{Lemme}
Let $A$ be a type in $\lambda\mu2$. There exists a canonical form
$A'$ such that $A\simeq_\varepsilon A'$.
\end{Lemme}

\textsc{Proof :}
Because of the associativity of $×$, $\ppar$ and $\forall$ in $\eq$, we can restrict ourselves to products, disjunction and 
quantifications of arity $n$.
Then, modulo $\alpha$-equivalence, canonical forms are the normal forms of the
following rewriting system:

\[\begin{array}{c@{\qquad}c}
(A× B)\ppar C\reec (A\ppar B)× (B\ppar C)&A\ppar\bot\reec A\\
(A\fl B)\ppar C\reec A\fl(B\ppar C)&\bot\ppar A\reec A\\
A\fl(B× C)\reec (A\fl B)×(A\fl C)&\top\ppar A\reec \top\\
A\fl(B\fl C)\reec (A× B)\fl C&A\ppar\top\reec\top\\
(\forall X.A)\ppar B\reec\forall X.(A\ppar B)&A× \top\reec A\\
\forall X.(A× B)\reec(\forall X.A)×(\forall X.B)&\top× A\reec A\\
A\fl\forall X.B\reec\forall X.(A\fl B)&A\fl\top\reec \top\\
\forall X.\top\reec\top&
\end{array}\]

This rewriting system is coherent with $\eq$: this means that if
$A\reec A'$ then $A\eq A'$. To show that this system terminates, we
define a function $\psi$ which associates to each second order
type $A$ a natural number $\psi(A)\geq2$:
\begin{gather*}
\psi(A× B)=\psi(A)+\psi(B)+1\\
\psi(\forall X.A)=2\psi(A)\\
\psi(A\fl B)=\psi(A)\psi(B)+1\\
\psi(A\ppar B)=2^{\psi(A)\psi(B)}\\
\psi(\top)=\psi(\bot)=\psi(Y)=2\\
\end{gather*}
where $Y$ stands for any type variable.

For each rewriting rule $A\reec A'$, we have $\psi(A)>\psi(A')$.\cqfd

\begin{Prop}
If $A$ and $B$ are two types built on the grammar of $\lambda\mu2$ such that $A^*$ and
$B^*$ are isomorphic, then $A\simeq_\varepsilon B$.
\end{Prop}

\textsc{Proof:} In this proof we are interested by the hyperforest structure of $A^*$ and $B^*$ rather than 
their arborescence.

Let $g$ and $\psi$ be the bijections which
characterize the isomorphism between $A^*$ and $B^*$. Suppose that
$A$ and $B$ are already in a canonical form, we will show that
these two forms are equal modulo $\eq$ by induction on the structure of
$A^*$:
\begin{minilist}
\item If $A^*$ is empty, then $B^*$ is empty and $A^*\eq B^*$.
\item If $A^*$ is a tree such that no hyperedge has the root
as target, then $B^*$ is a tree such that no hyperedge has the
root as target. Then $A\eq A'\fl(X_{i_1}\ppar\dots X_{i_n})$ 
(this is indeed the only normal form which can be interpreted by such a
tree) with $X_i$ free type variable and we have in this case, because of the bijection $g$ which sends $\DD_A$ to $\DD_B$,
$B\eq B'\fl(X_{i_{\sigma(1)}}\ppar\dots X_{i_{\sigma(n)}})$ where $\sigma$ is a permutation of $\{1,\dots,n\}$.
We obtain
$(A')^*$ (resp. $(B')^*$) by suppressing the root from $A^*$ (resp
from $B^*$), so $(A')^*$ and $(B')^*$ are isomorphic. Finally, by
induction hypothesis, $A'\simeq_\varepsilon B'$, so
$A\simeq_\varepsilon B$.
\item if $A^*$ is a tree (whose root is denoted $r$) with some
hyperedges $b_1,\dots,b_n$ such that $\mc{T}(b_i)=r$ for
$i\in[1,n]$, then $B^*$ is also a tree (whose root is denoted
$r'$) with some hyperedges $b'_1,\dots,b'_n$ such that
$\mc{T}(b'_j)=r$ for $j\in[1,n]$. Then we have
$A\simeq_\varepsilon\forall X_1\dots\forall X_n.A'$ (this is the
only possible representation by a normal form) and
$B\simeq_\varepsilon\forall X'_1\dots\forall X'_n.B'$, where each
$X_k$ is associated with some $b_i$ and each $X'_k$ is associated
with some $b'_j$ (the variables $X_i$ and $X'_i$ are chosen fresh
with respect to other free variables already occurring in $A$ and
$B$). By $\alpha$-renaming, we can choose the variables $X'_k$
such that: if $X_k$ is the variable associated to the hyperedge
$b_i$, then the variable associated to $\psi(b_i)$ is $X_k$.
$(A')^*$ (resp. $(B')^*$) is obtained from $A^*$ (resp. $B^*$) by
suppressing all hyperedges $b_1,\dots,b_n$ (resp.
$b'_1,\dots,b'_n$) and by decorating with $X_i$ each node $c$ such
that $c\in\mc{S}(b_i)$ (resp. $c\in\mc{S}(b'_i)$). By using the
property $\mc{S}\circ\psi=g\circ\mc{S}$, we see that $(A')^*$ and
$(B')^*$ are isomorphic, so $A'\simeq_\varepsilon B'$ and finally
$A\simeq_\varepsilon B$ by commutativity of quantifications.
\item If $A^*$ contains $k\geq2$ trees, then $B^*$ also contains $k\geq2$
trees and $A$ is obtained from $k$ formulas $A_1,\dots,A_k$ by
using the connector $×$, so by associativity
$A\simeq_\varepsilon((A_1× A_2)× A_{k-1})× A_k$,
where each $A_i^*$ is a tree of $A^*$ (this is indeed the only
representation by a normal form). In the same way, we have
$\simeq_\varepsilon((B_1× B_2)× B_{k-1})× B_k$
where each $\llbracket B_i\rrbracket$ is a tree of $\llbracket
B\rrbracket$. As $A^*$ and $B^*$ are isomorphic, one can find a
permutation $\phi$ of the trees of $A^*$ such that, for every
$1\leq i\leq k$, $A_{\phi(i)}^*$ and $B_i^*$ are isomorphic. By
induction hypothesis, this implies $A_{\phi(i)}\simeq_\varepsilon
B_i$, so by commutativity of $×$ we have $A\simeq_\varepsilon
B$.
\end{minilist}\cqfd

\begin{Th}
Two formulas $A$ and $B$ are isomorphic in $\lambda\mu2$ if and only if
$A\simeq_\varepsilon B$ (with $\eq$ defined on figure~\ref{isolm2} p. 4).
\end{Th}

\textsc{Proof:} If
$A\simeq_\varepsilon B$ then $A$ and $B$ are isomorphic in the
$\lambda\mu2$-calculus: to prove it we just have to give a couple of terms realizing the isomorphism. As an example we give
the isomorphism between $A\ppar(B\ppar C)$ and $(A\ppar B)\ppar C$:
$$\begin{cases}\vdash t:A\ppar(B\ppar C)\fl(A\ppar B)\ppar C&\\
\vdash u:(A\ppar B)\ppar C\fl A\ppar(B\ppar C)&\end{cases}$$
with
$$t=\lambda x^{A\ppar(B\ppar C)}.\mu(\alpha_2^{A\ppar B},\beta^C_1).[\alpha_2]\mu(\alpha_0^A,\alpha_1^B).[\alpha_1,\beta_1]\mu\beta_0^{B\ppar C}.[\alpha_0,\beta_0]x$$
and $$u=\lambda x^{(A\ppar B)\ppar C}.\mu(\alpha_1^A,\alpha_2^{B\ppar C}).[\alpha_2]\mu(\beta_1^B,\beta_0^C).[\alpha_1,\beta^1]\mu\alpha_0^{A\ppar B}.[\alpha_0,\beta_0]x$$

For the other implication, suppose there are two terms $u:A\fl B$
and $v:B\fl A$ such that $u\circ v=id_B$ and $v\circ u=id_A$. In
the uniform model, their respective interpretations $\sigma_u$ and
$\sigma_v$ are such that $\sigma_v;\sigma_u=id_B$ and
$\sigma_u;\sigma_v=id_A$. We then have a game isomorphism between
the arenas $A^*$ and $B^*$, so $A^*$ and $B^*$ are isomorphic, so
that $A\simeq_\varepsilon B$.\cqfd

\begin{Corollaire}
If we consider the system $\lambda\mu2'$ obtained by suppressing the constructors $[\alpha,\beta]t$ and 
$\mu(\alpha^A,\beta^B).t$ from the grammar of terms (as well
as their associated inference rules and reduction rules) and the constructor $\ppar$ from the grammar of types, then type isomorphisms 
in $\lambda\mu2'$ are characterized by the equational system $\simeq_{\varepsilon}'$ given on figure~\ref{isof}.
\end{Corollaire}

\begin{figure}[h!]
\[\begin{array}{c@{\quad}c@{\quad}c}
A× B\eq' B× A& A×\top\eq' A& \forall X.\forall Y.A\eq'\forall Y.\forall X.A\\
A× (B× C)\eq'(A× B)× C&\forall X.\top\eq'\top& \forall X.(A× B)\eq'\forall X.A×\forall X.B\\
A\fl (B\fl C)\eq'(A× B)\fl C& \top\fl A\eq' A& A\fl \forall X.B\eq'\forall X.(A\fl B)\\
A\fl (B× C)\eq'(A\fl B)×(A\fl C)& A\fl\top\eq'\top& \qquad\quad\text{\small if $X$ does not appear free in $A$}

\end{array}\]
\caption{Equational system for type isomorphisms in $\lambda\mu2'$ and in system F}
\label{isof}
\end{figure}

\textsc{Proof:} As this new system is included in $\lambda\mu2$, our
model is necessarily also a model of $\lambda\mu2'$. Thanks to the
theorem~\ref{isos}, we only need to check that: if $A$ and $B$ are two
types built on the grammar of $\lambda\mu2'$ such that $A^*$ and $B^*$
are isomorphic, then $A\simeq_{\varepsilon}' B$ (and also that there
exist terms in $\lambda\mu2'$ realizing the equations of
$\simeq_{\varepsilon}'$).  \cqfd

Finally, we also recover the results of Roberto Di Cosmo~\cite{isotypes}:

\begin{Corollaire}
Type isomorphisms for system F are given by the equational system $\simeq_{\varepsilon}'$.
\end{Corollaire}

\textsc{Proof:} System F is $\lambda\mu2'$ where we have suppressed
the constructors $[\alpha]t$ and $\mu\alpha^A.t$. The type system is
the same as the one of $\lambda\mu2'$, so we conclude by the same
reasoning as above.\cqfd

\section{Conclusion and further directions}

Game semantics has allowed us not only to retrieve semantically the results of Robert Di Cosmo concerning type isomorphisms in system F, 
but also to give a characterization of type isomorphisms for an extension of this system to a calculus with control.

However, although it has led us to a good characterization of type
isomorphisms, uniformity is still a very ad hoc property. We suspect that there is an analogy between
uniformity and innocence: maybe this link could help us understanding uniformity better.

Concerning extension, one can easily adapt our model to a calculus with a fixpoint operator: it only requires to 
enrich the structure of the model with an complete partial order on
strategies (which will simply be the inclusion of strategies). Note that it is possible only because we did not require the totality of strategies
in our model (as we were able to prove the totality of strategies realizing game isomorphisms). 
Moreover, our strategies are considered here as innocent,
but in fact it would suffice to have them well-threaded and
visible: in this case, our model would appear to be a good
candidate to interpret second-order Idealized Algol~\cite{iagames}
and to characterize type isomorphisms in this system by a similar
equational system. Many other programming features may also be
treated using this kind of model, like non-determinism,
probabilities, concurrency, \dots Taking the same approach for
Curry-style system F or for ML should also be possible, but it
will require to build a brand new model, because the model we give here does not suit these systems.

Finally, our approach needs to be tested on retractions, \ie\ in
the case where we have $f:A\fl B$ and $g:B\fl A$ such that $f\circ
g=id_B$ but not necessary $g\circ f=id_A$. Retractions can be
understood as a subtyping notion, and they are useful when dealing
with code reuse (because, schematically, it is no big deal to use
a function having a "more liberal" type than the one we expected).
In this domain few results~\cite{retract2, retract1,retract3} are
known, even in a propositional setting, so game semantics may give
a new enlightenment on this problem.

\bibliographystyle{alpha}
\bibliography{ll}

\newpage

\appendix

\section{Soundness of the interpretation of $\lambda\mu2$ in a control hyperdoctrine} 

\setcounter{Th}{0}

\begin{Th}[soundness]
The interpretation of second-order $\lambda\mu$-terms in a control hyperdoctrine is sound: for any couple of terms $t,u$ such that $t=u$,
we have $\trad{t}=\trad{u}$.
\end{Th}

\textsc{Proof:} We prove successively the soundness of the
interpretation for every reduction rule.  Most of the required
equalities are already valid (with the same proof) in a control
category or in a hyperdoctrine.  That is why, in many cases, we will
only give a sketch of the proof: we do not detail the commutativity of
the diagrams and the validity of substitution lemmas.

\begin{itemize}

\item[($\top$)] It suffices to recall that $1\ppar_I\Delta$ is isomorphic to $1$, and $1$ is a terminal object.

\item[($\pi_i$)] As $d^{-1}_{A,B,\Delta}=(\pi_1\ppar_I\Delta,\pi_2\ppar_I\Delta)$, one has:
$$
\xymatrix{
\Gamma \flt{(\trad{t},\trad{u})}(A\ppar_I\Delta)×(B\ppar_I\Delta)\ar[r]^{d}\ar[dr]^{\id}&(A× B)\ppar_I\Delta\ar[r]^{\pi_1\ppar\Delta}\ar[d]_{(\pi_1\ppar\Delta,}^{\pi_2\ppar\Delta)}&A\ppar_I\Delta\\
&(A\ppar_I\Delta)×(B\ppar_I\Delta)\ar[ur]_{\pi_1}&
}
$$

\item[($×$)] This comes directly from $d^{-1}_{A,B,\Delta}=(\pi_1\ppar_I\Delta,\pi_2\ppar_I\Delta)$.

\item[($\beta$)] Let us introduce the \textbf{linear distributivity} $\ld:A×(B\ppar_I C)\flt{w×\id}(A\ppar_I C)× (B\ppar_I C)\flt{d}
(A× B)\ppar_I C$. 

The first step is to prove that the following diagram commutes:
$$
\xymatrix{
\Gamma\ar[d]_{(\id,\id)}\ar[r]^{g\ \ \ \ \ \ \ \ \ \ \ \ \ \ \ \ \ \ \ }&(B^A\ppar_I\Delta)×(A\ppar_I\Delta)\ar[r]^{\ \ \ d}&(B^A× A)\ppar_I\Delta\ar[r]^{\ \ \ \ \ ev\ppar\Delta}&B\ppar_I\Delta\\
\Gamma×\Gamma\ar[r]^{\id×\trad{u}\ \ \ \ \  \ \ \ }&\Gamma×(A\ppar_I\Delta)\ar[r]^{\ld}&(\Gamma× A)\ppar_I\Delta\ar[r]^{\trad{t}\ppar\Delta}&B\ppar_I\Delta\ppar_I\Delta\ar[u]_{B\ppar\nabla}
}
$$
with $g=(\Lambda(\trad{t});s^{-1},\trad{u})$.

Then, one has to prove the following substitution lemma (by induction on $t$):
$$(id,id);id×\trad{u};\ld;\trad{t}\ppar_I\Delta;\id\ppar_I \Delta=\trad{t[u/x]}$$

\item[($\eta$)] What we want to show is: 
$$\Lambda((\trad{t},\pi_1);d;\epsilon\ppar_I\Delta);s^{-1}=\trad{t}$$
And this precisely means:
$$
\xymatrix{
(B\ppar_I\Delta)^A× A\ar[r]^{\ \ \ \ \ eval}&B\ppar_I\Delta\\
\Gamma× A\ar[u]^{(\trad{t};s)×\id}\ar[ur]_{\ \ \ \trad{t}×\id;d;\epsilon\ppar\Delta}
}
$$
which is straightforward.

\item[($\mu$)] For ($\mu^\fl$), the substitution lemma to prove by induction on $t$ is (modulo some trivial morphisms):
$$\id\times\trad{u};\trad{t}\times\id;d;\epsilon\ppar\Delta=\trad{t[[\beta](-)u/[\alpha](-)]}$$
The most significant case of the induction is the case where $t=[\alpha]t'$: it consists in this case in proving that (informally) 
$\nabla;(d;\epsilon\ppar\Delta)=(d;\epsilon\ppar\Delta);\nabla$: it can be done using 
curryfication and decurryfication (using the fact that $B^v$ is central for any morphism $v$).

For ($\mu^\forall$), first remark that:
$$(\kappa^{-1}(\trad{t};p))[U^n,B]=(\trad{t})_{U^n}^{U^{n+1}}[U^n,B];\kappa^{-1}(p)[U^n,B]$$
Then the substitution lemma to prove by induction on $t$ is (modulo
some trivial morphisms):
$$\trad{t};\kappa^{-1}(p)[U^n,B]=\trad{t[[\beta](-)\{B\}/[\alpha](-)]}$$
The most significant case of the induction is the case where
$t=[\alpha]t'$: it consists in this case in proving that (informally)
$\nabla;\kappa^{-1}(p)[U^n,B]=\kappa^{-1}(p)[U^n,B]\ppar\kappa^{-1}(p)[U^n,B];\nabla$:
this is ensured by focality of $\kappa^{-1}(p)[U^n,B]$ (which is due
to the centrality of this morphism).

The rules ($\mu^×$) and ($\mu^\ppar$) can be treated similarly.

\item[($\rho$)] The three rules $(\rho^\mu)$, $(\rho^\ppar)$ and $(\rho^{\bot})$ work on the same scheme. If we look for example to $(\rho^\mu)$,
one has:
$$\begin{array}{cl}
\trad{[\alpha']\mu\alpha^A.t}&=\Gamma\flt{\trad{t}}\bot_I\ppar_I A\ppar_I\Delta\flt{\cong}A\ppar_I\Delta\flt{w\ppar\Delta}\Delta\ppar_I\Delta\flt{\nabla}\Delta\flt{\cong}\bot_I\ppar_I\Delta\\
&=\Gamma\flt{\trad{t}}\bot_I\ppar_I A\ppar_I\Delta\flt{c;\Delta}\bot_I\ppar_I\Delta\\
&=\trad{t[\alpha'/\alpha]}\end{array}$$

\item[($\theta$)] If we focus on $(\theta^\mu)$ for example, note that the derivations we want to compare are the following ones:
\bec
\AxiomC{$\sigma$}
\UnaryInfC{$\XX,\Gamma\vdash t:A\barre\alpha:A,\Delta$}
\DisplayProof\qquad
\AxiomC{$W(\sigma)$}
\UnaryInfC{$\XX,\Gamma\vdash t:A\barre\alpha:A,\alpha':A,\Delta$}
\UnaryInfC{$\XX,\Gamma\vdash [\alpha]t:\bot\barre\alpha:A,\alpha':A,\Delta$}
\UnaryInfC{$\XX,\Gamma\vdash \mu\alpha^A.[\alpha]t:A\barre\alpha':A,\Delta$}
\DisplayProof\eec
where $W(\sigma)$ is obtained from $\sigma$ by a weakening lemma.

Hence, we have to compare $$\Gamma\flt{\trad{t}} A\ppar_I A\ppar_I\Delta$$ with 
$$\trad{\mu\alpha^A.[\alpha]t}=\Gamma\flt{\trad{W(t)}}A\ppar_I (A\ppar_I A\ppar_I\Delta)\flt{w}(A\ppar_I A\ppar_I\Delta)\ppar_I(A\ppar_I A\ppar_I\Delta)\flt{\nabla}A\ppar_I A\ppar_I\Delta$$

By induction on the proof $\sigma$, one can show that actually 
$\trad{W(t)}=\Gamma\flt{\trad{t}}A\ppar_I (A\ppar_I\Delta)\flt{w}A\ppar_I (A\ppar_I A\ppar_I\Delta)$. 
Besides, as $\alpha\notin FN(t)$, one can also prove by induction on $\sigma$ that 
$\trad{t}=\Gamma\flt{\trad{t'}}A\ppar_I \Delta\flt{w}A\ppar_I (A\ppar_I\Delta)$ for some $t'$.

Hence, we have two weakening composing with contractions that give us the identity, and $\trad{t'}$ is composed with the last weakening to give
us $\trad{t}$.

\item[($\beta2$)] 
$\trad{\Lambda X.t}\{B\}=\trad{t}[U^n,B]$, so one only needs to prove a substitution lemma:
$$\trad{t}[U^n,B]=\trad{t[B/X]}$$
which is ensured by the fact that specialization functors are strict functors of control categories (indeed, they are strict functors of pre-control
categories, and the strictness for other structural morphisms is automatically true).

\item[($\eta2$)] This is immediate, since:
$$\kappa((\kappa^{-1}(\trad{t};p))[U^{n},U]);p^{-1}=\trad{t}$$

\end{itemize}

\end{document}